\documentclass[final]{siamltex}
\usepackage[utf8x]{inputenc}
\usepackage{amsmath,amssymb,graphicx}
\usepackage{geometry}

 \newtheorem{remark}[theorem]{Remark}
 \newtheorem{conjecture}[theorem]{Conjecture}

\title{Numerical Study of the semiclassical limit of the 
Davey-Stewartson II equations}%Parallel computing for the study of the Davey-Stewartson II equation in small dispersion limit.}
\author{C.~Klein\thanks{Institut de Math\'ematiques de Bourgogne,
		Universit\'e de Bourgogne, 9 avenue Alain Savary, 21078 Dijon
		Cedex, France
    ({\tt christian.klein@u-bourgogne.fr})}
\and
K.~Roidot\thanks{Fakult\"{a}t f\"{u}r Mathematik, Universit\"{a}t Wien - Wien Rossau,
Oskar-Morgenstern-Platz 1, 1090 Wien, \"{O}sterreich 
    ({\tt kristelle.roidot@univie.ac.at})}
}

\date{\today}

\begin{document}

\maketitle

\begin{abstract}
    We present the first detailed numerical study of the 
    semiclassical limit of the Davey-Stewartson II equations both for the focusing and the defocusing 
    variant. We concentrate on rapidly decreasing initial data with a single 
    hump. The formal limit of these equations for vanishing 
    semiclassical parameter $\epsilon$, the semiclassical equations, 
    are numerically integrated up 
    to the formation of a shock. The use of parallelized algorithms 
    allows to determine the critical time $t_{c}$ and the critical solution for 
    these $2+1$-dimensional shocks. It is shown 
    that the solutions generically break in isolated 
    points similarly to the case of the $1+1$-dimensional cubic 
    nonlinear Schr\"odinger equation, i.e., cubic singularities in the 
    defocusing case and square root singularities in the focusing case. 
    For small values of $\epsilon$, 
    the full Davey-Stewartson II equations are integrated for the 
    same initial data up to the critical time $t_{c}$. The scaling in $\epsilon$ of the 
    difference between these solutions is found to be 
    the same as in the $1+1$ dimensional case, proportional to 
    $\epsilon^{2/7}$ for the defocusing case and proportional to 
    $\epsilon^{2/5}$ in the focusing case. 
    
    We document the Davey-Stewartson II solutions for small 
    $\epsilon$ for times much 
    larger than the critical time $t_{c}$. 
    It is shown that zones of rapid modulated oscillations are formed 
    near the shocks of the solutions to the semiclassical equations. 
    For smaller $\epsilon$, the oscillatory zones become smaller 
    and more sharply delimited to lens shaped regions. Rapid 
    oscillations are also found in the focusing case for initial data 
    where the singularities of the solution to the semiclassical equations 
    do not coincide. If these singularities do coincide, 
    which happens when the initial data are symmetric with respect to 
    an interchange of the spatial coordinates, no such zone is 
    observed. Instead the initial hump develops into a blow-up of the 
    $L_{\infty}$ norm of the solution. We study the dependence of the 
    blow-up time on the semiclassical parameter $\epsilon$.  
 \end{abstract}

\section{Introduction}

Nonlinear Schr\"odinger (NLS) equations have many applications, e.g.~in 
hydrodynamics, plasma physics and nonlinear optics where they can be 
used to model the amplitude modulation of weakly nonlinear, strongly 
dispersive waves. They can be cast into the form
\begin{equation}
i \epsilon  \Psi_t + \frac{\epsilon^2}{2} \Delta \Psi -\frac{\rho}{\sigma}\left|\Psi \right|^{2\sigma}\Psi = 0, \,\, x \in \mathbb{R}^d
\,\, ,t \in \mathbb{R},
\label{NLSeqgen}
\end{equation}
where $\Psi$ is a complex valued function, $\Delta$ is the 
Laplace operator in $d$ dimensions, and $\epsilon\ll 1$ is a 
real positive parameter. 
The exponent $1<2\sigma<\infty$ represents the power of the nonlinearity, 
and the parameter $\rho = \pm1$ determines whether this nonlinearity 
has a focusing ($\rho = -1$) or a defocusing ($\rho =  1$) effect.
The cubic (i.e., $\sigma=1$) $1+1$-dimensional NLS is known to be 
completely integrable \cite{ZS72} which implies that many exact 
solutions as solitons and breathers can be given in explicit form.

Since the parameter $\epsilon$ in (\ref{NLSeqgen}) has the 
same role as the Planck constant $\hbar$ in the classical 
Schr\"odinger equation in  quantum mechanics, the limit 
$\epsilon\to0$ is also referred to as the \textit{semiclassical limit}. This 
limit is mathematically challenging since it complements the well 
known difficulties of the semiclassical limit in quantum mechanics 
(highly oscillatory functions)
with the nonlinearity of the NLS equation (\ref{NLSeqgen}), which 
typically leads to strong gradients. Note that 
the parameter $\epsilon$ can be introduced into the dimensionless 
form of the NLS equation (eq.~(\ref{NLSeqgen}) with 
$\epsilon=1$) via a rescaling of the coordinates of the form 
$x\to x/\epsilon$, and $t\to t/\epsilon$. 
Thus the introduction of a small $\epsilon$ can be seen as 
equivalent to studying the solution of a Cauchy problem for initial data with support on scales of order 
$1/\epsilon$ for long times of order $1/\epsilon$. 
With the well known Wenzel-Kramer-Brillouin (WKB) ansatz, also known 
as Madelung transform \cite{Mad27} in this context,
\begin{equation}
\Psi = \sqrt{u} e^{i S/ \epsilon}, 
\label{inidisp}
\end{equation}
for $u$ and $S$ being real valued functions, the initial value problem $\Psi(x,0)=\Psi_{0}(x)$ for the  NLS equation 
(\ref{NLSeqgen}) is equivalent to
\begin{equation}
\left\{ \begin{array}{ccc} 
 u_t + \mbox{div}(u\cdot w) & = & 0 \\
 w_t  + w\cdot \nabla w +\rho \nabla u^{2\sigma}  & = & 
 \frac{\epsilon^{2}}{2} \nabla
 \left(\frac{\Delta (\sqrt{u})}{\sqrt{u}}            \right),\\
  u(x,0)=u_0(x), \,\, w(x,0)=w_0(x)
 \label{dnlseps}
\end{array}
\right.
\end{equation}
where $w=\nabla S$. This form of the NLS equation is also referred to as 
the \textit{hydrodynamic form}\footnote{Note that in (\ref{dnlseps})  the 
term with $\epsilon^{2}$ is also referred to as \textit{quantum 
pressure} in the context of the linear Schr\"odinger equation.} due to its similarity with the 
compressible
Euler equation which are obtained in the formal limit $\epsilon=0$,
\begin{equation}
\left\{ \begin{array}{ccc} 
 u_t + \mbox{div}(u\cdot w) & = & 0 \\
 w_t  +  w\cdot \nabla w +\rho \nabla u^{2\sigma} & = & 0 \\
  u(x,0)=u_0(x),\,\, w(x,0)=w_0(x)
 \label{dnls1}
\end{array}
\right.
\end{equation}
and that we call the \emph{semiclassical NLS system} in the 
following.

In the defocusing $1+1$-dimensional case ($\rho=1$), this system is hyperbolic, and the corresponding initial value problem is well-posed.
It describes an isentropic gas which can develop a gradient catastrophe 
in finite time similar to the shock formation in solutions to the 
Hopf equation $u_t + uu_x = 0$. The generic behavior of the solutions 
at the critical points is given by a cubic singularity. The 
defocusing NLS equation for finite small $\epsilon$ can be seen as a 
dispersive regularization of the system (\ref{dnls1}).  Its solutions 
have rapid modulated oscillations in 
the vicinity of the gradient catastrophe of solutions of the system 
(\ref{dnls1}) for the same initial data called \emph{dispersive shocks}. 
Using the complete integrability, Jin, Levermore and McLaughlin \cite{JSLDMD} gave 
an asymptotic description of the oscillatory zone in solutions to the defocusing
cubic NLS. The situation is 
very similar to the Korteweg-de Vries equation (KdV) for which a 
complete asymptotic theory of dispersive shocks
was developed in \cite{LL,Ven,DVZ}. A first numerical implementation 
of the asymptotic description for KdV was presented in \cite{GK}. No such theory 
exists for non-integrable cases. Before the critical time $t_{c}$ of the 
semiclassical solution and in the exterior of the oscillatory zone, 
the solution of the system (\ref{dnls1}) for the same initial data 
gives an asymptotic description of the NLS solution for 
$\epsilon\to0$. The behavior for $t\sim t_{c}$ has been addressed in 
\cite{DGK13} for a large class of two-component systems including 
$1+1$-dimensional NLS equations, also for non-integrable cases. It is 
conjectured that the solution near $t_{c}$ to such equations is given 
in order $\epsilon^{2/7}$ by a rescaled unique solution to an 
ordinary differential equation (ODE), the second 
equation in the Painlev\'e I hierarchy (PI2). The conjecture in 
\cite{DGK13} essentially states that the formation of a dispersive 
shock close to $t_{c}$ is equivalent to the corresponding situation 
in KdV solutions for which Dubrovin presented a conjecture in 
\cite{Dub06} for a large class of scalar equations containing KdV. 
The conjecture was 
proven for KdV with Riemann-Hilbert techniques by Claeys and Grava in 
\cite{CG}. 
The existence of the conjectured PI2 solution regular on the whole 
real line was proven in \cite{CV07}. Note that the conjecture in 
\cite{DGK13} postulates a universality property of hyperbolic 
dispersive shocks in the sense that the solutions near the break-up 
of the dispersionless or semiclassical limit are asymptotically given 
in terms of the PI2 transcendent for  a large class of 
dispersive partial differential equations (PDEs) and a large class of initial data.

The situation in the $1+1$-dimensional focusing case ($\rho=-1$) 
is more involved since the system (\ref{dnlseps}) is elliptic which 
implies an ill-posed Cauchy problem. The generic singularities 
forming in solutions to (\ref{dnls1}) for analytic initial data are 
of elliptic umbilic type as was shown in \cite{DGK}. In the 
semiclassical limit, zones of rapid modulated oscillations again form 
near such a singularity. In contrast to the defocusing NLS equation, 
an asymptotic description of the oscillatory zone has been given only 
for certain classes of initial data in \cite{JLM,KMM,TVZ}. The 
behavior close to the critical time $t_{c}$ of the semiclassical 
system (\ref{dnls1}) was addressed in \cite{DGK} for the 
integrable cubic case, and has been recently generalized to a large 
class of two-component systems including generalized NLS equations in 
\cite{DGK13}. It is conjectured that the NLS solution for $t\sim 
t_{c}$ is given by a rescaled \emph{tritronqu\'ee} solution 
\cite{Bou13} of the 
Painlev\'e I equation in order $\epsilon^{2/5}$. A partial proof of 
this conjecture for the integrable cubic case was given in 
\cite{BertTob}.

The questions discussed above for the $1+1$-dimensional case are 
mostly wide open for $2+1$-dimensional settings.  First 
numerical attempts in this direction were presented in \cite{KSM,KR,Roid1}. 
In \cite{dkpsulart}  a detailed numerical study of the formation of 
dispersive shocks in  the Kadomtsev-Petviashvili (KP) equation (a $2 
+ 1$-dimensional generalization of KdV) was given. 
Here we present a numerical study of the 
semiclassical limit of NLS equations in $2+1$ dimensions. Since the 
strongest results in the asymptotic description  of $1+1$-dimensional 
PDEs have been obtained with techniques from the theory of integrable 
systems, we concentrate here on an integrable nonlocal NLS equation 
in $2+1$ dimensions, the Davey-Stewartson (DS) system,
\begin{equation}
\label{DSgen}
\begin{array}{ccc}
i\epsilon 
\Psi_{t}+\epsilon^{2} \left(\Psi_{xx}-\alpha \Psi_{yy}\right)+2\rho\left(\Phi+\left| \Psi \right|^{2}\right)\Psi & = & 0,
\\
\Phi_{xx}+\beta \Phi_{yy}+2\left| \Psi \right|_{xx}^{2} & = & 0,
\end{array}
\end{equation}
where $\alpha, \beta$ and $\rho$ take the values $\pm1$, 
 $\epsilon\ll1$ is a small dispersion
parameter, and $\Phi$ is a mean field.
These systems   describe the amplitude modulation of weakly 
nonlinear, strongly dispersive $2 + 1$-dimensional waves in 
hydrodynamics and nonlinear optics, and appear also in plasma physics to describe the evolution of a plasma under the action of a magnetic field.
They have been classified in \cite{GS} as elliptic-elliptic, 
hyperbolic-elliptic, elliptic-hyperbolic and hyperbolic-hyperbolic, 
according to the signs of $\alpha$ and $\beta$. The DS system is known to 
be completely integrable  when $\alpha=\beta$ \cite{AH75}. The case 
$\alpha=\beta=-1$ is also called DS I, the case $\alpha=\beta=1$, DS 
II. We concentrate here on the latter where
the mean field $\Phi$ is governed by an elliptic equation which can 
be solved uniquely with some fall off condition at infinity. Then 
%\begin{equation}
$\Phi = \mathcal{M}\left( |\Psi|^2  \right)$, where the operator $\mathcal{M}$ is defined in Fourier space by  
$$ \widehat{\mathcal{M}(f)} = \frac{-2 k_x^2}{k_x^2 + k_y^2} 
\widehat{f} (k_x, k_y),$$ where $k_x$ and $k_y$ represent the wave 
numbers, in the $x$ and $y$ directions, respectively, and where 
$\hat{f}$ denotes the Fourier transform of a function $f$. With the 
operators $\mathcal{D}_{\pm}=\partial_{x}^{2}\pm \partial_{y}^{2}$, 
the DS II equations can be written in the form 
\begin{equation}
    i\epsilon \Psi_{t}+\epsilon^{2}\mathcal{D}_{-} \Psi-2\rho 
    (\mathcal{D}_{+}^{-1}\mathcal{D}_{-}|\Psi|^{2})\Psi=0
    \label{DSnonlocal},
\end{equation}
where $\mathcal{D}_{+}^{-1}$ is defined as above by its Fourier 
symbol. Thus DS II can be seen as an NLS equation with a 
nonlocal (due to the operator $\mathcal{D}_{+}^{-1}$) cubic nonlinearity. 
Similarly to the NLS equation, 
the latter admits a focusing ($\rho=-1$) and a defocusing $\rho=1$ version.
%
%the minus sign of $\rho$ in (\ref{DSnonlocal}) is referred to as the focusing case,  
%and the plus sign as the defocusing case.

Note, however, that the operator $\mathcal{D}_{-}$ leads to a 
different dynamics compared to the standard NLS equation 
(\ref{NLSeqgen}) with a Laplace operator.  
Therefore many PDE techniques successful for NLS could not be applied 
to the DS II equation. Using integrability,
Fokas and Sung \cite{FokS,Sun} studied the existence and long-time 
behavior of the solutions of the initial value problem for DS II (for 
$\epsilon=1$, and $\Psi(x,y,0)=\Psi_0$). They proved the following
\begin{theorem} \label{theosung}
If $\Psi_0$ belongs to the Schwartz space $\mathcal{S}(\mathbb{R}^2)$, 
then there exists  in the defocusing case ($\rho=1$) a unique global 
solution $\Psi$  to DS II such that $t\mapsto \Psi(\cdot,t)$ is a 
$C^\infty$ map from $ \mathbb{R}\mapsto \mathcal{S} (\mathbb{R}^2) $.
The same holds for the focusing case ($\rho=-1$) if
the initial data $\Psi_0 \in L_{q}$ for some $q$ with $1\leq q < 2$ 
have   a Fourier transform $\widehat{\Psi_0} \in L_{1} \cap L_{\infty}$ 
such that 
  $  \|  \widehat{\Psi_0} \|_{L_{1}} \|\widehat{\Psi_0}\|_{L_{\infty}}
    <\frac{\pi^{3}}{2}\left(\frac{\sqrt{5}-1}{2}\right)^{2}$. The 
    unique global solution $\Psi$ to DS II  satisfies the decay 
    estimate $\| \Psi(t) \|_{L_{\infty}} < \frac{const}{t}$.\\
Furthermore, if $\Psi_0$ belongs to the Schwartz space $\mathcal{S}(\mathbb{R}^2)$, then there is an infinite number of conserved quantities.  
\end{theorem}

The first in a hierarchy of conserved quantities are the wave energy $N := \int_{\mathbb{R}^2} |\Psi(x,y,t)|^2 dx dy$, the linear momentum $P:=\int_{\mathbb{R}^2} i \left( \Psi^{\ast} \nabla \Psi - \Psi \nabla \Psi^{\ast}
 \right) dx dy$, and the energy
\begin{align}
    E & := \frac{1}{2} \int_{\mathbb{R}^2} \bigg[ \epsilon^2|\Psi_x|^2 - \epsilon^2
|\Psi_y|^2  -\rho\left(|\Psi|^{4}-\frac{1}{2}\left(\Phi^{2}+(\partial_{x}^{-1}\Phi_y)^{2}\right) \right)
  \bigg] d x d y.\label{hamil}
 \end{align}

The smallness condition in Theorem \ref{theosung} indicates that 
there might be a blow-up, i.e., a loss of regularity with respect to 
the initial data, in solutions to the focusing DS II equations. 
In fact,  for 
focusing cubic NLS equations,
$2+1$ dimensions constitute the critical dimension where 
blow-up can occur. However due to the operator $\mathcal{D}_{-}$ in 
(\ref{DSnonlocal}) this cannot be directly generalized to DS II. Therefore it is important 
in this context
that  Ozawa gave an exact blow-up solution in \cite{Oza}. The 
solution is similar to  the well known lump solutions \cite{APPDS}, 
travelling solitonic wave solutions with an algebraic fall off at 
infinity.  Note that Theorem 1.1 does not hold for lumps since they 
are not in $L_{1}(\mathbb{R}^{2}$). It is thus not 
known whether there is generic blow-up for initial data not satisfying 
this condition, nor whether the condition is optimal. Numerical 
studies in \cite{MFP,KRM} indicate, however, that blow-up can 
occur in perturbations of the lump and the Ozawa solution and is thus 
a generic feature of solution to the focusing DS II. In fact it was conjectured 
in \cite{MFP} that generic localized initial data are either just radiated 
away to infinity or blow up for 
large $t$. 

Therefore it is not obvious whether dispersive shocks can be observed 
at all in focusing DS II systems, or whether the solutions blow-up 
directly. Numerically it was shown in 
\cite{KR,Roid1} that dispersive shocks can indeed be seen.
In the semiclassical limit ($\Psi= \sqrt{u} e^{i S/ \epsilon}, \,\, \epsilon \to 0$), DS II reduces to the following system 
\begin{equation}
\left\{
\begin{array}{ccc}
    S_t + S^{2}_{x} - S^{2}_{y} + 2\rho 
    \mathcal{D}_{+}^{-1}\mathcal{D}_{-}(u) & = & \frac{\epsilon^2}{2} \left(
    \frac{u_xx}{u} -  \frac{u_x^2}{u^2} -  \frac{u_yy}{u} +  \frac{u_y^2}{u}
    \right)
     \\
u_t + 2\left(S_{x}u \right)_x -2\left(S_{y}u\right)_y & = & 0 \\
\end{array}
\right.
\label{disDSeps},
\end{equation}
with the formal limit $\epsilon\to0$
\begin{equation}
\left\{
\begin{array}{ccc}
    S_t + S^{2}_{x} - S^{2}_{y} + 2\rho 
    \mathcal{D}_{+}^{-1}\mathcal{D}_{-}(u) & = & 0 \\
u_t + 2\left(S_{x}u \right)_x -2\left(S_{y}u\right)_y & = & 0 \\
\end{array}
\right.
\label{disDSs},
\end{equation}
which will be referred to as the \emph{semiclassical DS II system} in the 
following.
This system is not  integrable in the usual sense that it 
can be associated to a standard Riemann-Hilbert problem (RHP), but it 
is integrable in the sense of hydrodynamic reductions \cite{Kon07}. It 
is an open question whether it can be treated with the nonlinear RHP 
approach proposed in  \cite{MS06h1,MS07,MS09Tod} for 
dispersionless $2+1$-dimensional PDEs as the dispersionless KP (dKP) or 
the two-dimensional Toda equation in the long wavelength limit. 
It is also not clear whether it can be treated via 
infinite Frobenius manifolds as in \cite{CDM,Rai12} for these two PDEs.

Therefore we will present in this paper the first comprehensive 
numerical study of the  semiclassical limit of DS II equations. 
This is a highly nontrivial task already 
in $1+1$ dimensions, and will be even more so with an additional 
spatial dimension. To obtain the necessary resolution, we will use 
parallel computing.  Firstly we have to integrate the semiclassical DS 
II system up to a break-up of the solution, i.e, up to the formation 
of a singularity 
of the solution which is numerically extremely challenging. Since we want to study 
various scalings in $\epsilon$ for DS II solutions close to the 
critical time $t_{c}$, the latter has to be reliably identified. 
A careful numerical investigation  in \cite{dkpsulart} for the 
dKP equation allowed to study the small 
dispersion limit of KP solutions close to the break-up of the 
corresponding dKP solution. In particular, it could be shown that the 
difference between the KP and dKP solutions for the same initial data 
shows the same characteristic scaling in $\epsilon$ as the one-dimensional model (KdV/Hopf). 
The main technique used was asymptotic Fourier analysis as first 
applied  numerically in \cite{SSF} to trace singularities in the 
complex plane. In this framework the singularity of the real solution 
appears when one of the singularities in the complex plane hits the 
real axis. This method will also allow here 
to identify both the critical time $t_{c}$ of solutions to the semiclassical DS 
II system (\ref{disDSs}) and the critical solution.  We obtain the 
following\\
\begin{conjecture}
Consider rapidly decreasing smooth initial data in $L_{2}(\mathbb{R}^{2})$ with a single 
maximum. Then
\begin{itemize}
    \item  Solutions to the defocusing variant of the 
    semiclassical DS II equation (\ref{disDSs}) show the 
    same type of break-up as for the corresponding limit of the 
    $1+1$-dimensional NLS equation: the solutions have two break-up 
    points in each spatial direction (not necessarily on the 
    coordinate axes and at the same 
    time) which are generically of cubic type as for generic 
    solutions to the Hopf 
    equation.

    \item  Solutions of the focusing variant of the semiclassical DS 
    II equation (\ref{disDSs}) have 
    in general two break-up points of the same type as solutions of 
    the focusing $1+1$-dimensional NLS equation, a square root cusp 
    for each spatial direction. For initial data with a symmetry with 
    respect to an interchange of the spatial coordinates, these cusps 
    appear at the same time and location. 
\end{itemize}
\end{conjecture}

Secondly, the DS II equation is a purely 
dispersive equation. For such equations, the introduction of 
numerical dissipation has to be limited as much as possible, to avoid 
the suppression of dispersive effects. Therefore we use Fourier 
spectral methods for the spatial dependence of the solution as well.  
In addition, focusing NLS and DS 
equations are known to have a  \emph{modulational 
instability}, i.e., self-induced amplitude modulation of a continuous 
wave propagating in a nonlinear medium, which has dramatic consequences in numerics, see e.g.~\cite{ckkdvnls}, if not sufficient 
spatial resolution is used. 
Moreover an efficient time integrator of high accuracy is needed 
in order not to pollute the Fourier coefficients. As will be shown in 
the paper, the applied methods allow to solve the DS II equation and 
to identify the scaling in $\epsilon$ of the difference between DS II (\ref{disDSeps})
and semiclassical DS II (\ref{disDSs}) solution at the critical time. The main results of the study for 
rapidly decreasing (in both spatial directions) initial data with a 
single hump for the DS II equations can 
be summarized in the following\\
\begin{conjecture}
Consider rapidly decreasing smooth initial data in $L_{2}(\mathbb{R}^{2})$ with a single 
maximum. Then
\begin{itemize}
    \item  The difference of solutions to the defocusing 
    semiclassical DS II equation 
    (\ref{disDSs}) at the critical time $t_{c}$ and the solutions to the 
     defocusing   DS II equation for the same initial data and 
     different values of $\epsilon$ scales as $\epsilon^{2/7}$ as in 
     the case of the $1+1$-dimensional defocusing NLS equation. 

    \item  The difference of solutions to the focusing semiclassical 
    DS II equation 
    (\ref{disDSs}) at the critical time $t_{c}$ and the solutions to the 
     focusing   DS II equation for the same initial data and 
     different values of $\epsilon$ scales as $\epsilon^{2/5}$ as in 
     the case of the $1+1$-dimensional focusing NLS equation. 

    \item  For times $t\gg t_{c}$, solutions to the defocusing DS II 
    equation show for small $\epsilon$ zones of rapid modulated 
    oscillations in the vicinity of the critical points of the 
    solutions of (\ref{disDSs}) for the same initial data.

    \item  For times $t\gg t_{c}$ and initial data where the 
    two cusps do not appear at the same time and location, solutions to the focusing DS II 
    equation show for small $\epsilon$ zones of rapid modulated 
    oscillations: the initial hump breaks up for $t>t_{c}$ into an 
    array of smaller humps forming a cusped zone in the $x,y,t$ space.
    
    \item For the focusing DS II equation, initial data with a symmetry with respect to an interchange 
    of $x$ and $y$ lead to a blow-up of the $L_{\infty}$ norm of the 
    solution in finite time $t^{*}$ for sufficiently small $\epsilon$.
        There is no oscillatory zone for 
    $|\Psi|^{2}$ in this case, the initial hump evolves directly into a 
    singularity. The difference between blow-up time $t^{*}$ and 
    break-up time $t_{c}$ scales roughly as, $t^{*}-t_{c}\propto \epsilon$. 
    The blow-up time is always larger than the break-up time.
\end{itemize}
\end{conjecture}

The paper is organized as follows: in section 2, we describe the 
various numerical methods used. In section 3, we 
illustrate the use of asymptotic Fourier analysis for the 
semiclassical $1+1$-dimensional NLS system, for which explicit 
results are known.  Then we apply these methods to the semiclassical 
DS II system (\ref{disDSs}).  In section 4, we study the behavior of the solutions 
of DS II for small $\epsilon$, and establish scaling laws in 
$\epsilon$ for the difference between semiclassical DS II and DS II solution 
at break-up. In section 5 we investigate blow-up in DS II solutions 
for small $\epsilon$. We add some concluding remarks in section 6.

\section{Numerical Methods}

In this section we summarize the numerical approaches used in this 
paper. The task is to study numerically two different kinds of systems, 
the first one being a coupled system of nonlinear dispersionless 
equations (\ref{disDSs}), the second one being a nonlinear dispersive PDE of NLS type  (\ref{DSnonlocal}). 
For both, we will consider a periodic setting for the spatial coordinates, which allows
the use of a Fourier spectral method for the space discretization. We 
treat the rapidly decreasing functions we are studying as essentially 
periodic analytic functions within the finite numerical precision. 
For such functions, spectral methods are known for their excellent 
(in practice exponential) approximation properties, see for instance 
\cite{can,tref}. In addition they 
introduce only very little numerical dissipation which is important 
in the study of dispersive effects. Last but not least we use the Fourier 
coefficients of the solutions to the semiclassical systems to identify the break-up of 
the solution as in \cite{SSF,dkpsulart}.

In all cases,
the numerical precision is controlled via the numerically computed energy for each system considered.
More precisely, given $E$,
 a conserved quantity of the system, the numerically computed $E$ will  depend on time due to unavoidable numerical errors. 
It was shown for instance in \cite{ckkdvnls,KR} that the conservation 
of $E$ in the form of the quantity 
\begin{equation}
\Delta_E = \left|\frac{E(t)}{E(0)} - 1\right|
\label{delE}
\end{equation}
can be used as a reliable indicator of numerical accuracy, provided that 
there is sufficient spatial resolution (generally the accuracy of the 
numerical solution is overestimated by two orders of magnitude).
We always aim at a $\Delta_E$ smaller than $10^{-6}$ to ensure an accuracy well beyond the plotting accuracy $\sim 10^{-3}$.

\subsection{Dispersionless Systems}

The most difficult task in the solution of the dispersionless systems 
is to identify numerically the break-up of the solution with 
sufficient accuracy to allow the scaling studies we are interested 
in. To this end we have to compute the solution  up to the time of 
gradient catastrophe, and both this time and the solution should be 
found with sufficient 
accuracy. To do so, we will use asymptotic Fourier analysis as  first 
applied numerically by Sulem, Sulem and Frisch in \cite{SSF}. The 
basic idea of this method is that functions  analytic in a 
strip around the real axis in the complex plane have a characteristic 
Fourier spectrum for large wave numbers. Thus it is in principle 
possible to obtain the  
 width of the analyticity strip from the asymptotic behavior of the 
 Fourier transform of the solution (in one spatial dimension), or 
 from the angle averaged energy spectrum in higher dimensions. This 
 allows in particular to identify the time when a singularity in the 
 complex plane hits the real axis and thus leads to a singularity of 
 the function on the real line. 
Singular solutions to the two-dimensional cubic NLS equation have been studied with this approach in
\cite{SSP}, and an application of the method to the two-dimensional Euler equations can be found 
in \cite{FMB, MBF}. The method has also been applied to 
the study of complex singularities of the three-dimensional Euler equations
in \cite{CR}, in thin jets with surface tension \cite{PS98}, the 
complex Burgers' equation \cite{SCE96} and the Camassa-Holm equation \cite{RLSS}. 
More recently, we investigated its efficiency quantitatively for the 
Hopf equation and showed that the method can be efficiently used in 
practice to describe the critical behavior of solutions to dispersionless equations. 
As an example, a study of the break up of dKP solutions for certain 
classes of initial data has been presented in \cite{dkpsulart}.
%More precisely, the method is based on the following analytical result \cite{asymbook}.
%\begin{theorem}
% Let $u(z)$ an analytic function of one variable $z\in\mathbb{C}$ such that $|u(z)| \to 0$ uniformly as $|z| \to \infty$.
% Assuming that the singularities of $u(z)$  are isolated one from
%another and are of one of the following type: pole, algebraic- or logarithmic-
%branch point, then the behavior of the Fourier transform of $u$, denoted by $\hat{u}$, is asymptotically
%(when $k \to \infty$) governed by the singularity of the lower half-space closest to
%the real domain that is not a multiple pole.
%If this singularity is located at $z_{j}=\alpha_{j}-i\delta_{j}$ , with $\delta_{j} \geq 0$, 
%and has an exponent
%$\mu_{j}\notin \mathbb{Z}$, such that in a neighborhood of $z_{j}$, $u(z)\sim 
%(z-z_{j})^{\mu_{j}}$, then
%\begin{equation}
%    \hat{u}\sim 
%    \sqrt{2\pi}\mu_{j}^{\mu_{j}+\frac{1}{2}}e^{-\mu_{j}}\frac{(-i)^{\mu_{j}+1}}{k^{\mu_{j}+1}} e^{-ik\alpha_{j}-k\delta_{j}}.
%    \label{fourierasym}
%\end{equation}
%\end{theorem}
%
%
%
If $u(z)$ is an analytic function of one variable $z\in\mathbb{C}$ such that $|u(z)| \to 0$ uniformly as $|z| \to \infty$, and 
if the singularities of $u(z)$ are isolated and  of the form $u\sim 
(z-z_{j})^{\mu_{j}}$, $\mu_{j}\notin \mathbb{Z}$, 
with $z_{j}=\alpha_{j}-i\delta_{j}$, 
$\alpha_{j},\delta_{j}\in\mathbb{R}$ in the lower half 
plane ($\delta_{j}\geq 0$), a steepest descent argument  for 
$k\to\infty$  implies
the following asymptotic behavior of the Fourier 
coefficients (for a detailed derivation see e.g.~\cite{asymbook}),
\begin{equation}
    \hat{u}\sim 
    \sqrt{2\pi}\mu_{j}^{\mu_{j}+\frac{1}{2}}e^{-\mu_{j}}\frac{(-i)^{\mu_{j}+1}}{k^{\mu_{j}+1}} e^{-ik\alpha_{j}-k\delta_{j}},
    \label{fourierasym}
\end{equation}
where $\hat{u}$ is the Fourier transform of $u(x)\in L^{2}(\mathbb{R})$, defined as 
\begin{equation}
    \hat{u}(k) = \int_{\mathbb{R}}^{}u(x)e^{-ik x}dx
    \label{fourier}.
\end{equation}
Consequently for a single such singularity with positive $\delta_{j}$, the modulus of the Fourier 
coefficients decreases exponentially for large $k$. For 
$\delta_{j}=0$, i.e., a singularity on the real axis,  
the modulus of the Fourier coefficients has an algebraic dependence 
on $k$, and thus the location of 
singularities in the complex plane can be obtained from a given Fourier 
series computed on the real axis. 
If there are several singularities of this form at $z_{j}$, 
$j=1,\ldots,J$, there will be oscillations in the modulus of the 
Fourier coefficients for moderately large $k$.

To numerically compute a Fourier transform, it has to be approximated 
by a discrete Fourier series which can be done efficiently via a 
 fast Fourier transform (FFT), see e.g.~\cite{tref}. The discrete Fourier transform  of the 
 vector $\mathbf{u}$ with components $u_{j}=u(x_{j})$, where 
 $x_{j}=2\pi L j/N$, $j=1,\ldots,N$ (i.e., the Fourier transform on 
 the interval $[0,2\pi L]$ where $L$ is a positive real number) 
 will be always denoted by $v$ in 
 the following. There is no obvious analogue of 
 relation (\ref{fourierasym}) for a discrete Fourier series, but it 
 can be seen as an approximation of the former, which is also the 
 basis of the numerical approach in the solution of the PDE. It is possible to 
 establish bounds for the discrete series, see for 
 instance \cite{arnold}.
 \\
 \\
 According to (\ref{fourierasym}), $v$ is assumed to be of the form
$v(k,t) \underset{k \to \infty}{\sim} e^{A(t)} k^{-B(t)}e^{-\delta(t)k},
$
and one can trace the temporal behavior of $\delta(t)$ obtained 
via some fitting procedure in order to obtain evidence for the 
formation of a singularity on the real line
(the problem is reduced to check if $\delta(t)$ vanishes at a finite time $t_c$).
In order to determine $\delta(t)$ from direct numerical simulations, a least-square fit is performed on the logarithm of the
Fourier transform in the form
\begin{equation}
    \ln |v|\sim A- B\ln k-k\delta.
    \label{abd}
\end{equation}
The fitting is done for a given range of wave numbers $k_{min}<k<k_{max}$ (we only consider positive $k$), that have to be controlled, 
as explained in detail in \cite{dkpsulart}. The critical time 
$t_c$ is determined by the vanishing of $\delta$, and the type of the 
singularity is  given by the parameter $B(t_c)$ which is equal to $\mu_j+1$.
One can also determine the real part of the location of the singularity by doing a least square fitting on the 
 imaginary part of the logarithm of $v$ for which one has asymptotically
 \begin{equation}
    \phi:=\Im \ln v\sim C-\alpha k
    \label{phi}.
\end{equation}
Since the logarithm is 
branched in Matlab/Fortran at the negative real axis with jumps of $2\pi$, the computed $\phi$ 
will in general have many jumps. Thus one has first to construct a 
continuous function from the computed $\phi$, as explained in 
\cite{dkpsulart}. Then the location of the singularity on the real 
axis is given by $\alpha(t_c)$.

Obviously the choice of the fitting bounds $(k_{min}, k_{max})$ has an impact on the determination of the fitting parameters, and we carefully 
investigated this issue in \cite{dkpsulart}. There we found that in 
order to obtain reliable results, high space resolution has to be 
used (typically $2^{14}$ or more Fourier modes in each space 
directions). This implies in particular that for the study of 
2+1-dimensional problems, codes have to be parallelized, see below. 
Moreover, a procedure to obtain the `minimal error fitting bounds' 
whilst using at least half of the Fourier coefficients has been proposed. 
Obviously, only the values for which the Fourier coefficients have a 
greater modulus than the numerical error have to be considered (we choose this threshold to be $|v|>10^{-10}$).
Now, 
let $f_e$ denote a prescribed value for $\Delta=\|   \ln |v| - (A - B 
\ln k - k \delta)   \|_{\infty}$ referred to as \emph{fitting error} in the following. Then, 
one can determine the minimal fitting error $\min (f_e)$ that can be 
reached, by using at least half of the Fourier coefficients available 
for the studied problem, by
choosing a suitable lower threshold (which depends on the problem), and by varying the upper limit to reach $f_e$. 
 \begin{remark}\label{ms}
 The minimal 
 distance in Fourier space is $m:=2\pi L/N$ with $N$ being the number of 
 Fourier modes and $2\pi L$ the length of the computational domain in 
 physical space. Thus this defines the smallest distance which can be 
 resolved in Fourier space. All values of $\delta$ below this 
 threshold cannot be distinguished numerically from 0.  
 \end{remark}

In \cite{dkpsulart}, the identification of the break-up in dKP 
solutions could be done  by studying only the 
Fourier transform as in (\ref{fourierasym}) in one dimension,  since only a 
one-dimensional break-up was conjectured to occur. Here a true 
two-dimensional singularity is possible, and instead of the Fourier transform $\hat{u}(k,t)$, one 
can consider as in \cite{SSP} the angle averaged energy spectrum defined by
\begin{equation}
 \mathcal{E}(K,t) = \underset{K<|k'|<K+1}{\sum} |\hat{u}(k',t)|^{2},
 \label{enspec}
\end{equation}
where 
$|k'|=\sqrt{ k_{x}^2 + k_{y}^2 }$.
Slightly weaker estimates hold for $\mathcal{E}(K,t)$ for an analytic
function $u$ and thus, to apply the asymptotic 
fitting to the Fourier coefficients, one assumes
that 
$ \mathcal{E}(K,t)=e^{A(t)}K^{-\alpha(t)} e^{-\delta(t)K},
$
and performs the fitting on
\begin{equation}
    \ln |\mathcal{E}|\sim A_{2d}- B_{2d}\ln K-K\delta_{2d}.
    \label{abd2}
\end{equation}

Similarly to the one-dimensional case, the 
appearance of a real singularity implies that $\delta(t)$ vanishes at 
a finite time $t_c$, and the descriptions given above hold except for the determination of $\alpha$.

For the numerical integration of the semiclassical DS II system (\ref{disDSs}), we thus use a 
Fourier discretization which leads to a large system of ODEs. In 
principle any ODE solver can be applied for the time integration. 
Typically we use the explicit fourth order Runge-Kutta scheme (RK4), and study 
the asymptotic behavior of the Fourier coefficients as  
explained above. We also use a Krasny filter \cite{Krasny} with a 
prescribed error of $10^{-14}$, which means that Fourier coefficients 
with a modulus of $10^{-14}$ and smaller are put equal to 0. This 
allows to reduce computer roundoff errors and to perform accurate computations with a larger number of points.

\subsection{Dispersive PDEs}

The situation is more involved for the study of dispersive PDEs. We 
again use a Fourier discretization for the spatial coordinates for 
the reasons explained above. Approximating the spatial dependence via 
truncated Fourier series leads for the studied equations 
(\ref{NLSeqgen}) and (\ref{DSnonlocal}) to large \textit{stiff}\footnote{We use the word stiffness in this context 
to indicate that there are largely different scales to be resolved in 
this system of ODEs which make the use of explicit methods inefficient 
for stability reasons.} systems of 
ODEs in Fourier space of the form   
\begin{equation}
    v_{t}=\mathbf{L}v+\mathbf{N}(v,t)
    \label{utrans},
\end{equation}
where $v$ is again the discrete Fourier transform of $\Psi$, 
and where $\mathbf{L}$ and $\mathbf{N}$ denote linear and nonlinear 
operators, respectively. These systems of ODEs are classical examples 
of stiff equations where the 
stiffness is related to the linear part $\mathbf{L}$ (it is 
a consequence of the distribution of the eigenvalues of 
$\mathbf{L}$), whereas the 
nonlinear part contains only low order derivatives. 

There are several approaches to 
deal efficiently with equations of the form (\ref{utrans}) with a 
linear stiff part as implicit-explicit (IMEX), time splitting, 
integrating factor (IF) as well as 
sliders and exponential time differencing. 
By performing a comparison  of stiff integrators for the 
1+1-dimensional cubic NLS equation in the semiclassical limit 
(\ref{NLSeqgen}) in  \cite{ckkdvnls}, and for the semiclassical limit 
of the DS II equation in  \cite{KR}, 
it was shown that Driscoll's composite Runge-Kutta (DCRK) method 
\cite{Dris} is very efficient in this context. 
We thus use this scheme for the time integration here. 

The basic idea of the DCRK method is inspired by IMEX methods, i.e.,  
the use of a stable implicit method for the linear part of the 
equation (\ref{utrans}), which introduces  the stiffness into the 
system, and an explicit scheme for the nonlinear part which is assumed to be non-stiff. 
Classic IMEX schemes do not perform in general satisfactorily for 
dispersive PDEs \cite{KassT}.  Driscoll's \cite{Dris}  more 
sophisticated variant consists in splitting the linear part of the 
equation in Fourier space into regimes of high and low frequencies, 
and to use the fourth order RK integrator for the low frequencies and 
the nonlinear part, and the linearly implicit RK method of order three for the high frequencies. 
He showed that this method is in practice of fourth order over a wide range of step sizes. 

An additional problem here is the modulational instability of the 
focusing NLS equations, i.e., a self-induced amplitude modulation of 
a continuous wave propagating in a nonlinear medium, with subsequent 
generation of localized structures, see for instance \cite{AGR,CH,FL} 
for the NLS equation. This instability leads to an artificial 
increase of the high wave numbers which eventually breaks the code, 
if not enough spatial resolution is provided (see for instance 
\cite{ckkdvnls} for the focusing NLS equation). It is not possible to 
reach the necessary resolution on single processors which makes a 
parallelization of the codes obligatory. 

\subsection{Parallelization for 2+1-dimensional problems}

To be able to provide the high space resolution needed  for the DS II 
simulations (see above), the numerical codes for the 2+1-dimensional 
problems have been parallelized.  This can be conveniently done for 
two-dimensional Fourier 
transforms where the task of the one-dimensional FFTs is performed 
simultaneously by several processors. This reduces also the memory 
requirements per processor with respect to alternative approaches such as finite 
difference or finite element methods. We consider periodic (up to 
numerical precision) solutions in $x$ and $y$, i.e., solutions on $\mathbb{T}^2 \times \mathbb{R}$. The computations are carried out with $N_x \times N_y$ points for $(x, y) \in [-L_x\pi, L_x\pi] \times  [-L_y\pi, L_y\pi] $.
In the computations, $L_x = L_y$ is chosen large enough such that the 
numerical solution is of the order of machine precision ($\sim 
10^{-16}$ here) at the boundaries.

A prerequisite for parallel numerical algorithms is that sufficient 
independent computations can be identified for each processor,
that require only small amounts of data to be communicated between 
independent computations.
To this end,
we perform a data decomposition, which makes it possible to 
do basic operations on each object in the data domain (vector, matrix...) 
 to be executed safely in parallel by the available processors.
Our domain decomposition is implemented by developing a
code describing the local computations and local data structures for a single process. 
Global arrays are divided in the following way:
denoting by 
$x_n = 2 \pi n L_x/N_x,\,\, y_m = 2 \pi m L_y/N_y$, $n=-N_x/2,...,N_x/2, \,\, m=-N_y/2,...,N_y/2,$
the respective discretizations of $x$ and $y$ in the corresponding 
computational domain, 
$u$ (respectively $\Psi$) is then represented by a $N_x \times N_y$ matrix.
For programming ease and for the efficiency of the Fourier transform, 
$N_x$ and $N_y$ are chosen to be powers of two. The number  $n_p$ of processes is chosen to divide $N_x$ and $N_y$ perfectly, 
so that 
each processor $P_i, i=1...n_p$, will receive $N_x \times \frac{N_y}{n_p}$ elements of 
$u$ corresponding to the elements 
\begin{equation}
u\left(1:N_x, (i-1).\frac{N_y}{n_p}+1 : i.\frac{N_y}{n_p} \right) 
\end{equation}
in the global array, and then 
each parallel task works on a portion of the data.

While processors execute
an operation, they may need values from other processors. The above domain decomposition 
has been chosen such that the distribution of operations is balanced 
and that the communication
is minimized. 
The access to remote elements has been implemented
via explicit communications, using  sub-routines of the MPI (Message Passing Interface) library \cite{GTL}.

Actually, the only part of our codes that requires communications is 
the computation of the two-dimensional FFT and 
the fitting procedure for the Fourier coefficients.
For the former we use the transposition approach. The latter allows to 
use highly optimized single processor one-dimensional FFT routines, 
that are normally found in most architectures, and a transposition 
algorithm can be easily ported to different distributed memory 
architectures. We use the well known FFTW library because its 
implementation is close to optimal for serial FFT computation, see 
\cite{FJ}. Roughly speaking, a two-dimensional FFT does one-dimensional FFTs on all rows and 
then on all columns of the initial global array.        
We thus first transform in $x$ direction, each processor transforms all the dimensions of the data that are completely local to it, and
the array is transposed once this has been done by all processors. 
Since the data are evenly distributed among the MPI processes, this transpose is efficiently implemented using MPI ALLTOALL communications of the MPI library.

The
asymptotic fitting of the Fourier coefficients in one spatial 
direction requires in addition two local communications, and in two 
dimensions this 
is performed through only one global communication (MPI REDUCE, MPI 
SUM), all processors doing previously the computation of $\mathcal{E}(K, t)$ locally.

\section{Numerical study of the semiclassical systems}

In this section, we numerically solve semiclassical systems up to the 
time of gradient catastrophe. Of special importance is the accurate 
determination of the critical time $t_{c}$ since we are interested in 
the following in the scalings of DS II solutions with respect to 
semiclassical DS II solutions at this time. To this end we use the 
asymptotic (for large wave numbers) behavior of the Fourier 
coefficients of the numerical solution in dependence of time. We 
first test the method for examples from  1+1-dimensional 
semiclassical systems for NLS equations, which can be treated 
analytically. The question is how the quantitative approach used in 
\cite{dkpsulart} for the scalar Hopf equation performs in the context of a 
two-component system  (\ref{dnls1}). We find an increase of the 
numerical errors with respect to the scalar case, but mainly in the 
determination of the type of the singularity (i.e., the smallest 
$\mu_{j}$ in (\ref{fourierasym})). 
However, it appears that the critical time is still well identified. 
Then we study the semiclassical DS II system (\ref{disDSs}) with this 
method. Whereas  we performed only a one-dimensional study in 
\cite{dkpsulart}, since it was known that only the $x$-derivative would 
blow up,  we  also study here the formation of 
singularities with a  two-dimensional approach via the energy spectra 
(\ref{enspec}). We find, however, that the singularities are also 
one-dimensional in this case in the sense that only one component of 
the gradient of the solution blows up. The observed singularities are 
as in the $1+1$-dimensional case, cubic in the defocusing setting, 
square root behavior in the focusing case. 

\subsection{1+1-dimensional semiclassical cubic NLS}
In this subsection we numerically solve the semiclassical system 
(\ref{dnls1}) for 
the $1+1$-dimensional cubic NLS equation both in the focusing and 
defocusing case for initial data, for which the critical time and 
solution can be given analytically. This is used as a test for the 
numerical approach to determine the critical time.  
 To check the numerical accuracy, we compute the 
following conserved quantity of (\ref{dnls1}),
\begin{equation}
E[u, w](t) := \int_{\mathbb{T}} \left( u(x,t) w^2(x,t) - \rho u^2(x,t)   \right) dx.
\end{equation}

\subsubsection{Defocusing case}
First we consider an example for the defocusing case ($\rho=1$) which was 
numerically studied in \cite{DGK13} to which we refer the reader for details. 
We consider the initial data
\begin{equation}
u(x,0)=A^2_0\, \mbox{sech}^2 \, x,\quad w(x,0)=0,\quad A_{0}=const.
\label{dataspec}
\end{equation}
The solution to the defocusing semiclassical system (\ref{dnls1}) can be found in terms of 
the Riemann invariants $r_{\pm}=w\pm 2\sqrt{u}$ for given initial 
data $r_{\pm}(x,0)=\phi_{\pm}(x)$ in the form
\begin{equation}
    x = -\lambda_{\pm}t+\mu_{\pm}
    \label{sdNLSint},
\end{equation}
where
$$ \lambda_{+}=-\frac{1}{4}(3r_++r_-),\quad \lambda_{-} = 
-\frac{1}{4}(r_++3r_-),$$
and where 
\begin{eqnarray*}
\mu_{\pm}&=&-\log(\sqrt{2A_{0}+r_+}+\sqrt{2A_{0}+r_-})-\log(\sqrt{2A_{0}-r_+}+\sqrt{2A_{0}-r_-})+\log(r_+-r_-)\\
&&\pm\frac{1}{r_+-r_-}\left(\sqrt{(2A_{0}+r_+)(2A_{0}+r_-)}-\sqrt{(2A_{0}-r_+)(2A_{0}-r_-)})\right).
\end{eqnarray*}
The critical point for these initial data is given by
\[
r^c_+=\frac{A_{0}}{3}(6-\sqrt{33})\sqrt{2\sqrt{33}+6},\;\;r^c_-=-\frac{A_{0}}{3}\sqrt{2\sqrt{33}+6}, 
\]
\[
t_c=\frac{3\sqrt{2}}{32 A_{0}}\sqrt{69+11\sqrt{33}},\;\;x_c\sim-2.209395255.
\]
Equations (\ref{sdNLSint}) can be numerically solved as discussed in 
\cite{DGK13} with the optimization algorithm \cite{fminsearch} 
distributed with Matlab as \textit{fminsearch} to in principle 
machine precision. 

To numerically solve 
(\ref{dnls1}), we use a Fourier spectral method, as indicated in 
section 2, and the explicit fourth order Runge-Kutta scheme for the time 
integration.

\begin{figure}[htb!]
\centering
 \includegraphics[width=0.49\textwidth]{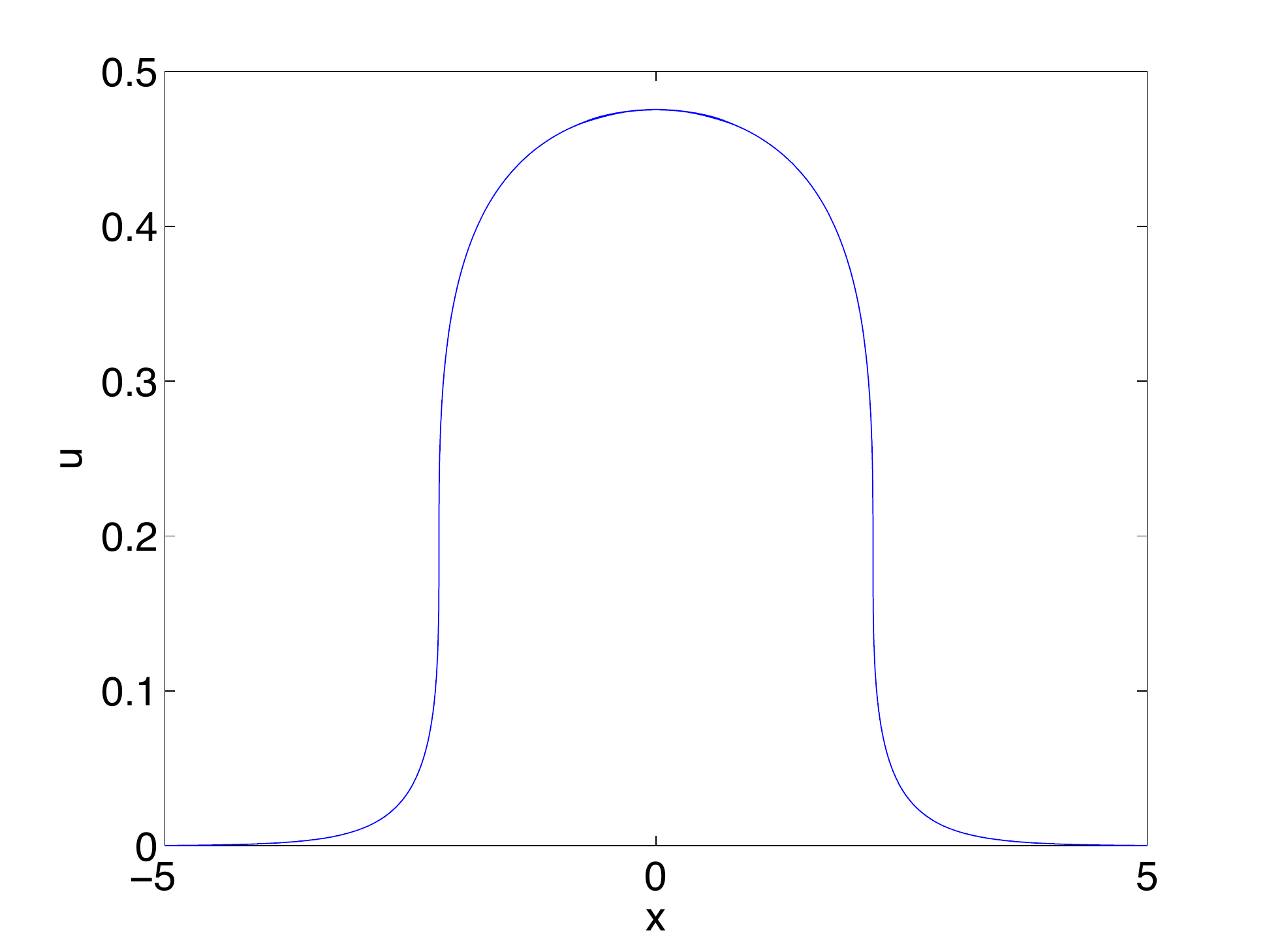} 
 \includegraphics[width=0.49\textwidth]{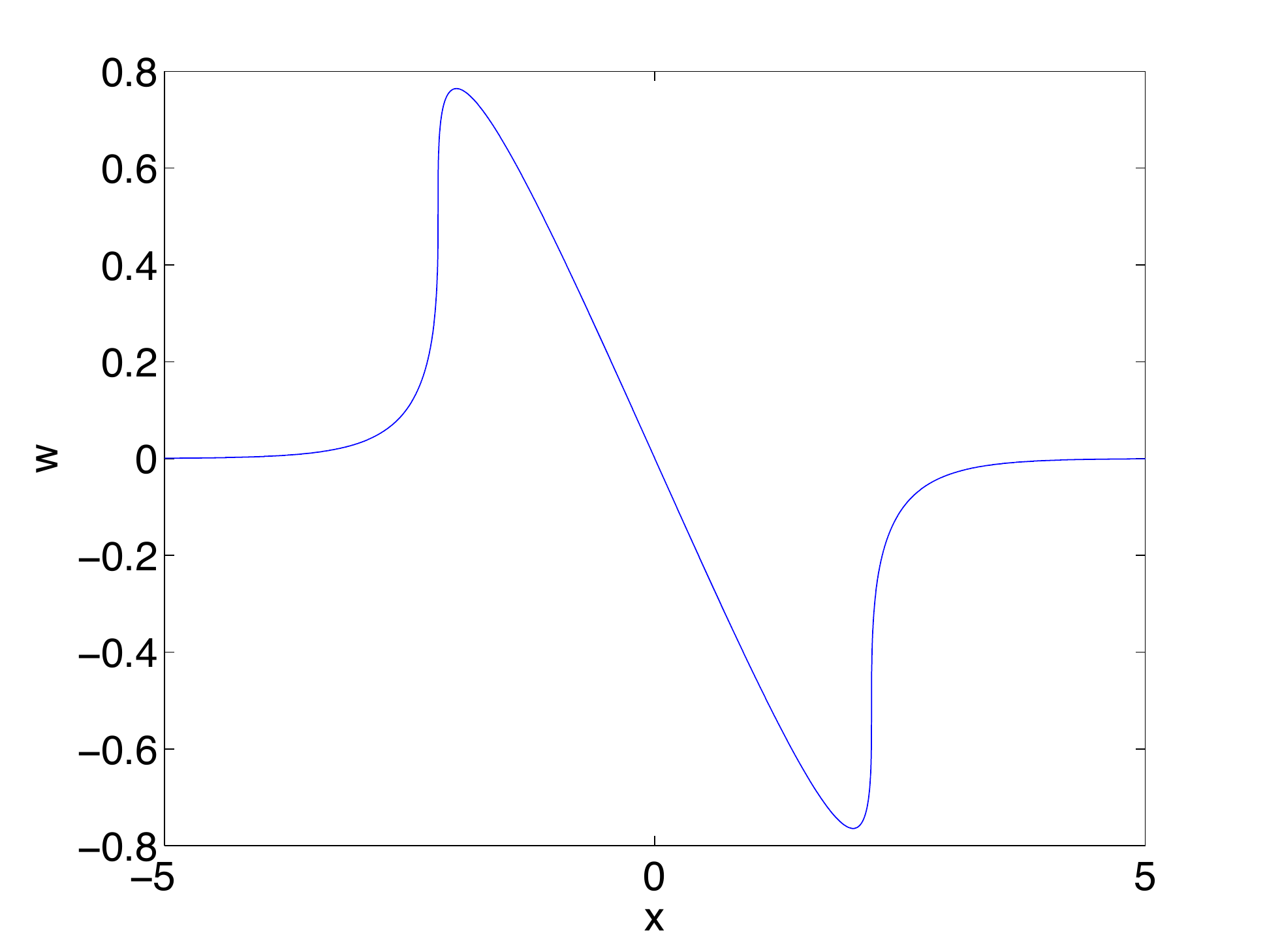} 
 \caption{Solution to the defocusing semiclassical system 
 (\ref{dnls1}) for the initial data 
 $u_{0} (x) = sech^{2}(x), \,\, w_0(x)=0$  at the critical time 
 $t_{c}$; the function $u$ on the left and the corresponding 
function $w$ on the right.}
 \label{dsNLSu}
\end{figure}

The computation is carried out with $N=2^{15}$ Fourier modes for $x \in 
[-5\pi, \, 5\pi]$ and $\Delta_{t}=1.5244*10^{-4}$ up to $t_{c}$ for initial data of the form (\ref{dataspec}) with 
$A_0=1$. The solution at the critical time can be seen in 
Fig.~\ref{dsNLSu}. The conservation of the numerically computed energy, $\Delta_E$ 
(\ref{delE}), typically used as an indicator of the quality of the 
numerics \cite{ckkdvnls, KRM, KR}, is of the order of $10^{-9}$ 
at the maximal time of computation. But as mentioned before, this 
quantity cannot indicate reliably a higher precision than the 
modulus of the Fourier coefficients for the highest wave numbers. 

\begin{figure}[htb!]
\centering
 \includegraphics[width=0.49\textwidth]{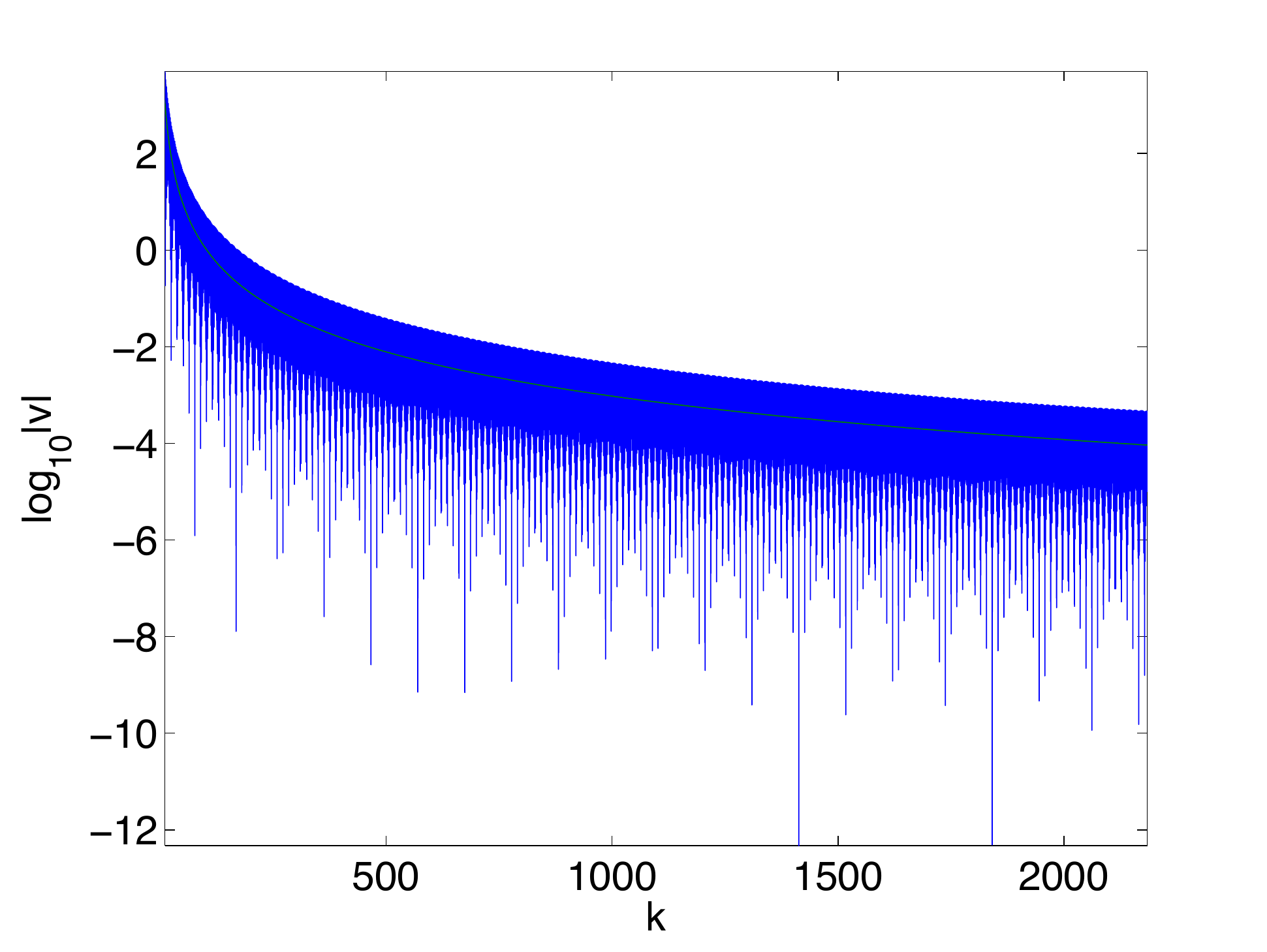} 
 \includegraphics[width=0.49\textwidth]{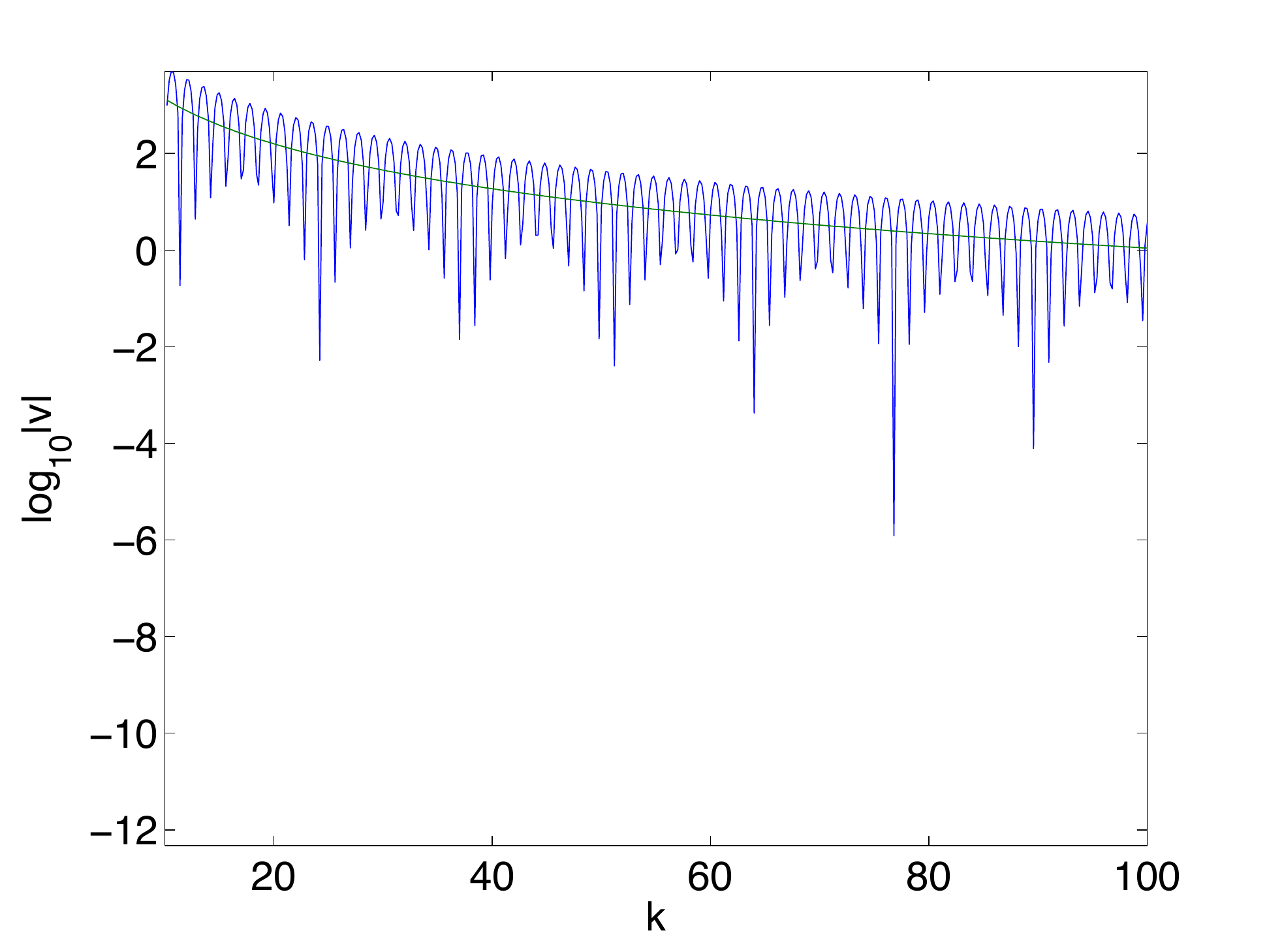} 
 \caption{Modulus of the Fourier coefficients of the solution to the 
 defocusing semiclassical system (\ref{dnls1}) for the initial data 
 $u_{0} (x) = sech^{2}(x), \,\, w_0(x)=0$  at the critical time 
 $t_{c}$ in blue and the fitted curve (\ref{abd}); a close-up of the 
 curve close to the origin is shown on the right.}
 \label{dsNLSfourier}
\end{figure}

We consider the initial data (\ref{dataspec}) since they can be 
treated analytically, see \cite{DGK13},
and thus provide a strong test for our methods. A problematic aspect 
of these data is that two singularities form at the same time $t_{c}$ 
at $\pm x_{c}$. Since the contribution of each singularity in the 
complex plane to the asymptotic behavior of the Fourier coefficients 
(\ref{fourierasym}) is given by a sum over all singularities, and 
since the two hitting the real axis at $t_{c}$ differ only in the 
parameter $\alpha_{j}=\pm x_{c}$, the Fourier coefficients are 
proportional to $\cos(x_{c}k)$. Thus there are strong oscillations in 
the coefficients, see Fig.~\ref{dsNLSfourier} which both affect the accuracy of the solution and 
impose some potential problems on the asymptotic fitting of the 
Fourier coefficients. The fitting of $\ln |\hat{u}|$ to 
(\ref{dsNLSu}) is done for 
$10<k<\max(k)*2/3$. We find the fitting parameters (\ref{abd}) to be $\delta=-3*10^{-5}$, $B=1.3426$, both 
very close to the theoretical values $0$ and $4/3$, and $A=-6.2224$.  
Thus the oscillations do not affect the quality of the fitting, but 
rule out the norm of the difference between $\ln |\hat{u}|$ and the fitted 
curve (\ref{abd}) as an indicator of the quality of the fitting, as 
was possible in the examples considered for the Hopf equation in 
\cite{dkpsulart}. The results of an analogous fitting for $\ln 
\hat{w}$ are almost identical indicating that both functions have a 
cubic singularity. Fitting the imaginary part of $\ln \hat{u}$ 
according to (\ref{phi}), we find $\alpha=2.5281$ compared to 
$x_{c}\sim 2.21$. 

\begin{figure}[htb!]
\centering
 \includegraphics[width=0.49\textwidth]{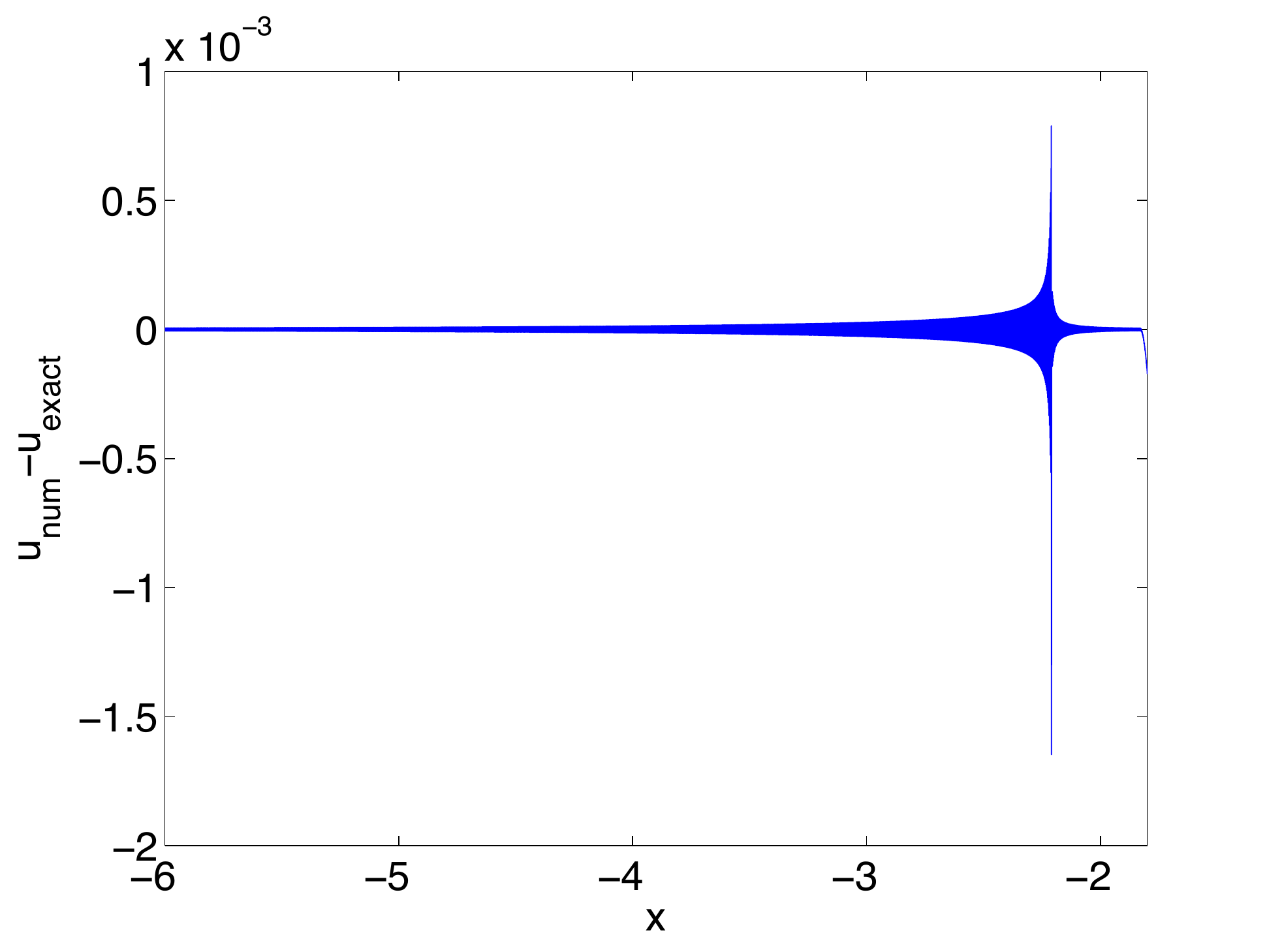} 
 \includegraphics[width=0.49\textwidth]{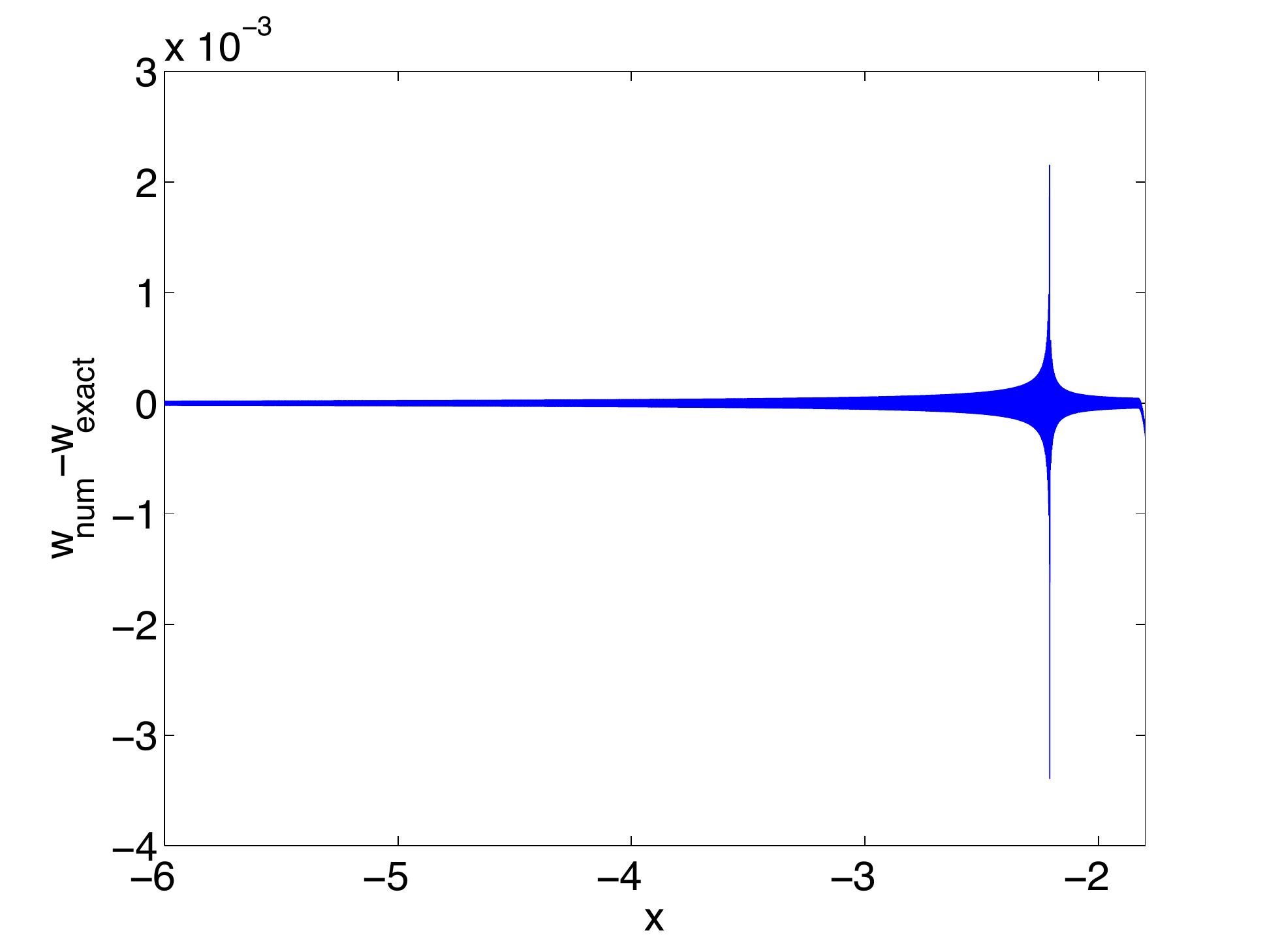} 
 \caption{Difference of the numerical and the exact solution to the 
 defocusing semiclassical system (\ref{dnls1}) for the initial data 
 $u_{0} (x) = sech^{2}(x), \,\, w_0(x)=0$  at the critical time 
 $t_{c}$; for the function $u$ on the left and  for the corresponding 
function $w$ on the right.}
 \label{dsNLSudiff}
\end{figure}

To check the quality of the numerical solution in Fig.~\ref{dsNLSu}, 
we compare it to the exact solution to the system (\ref{sdNLSint}) which can 
be computed in principle with machine precision. It can be seen in 
Fig.~\ref{dsNLSudiff}, where the vicinity of $-x_{c}$ is shown, that 
this difference is largest near the critical point where it is of the 
order of $10^{-3}$. We conclude that the solution to the 
semiclassical system can be obtained numerically with a precision of 
the order of $10^{-3}$ at the critical time, and that the fitting for 
the Fourier coefficients can be done with a similar accuracy. The 
difference of the fitting curve and the Fourier coefficients cannot 
be used as an indicator here since two singularities form at  the 
same time which leads to the oscillations in the Fourier coefficients 
in Fig.~\ref{dsNLSfourier}. 

\subsubsection{Focusing case}
For the focusing case we will again consider the initial data 
(\ref{dataspec}) for which an exact solution to the focusing cubic NLS was given by Satsuma and Yajima 
\cite{SatYa} for a sequence of positive values $\epsilon_{N}$ 
converging to 0. In \cite{ASK66}, an exact solution of the modulation 
equations for these data was given. 
More precisely, the  system (\ref{dnls1}) was introduced there
as a model for the self-focusing phenomenon in one transverse dimension. 
A system of two real equations was given implicitly defining two real unknowns 
as functions of $x$ and $t$, and leading to the formation of a 
finite-amplitude singularity (i.e., a gradient catastrophe) at the 
time $t = t_c = \frac{1}{2A_0}$. These data were also studied in 
\cite{KMM} for the semiclassical limit of the focusing NLS and numerically in \cite{MK2}.
In \cite{DGK}, the system (\ref{dnls1}) was reduced to a linear equation by 
hodograph techniques. For generic localized analytic initial data, the 
solutions of this system have an elliptic umbilic singularity. 
For symmetric initial data of the form (\ref{dataspec}),
the solution of the focusing system (\ref{dnls1}) can be found by solving the system, 
\begin{equation}\label{hodograph}
\left\{
\begin{array}{ccc}
 x & = & wt + f_u, \\
 0 & = & ut+f_w,
 \end{array}
 \right. 
\end{equation}
for  $u=u(x,t)$ and $w=w(x,t)$, where $f(u,w)$ has the explicit form
\begin{equation}
 f(u,w) = \mbox{Re} 
\left[   
\begin{array}{c}
 \left(  -\frac{w}{2} + i A_{0} \right) \sqrt{
u + \left(  -\frac{w}{2} + i A_{0} \right)^{2}} \hspace{3cm}\\
 \hspace{3cm} + u \log \frac{ \left( -\frac{w}{2} + i A_{0} \right) +
\sqrt{ \left(-\frac{w}{2} + i A_{0} \right)^{2} + u}    
}{\sqrt{u}}
\end{array}
\right].
\label{solzarb}
\end{equation}
The critical point\footnote{Note that in the limit $\epsilon$ goes to zero equations (\ref{thgc}) are also true for solutions of the full NLS flow system (\ref{dnlseps}) and for a wider class of potentials (\ref{dataspec}), see \cite{TVZ}. } is given by
\begin{equation}
 u_c = 2A_0, \,\,\, w_c = 0, \,\,\, x_c = 0, \,\,\, t_c = \frac{1}{2A_0}
 \label{thgc}.
\end{equation}
Near the critical point  $x_{c}$ for the critical time $t_c$, the solution 
has a cusp,  $u(x)\sim |x|^{1/2}$ and $w(x)\sim \mbox{sign}(x)|x|^{1/2}$.

% The computation is carried out with $2^{14}$ points, for $x \in 
% [-10\pi, \, 10\pi]$, for initial data of the form (\ref{dataspec}) with $A_0=2$.
% For a fitting of the Fourier coefficients in the range 
% $5<k<\max(k)/2$, $\delta$ vanishes at $t_c\sim 0.2506$.  In  Fig. 
% \ref{dcnlsuv}  we show the solution and the corresponding Fourier 
% coefficients at $t_c/2$ and at $t_c$. The cusp at $t_{c}$ corresponds 
% to a gradient catastrophe as can be also  seen in Fig.~\ref{dcnlsux}, 
% where the time evolution of the $L_{\infty}$ norm  of the gradient of 
% $u$ is shown.

\begin{figure}[htb!]
\centering
 \includegraphics[width=0.49\textwidth]{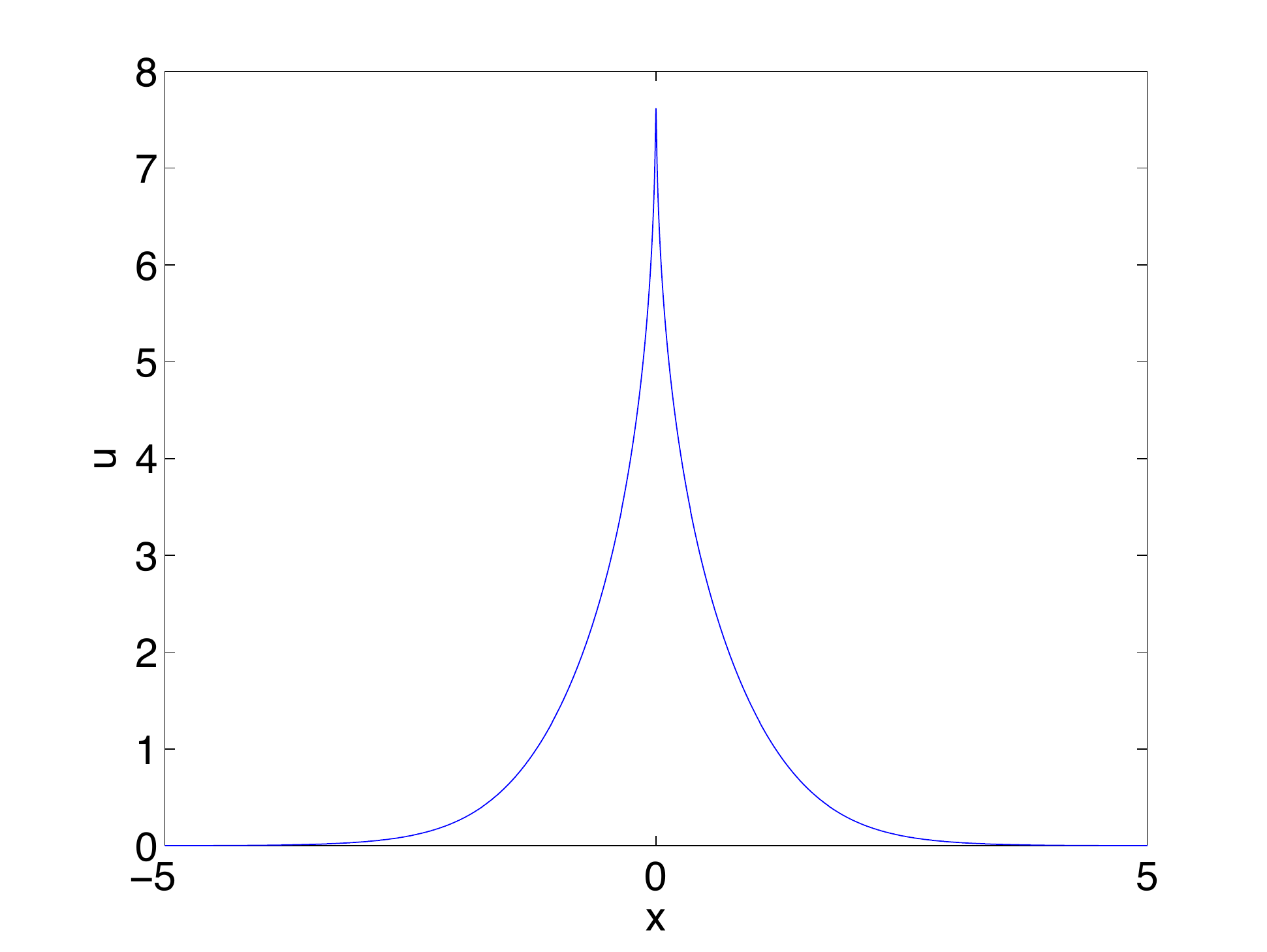} 
 \includegraphics[width=0.49\textwidth]{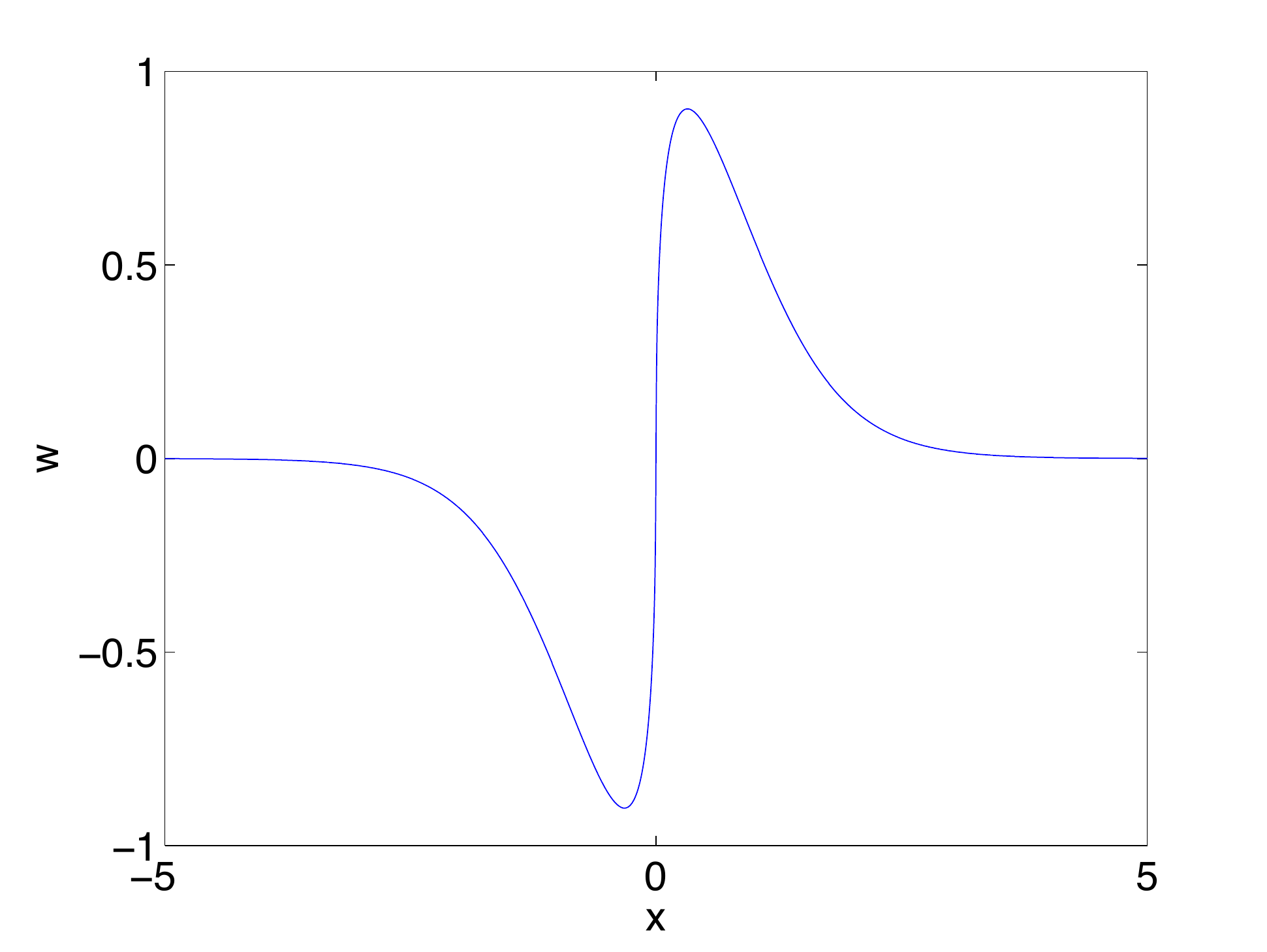} 
 \caption{Solution to the focusing semiclassical system (\ref{dnls1}) for the initial data 
 $u_{0} (x) = 4sech^{2}(x), \,\, w_0(x)=0$  at the critical time 
 $t_{c}=0.25$; the function $u$ on the left and the corresponding 
function $w$ on the right.}
 \label{fsNLSu}
\end{figure}

For initial data of the form (\ref{dataspec}) with $A_0=2$, the computation is carried out with $2^{14}$ points for $x \in 
[-5\pi, \, 5\pi]$ and with a time step $\Delta_{t}=2.5*10^{-5}$.
The solution at the time $t_{c}=0.25$ can be seen in 
Fig.~\ref{fsNLSu}. The numerically computed energy is conserved to 
the order of $10^{-14}$. It can be seen that the maximum at the cusp is not 
fully reached by the numerical solution (its maximum is roughly 
$7.62$ instead of 8). Note that this does not change much if a higher 
resolution in Fourier space is used (we reach $7.64$ with $2^{16}$ 
Fourier modes).

\begin{figure}[htb!]
\centering
 \includegraphics[width=0.49\textwidth]{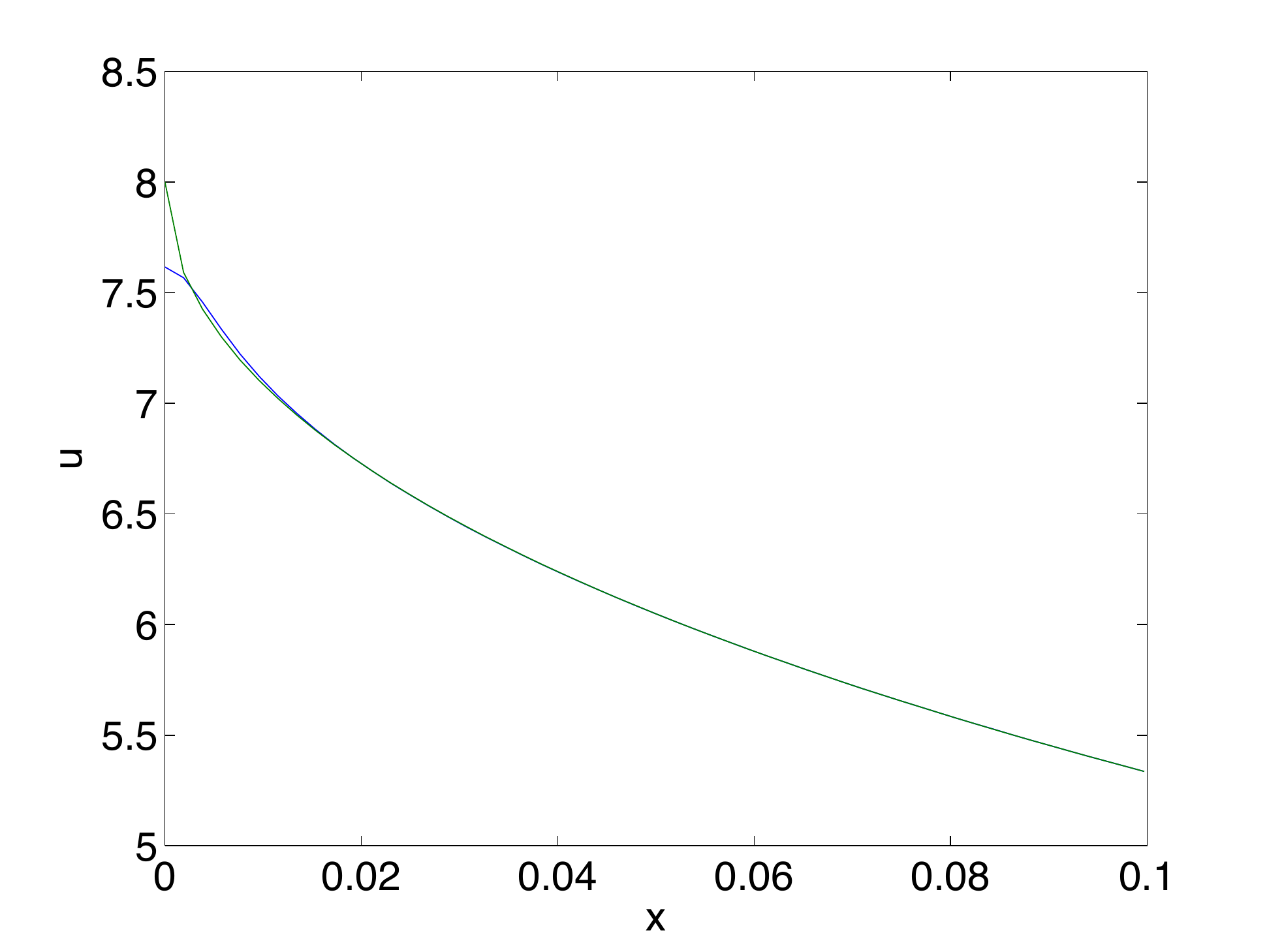} 
 \includegraphics[width=0.49\textwidth]{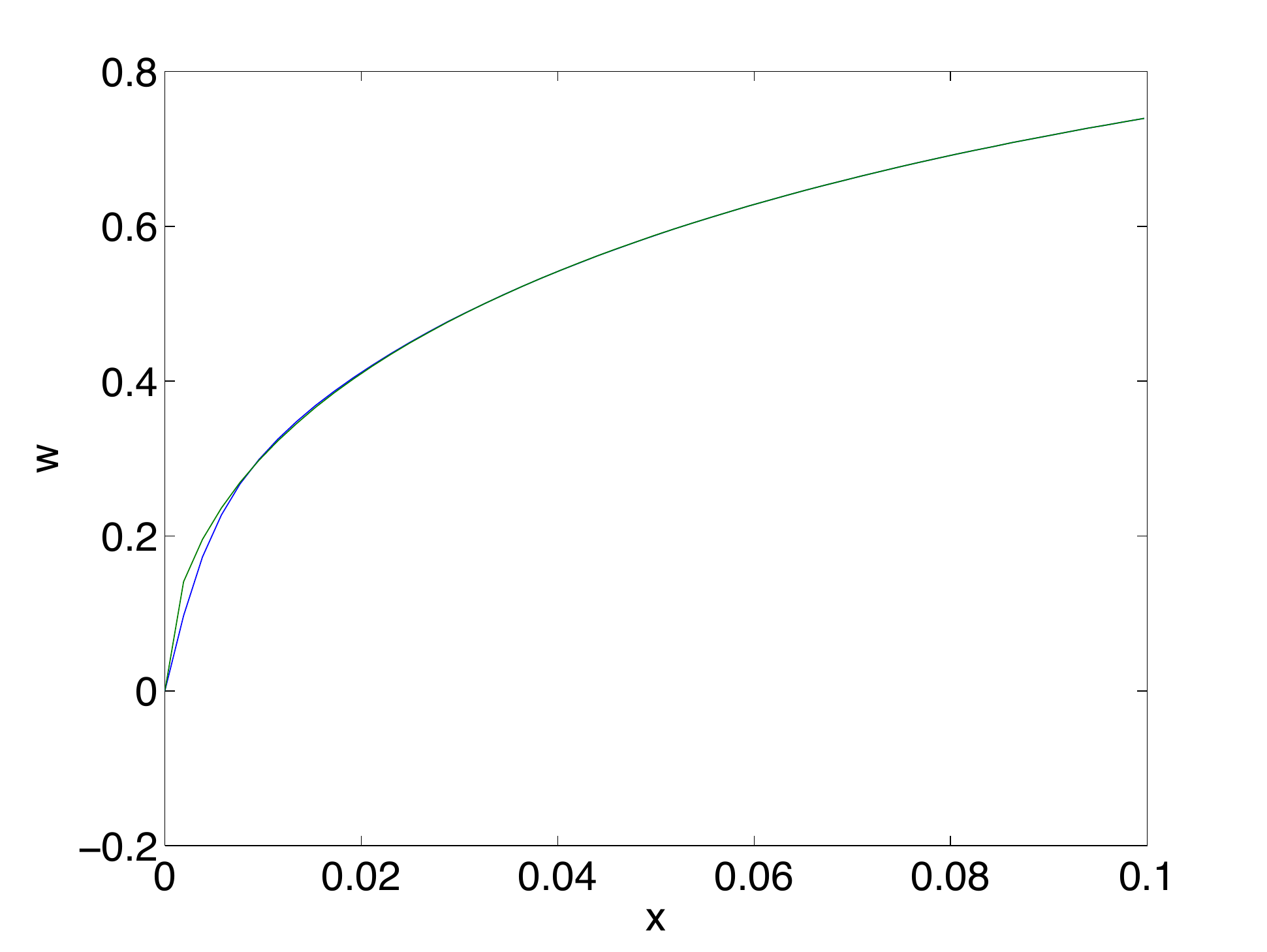} 
 \caption{Close-up of the situation in Fig.~\ref{fsNLSu} together with the 
 corresponding exact solution in green.}
 \label{fsNLSuexact}
\end{figure}

In Fig.~\ref{fsNLSuexact} we show a close-up of the numerical 
and the exact solution obtained by inverting (\ref{hodograph}). It 
can be seen that the agreement is excellent except for the immediate 
vicinity of the cusp, 
where there is only a small difference for $w$ which vanishes at the 
critical point for symmetry reasons, but a more pronounced one for 
$u$. 

\begin{figure}[htb!]
\centering
 \includegraphics[width=0.49\textwidth]{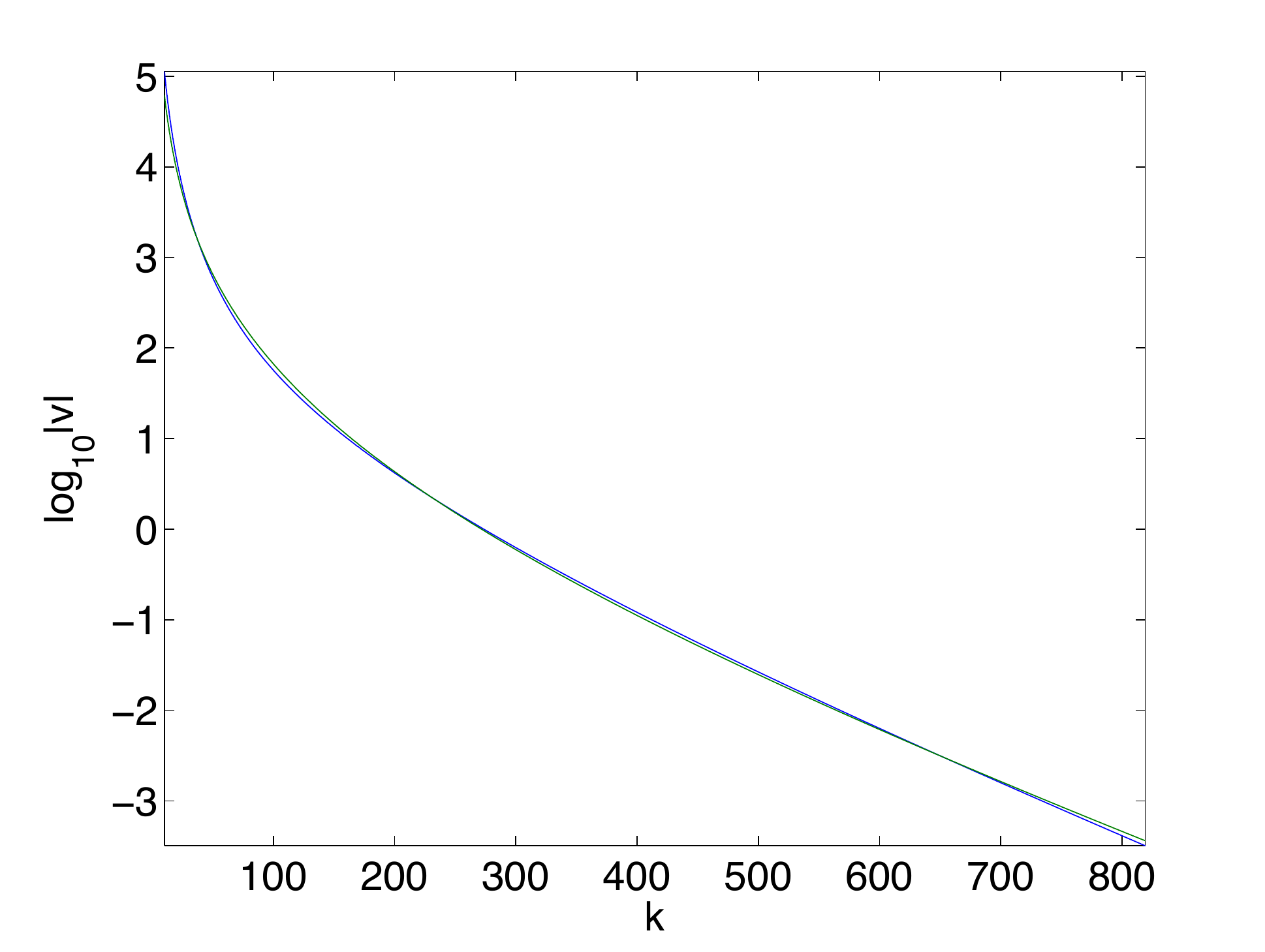} 
 \includegraphics[width=0.49\textwidth]{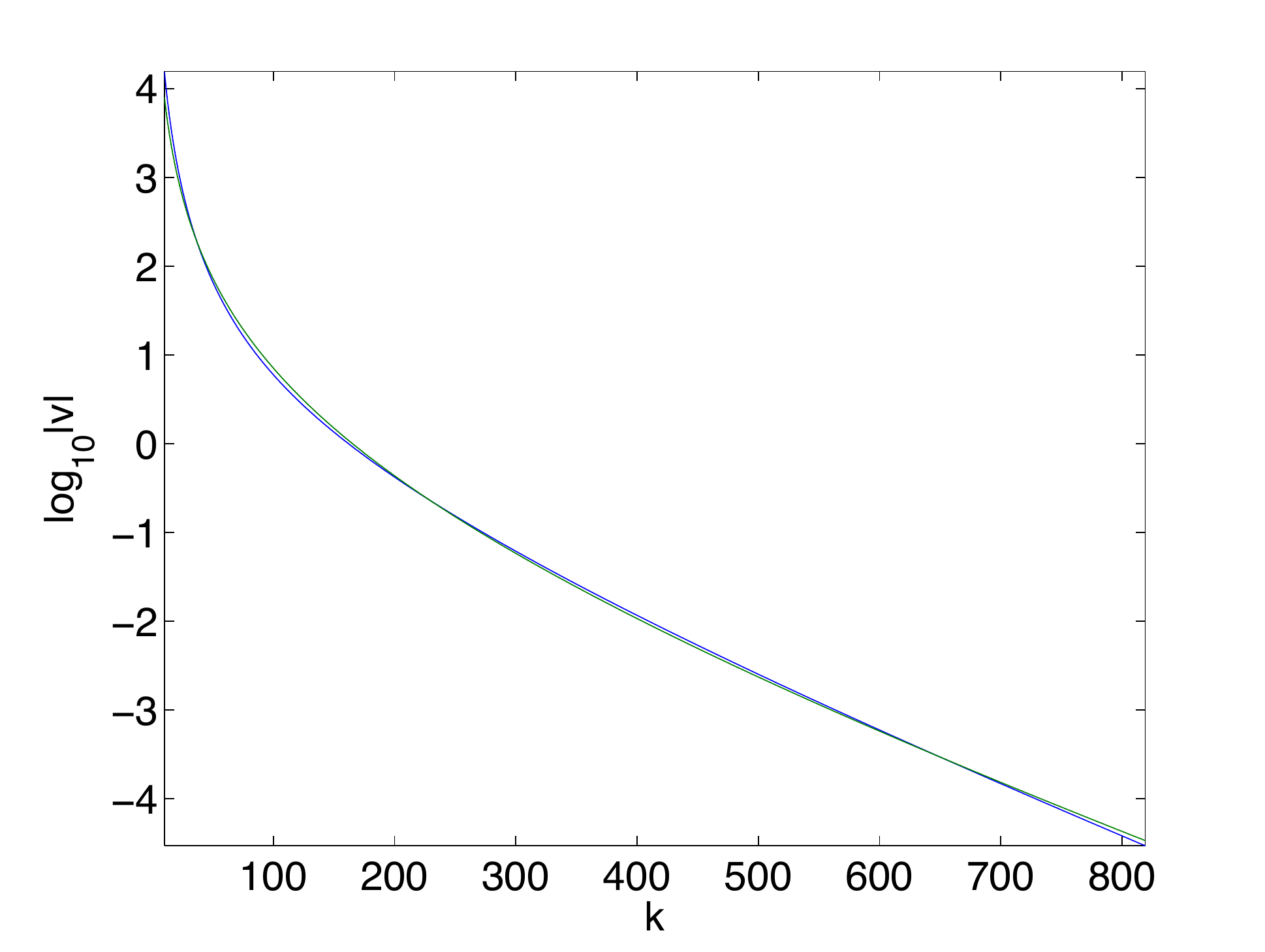} 
 \caption{Modulus of the Fourier coefficients of the solution to the 
 focusing system (\ref{dnls1}) for the initial data 
 $u_{0} (x) = 4sech^{2}(x), \,\, w_0(x)=0$  at the critical time 
 $t_{c}=0.25$ in blue and the fitted curve (\ref{abd}); the  Fourier 
 coefficients for $u$ on the left, the ones for $w$ on the right.}
 \label{fsNLSufourier}
\end{figure}

A fitting of the Fourier coefficients according to (\ref{abd}) for 
$10<k<\mbox{max}(k)/2$ gives $\delta=0.0042$, $B = 1.09$ and 
$A=7.28$, see Fig.~\ref{fsNLSufourier}. Visibly the square root singularity is more difficult to 
reproduce than the cubic root in the defocusing case. In addition 
the system (\ref{dnls1}) can be written in the defocusing case in 
Riemann invariant form as essentially two equations of Hopf type. 
Thus we can solve it with the same precision as in \cite{dkpsulart} 
for the Hopf equation. In the focusing case, the Riemann invariants 
are complex, and thus we face a true system in this case. Not 
surprisingly the cusp is not as well reproduced by the numerics as the 
cubic root, and this is reflected also by the Fourier coefficients. 
This is mainly true for the algebraic decrease given by the 
parameter $B$ (which should be equal to $3/2$) which is always 
more sensitive to this fitting procedure. In fact the fitting 
parameters are closer to the theoretically expected ones if the 
fitting is done for $1<k<\mbox{max}(k)/2$. In this case we get 
$\delta=0.0038$, $B=1.21$ and 
   $A=-7.77$ for the parameters in (\ref{abd}). The fitting error $\Delta$ is in both cases of the 
   order of $0.4$. A larger lower bound has only little effect on the 
   fitting. Note that the fitting parameters for the Fourier 
   coefficients $\hat{w}$ are very similar to what we get for 
   $\hat{u}$: for 
$10<k<\mbox{max}(k)/2$ we get $\delta=0.0039$, $B = 1.18$ and 
$A=6.69$, see Fig.~\ref{fsNLSufourier}.

\begin{figure}[htb!]
\centering
 \includegraphics[width=0.49\textwidth]{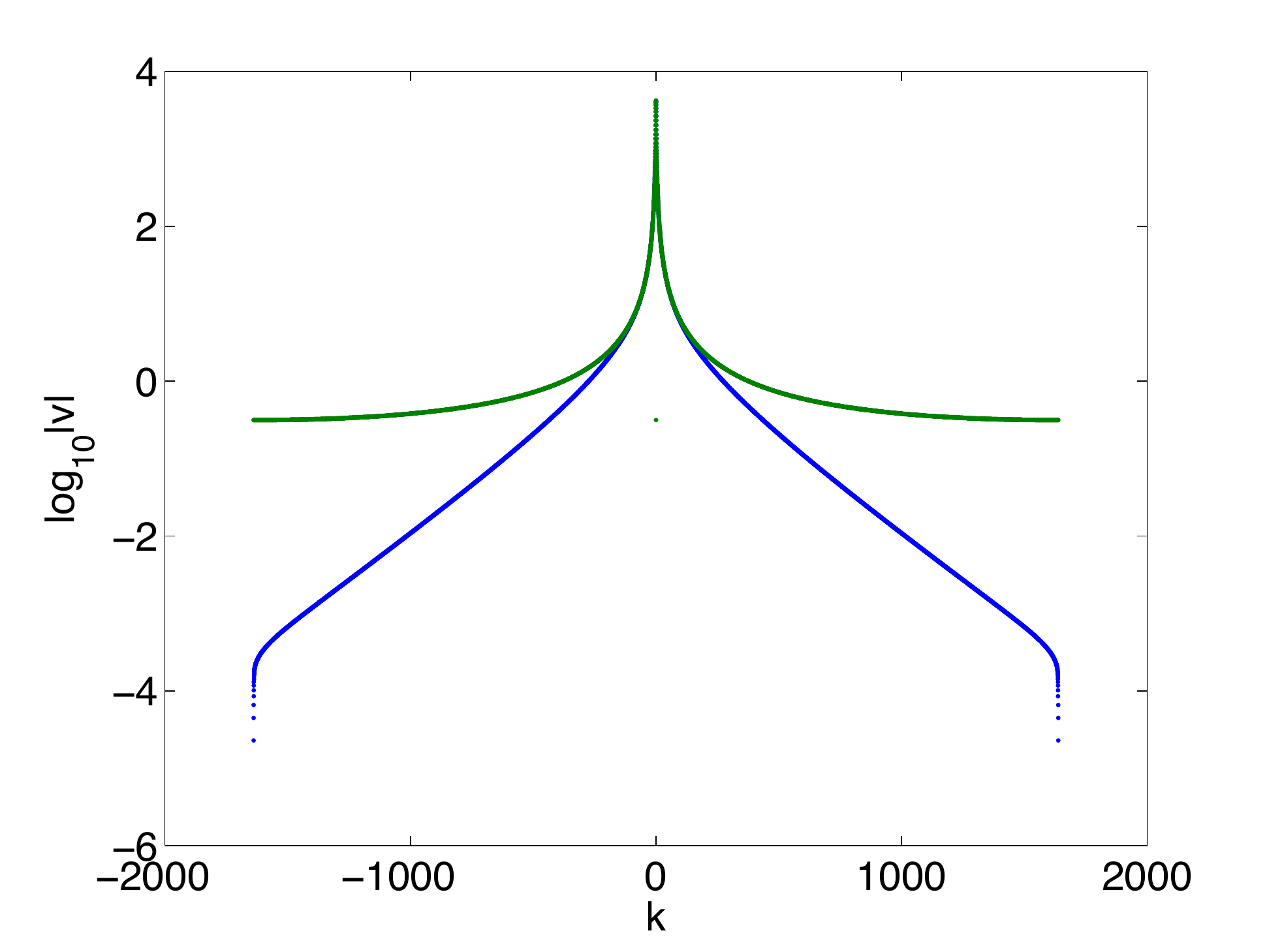} 
 \includegraphics[width=0.49\textwidth]{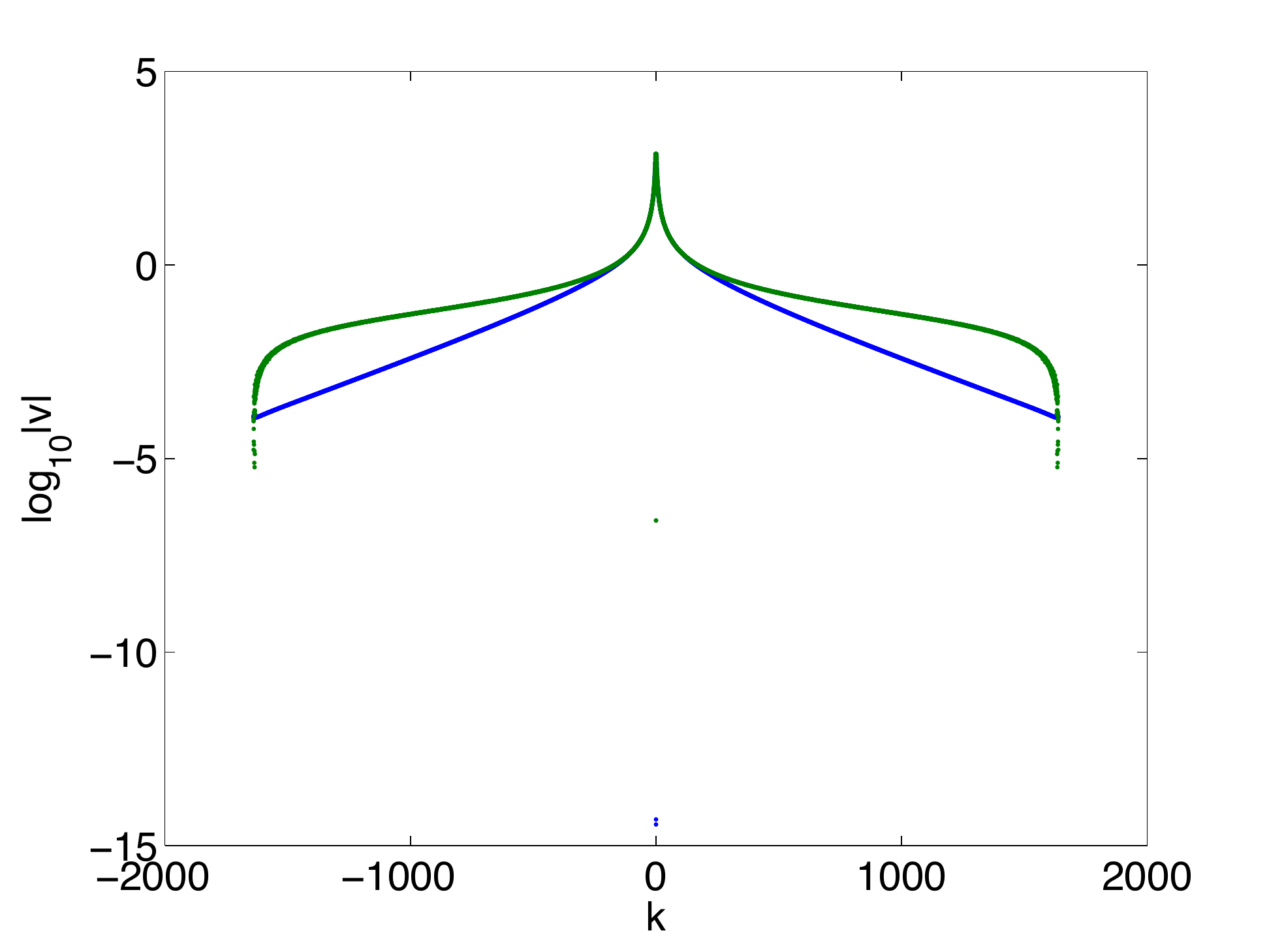} 
 \caption{Modulus of the Fourier coefficients of the solution to the 
 focusing system (\ref{dnls1}) for the initial data 
 $u_{0} (x) = 4sech^{2}(x), \,\, w_0(x)=0$  at the critical time 
 $t_{c}=0.25$ in blue and the coefficients of the exact solution in 
 green; the  Fourier 
 coefficients for $u$ on the left, the ones for $w$ on the right.}
 \label{fsNLSuexactfourier}
\end{figure}

The difference between the numerical and the exact solution in 
Fig.~\ref{fsNLSuexact} is of course the reason that the fitting 
parameters disagree from the theoretical ones. Since the  cusp is 
formed by the high wave numbers in Fourier space, a discrepancy 
there affects the asymptotic behavior of the Fourier coefficients as 
can be seen in Fig.~\ref{fsNLSuexactfourier}; the coefficients agree 
very well for small wave numbers, but disagree for high wave numbers. Again the agreement is 
better for $w$ than for $u$. Thus a fitting of the Fourier 
coefficients of the exact solution to (\ref{abd}) yields 
   $\delta = -0.0007$, $B=1.4935$ and 
   $A = -8.5846$, i.e., very good approximations to the theoretical 
   values. Similarly we get from the fitting of the Fourier 
   coefficients of $w$ the values  $\delta = 0.0003$, $B=1.4654$ and 
   $A = -7.5718$. 

\begin{figure}[htb!]
\centering
 \includegraphics[width=0.49\textwidth]{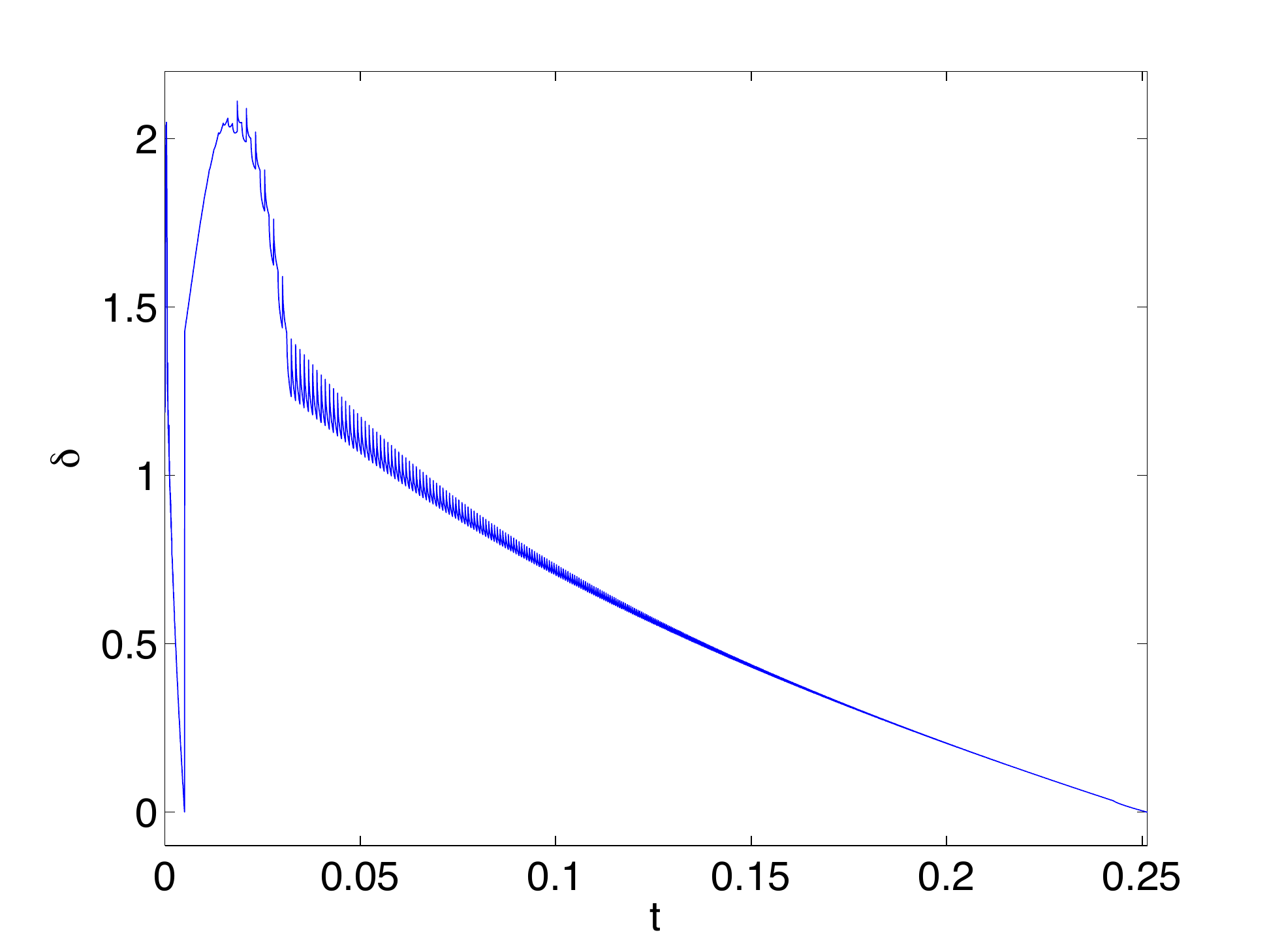} 
 \includegraphics[width=0.49\textwidth]{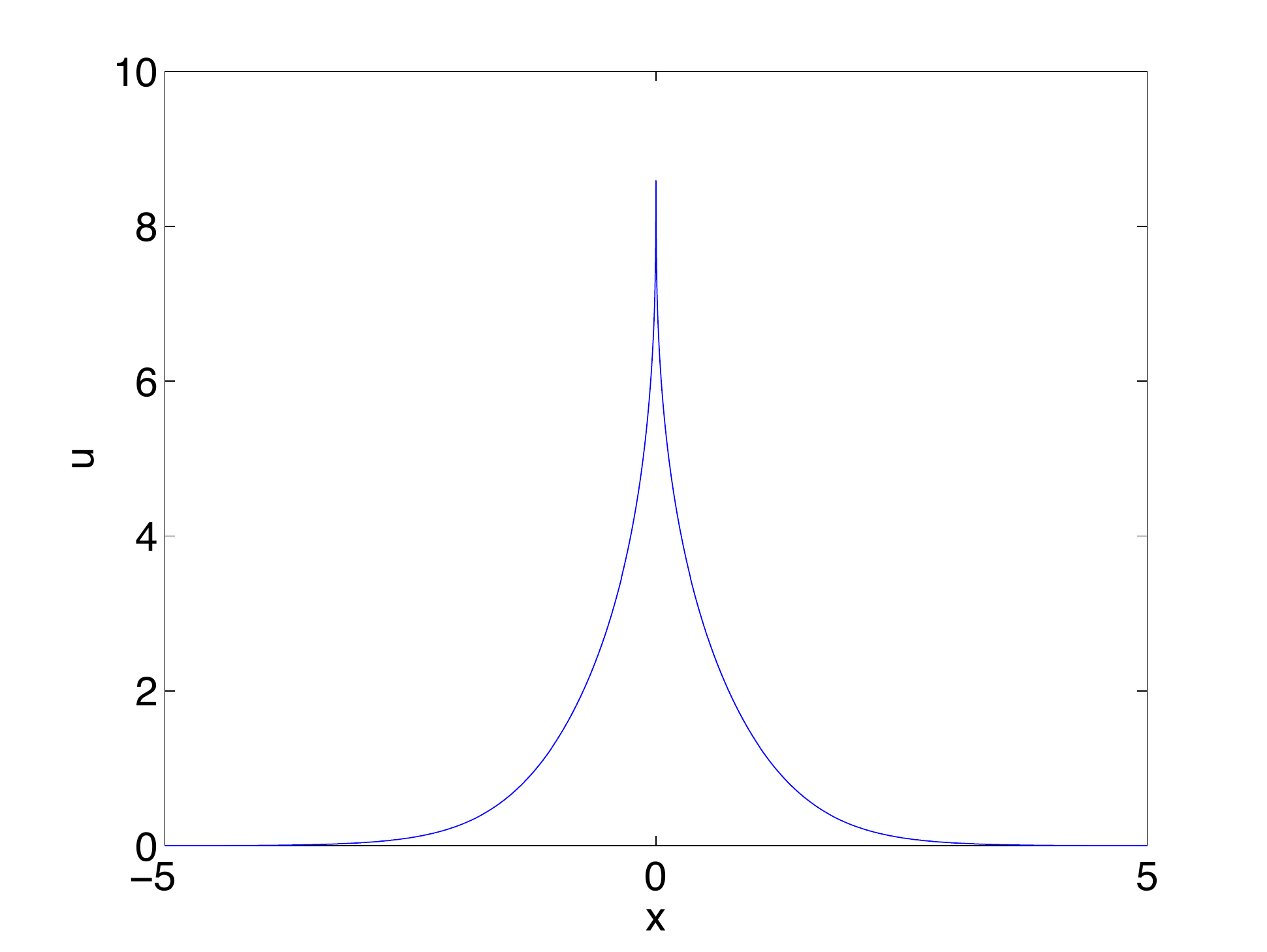} 
 \caption{Fitting parameter $\delta$ of the solution to the focusing  
 system (\ref{dnls1}) for the initial data 
 $u_{0} (x) = 4sech^{2}(x), \,\, w_0(x)=0$  in dependence of $t$ on 
 the left, and the solution $u$ at the last time with positive 
 $\delta$ on the right.}
 \label{fsNLSdelta}
\end{figure}

Since the main question in our context is whether the critical time 
$t_{c}$ can be identified from the asymptotic behavior of the Fourier 
coefficients, we let the code run until $t=0.255$ with the same 
parameters as before. The fitting of the Fourier coefficients is done 
at each time step for $10<k<\mbox{max}(k)/2$. The parameter $\delta$ in (\ref{abd})
vanishes at $t\sim 0.2512$ as can be seen in Fig.~\ref{fsNLSdelta}. 
The parameter $B$ at this time has the value $1.17$. The solution $u$ 
at this time can be seen in the same figure. The value of $u$ at the 
cusp is now roughly $8.5$, thus a bit larger than the correct value 
of 8. As stated in Remark \ref{ms}, 
the smallest distance in the used discrete Fourier space 
is equal to $2\pi L/N\sim0.0019$. 
In principle no distance below this threshold can be numerically 
distinguished from zero. If we stop the code when $\delta<0.0019$, we 
find a $t_{c}\sim 0.2506$ and a solution $u$ with a maximum of 
$7.93$, both very close to the theoretical values.

The difference between the defocusing and focusing case appears to be 
in the hyperbolicity respectively ellipticity of the semiclassical 
system. In the former case we can integrate up to a vanishing of the 
fitting parameter $\delta$ in (\ref{abd}), and both the critical time and solution 
are well approximated. In the latter case the ellipticity of the 
system implies the modulation instability which leads in particular 
to a pollution of the high wavenumbers. Since in addition the 
singularity is of higher order in this case (square root instead of 
cubic), we cannot get as close as in the hyperbolic case to the 
critical time. Thus the code has to be stopped as soon as $\delta$ is 
of the order of the smallest distance in Fourier space. This gives a 
slightly less accurate, but still satisfactory approximation to the 
critical time and solution than in the hyperbolic case.

The above example shows in fact that we find a critical time close to the
theoretical $t_{c}$ of the gradient catastrophe (\ref{thgc}), 
which indicates the fitting is reliable. 
However as already pointed out, the accuracy is much lower for the  
value of $B$ in (\ref{abd}), which is not close to the theoretical one ($1.5$). It was 
already discussed in \cite{dkpsulart} that the fitting procedure (\ref{abd}) is 
much more reliable for the exponential part, i.e., $\delta$ the 
vanishing of which gives the value of $t_{c}$. 
But it is less so for the algebraic dependence of the Fourier 
coefficients on $k$,  
and it appears that the  study of a two components system increases 
this effect. Since we are mainly concerned about the correct 
identification of the critical time, this is not  a problem in this 
context.

\subsection{Semiclassical DS II system}

We consider now the semiclassical DS system  (\ref{disDSs}) which is 
integrable in the sense of \cite{Kon07}.\\
% E = fft2(u.*(Sx.^2-Sy.^2)-rho*(u.^2-2*phi1.^2-2*phi2.^2));
To check the numerical accuracy, we use the energy,
\begin{equation}
E[u, S](t) := \int_{\mathbb{T}^2} \left(    u \left(  S_x^2 - S_y^2  \right) - \rho \left(  u^2 -2( \mathcal{P}^2(u) -2 \mathcal{Q}^2(u)  )    \right)                 \right) dxdy.
\end{equation}
where $\mathcal{P}$ and $\mathcal{Q}$ are defined in Fourier space by
\begin{equation*}
 \widehat{\mathcal{P}(f)} =  \frac{k_x^2}{k_x^2 + k_y^2} \widehat{f} (k_x, k_y), \,\,\,
  \widehat{\mathcal{Q}(f)} =  \frac{k_x k_y}{k_x^2 + k_y^2} \widehat{f} (k_x, k_y) .
\end{equation*}
\\
\\
To numerically solve the system, we use again a fourth order 
Runge-Kutta scheme, and a Krasny filter of order $10^{-14}$. In the 
following, the computations are carried out with $2^{14} \times 
2^{14}$ points for $x \times y \in [-5\pi, 5\pi] \times [-5\pi, 
5\pi]$. We always perform an asymptotic fitting of the Fourier 
coefficients in the $k_x$-direction as in \cite{dkpsulart} for the 
dKP equation, and in both spatial directions via the energy spectrum 
$\mathcal{E}(K)$ (\ref{enspec}) (see sect.~2). We will denote by $\delta_{1d}, B_{1d}$ the fitting parameters resulting from the one-dimensional study, for $ln (\hat{u}(k_x,0))$ in (\ref{abd}), and by $\delta_{2d}, B_{2d}$ those resulting from the two-dimensional study (\ref{abd2}). 
For the one-dimensional fitting, we consider in all cases the following range of 
the Fourier coefficients: $10<k_x<\max(k_x)/2$,  
and for the two-dimensional fitting, the corresponding range for 
$K=\sqrt{|k_x|^2 + |k_y|^2}$: $\sqrt{|k_{x_{min}}|^2 + 
|k_{y_{min}}|^2}<K<\sqrt{|k_{x_{max}}|^2 + |k_{y_{max}}|^2}$. For the 
resolution used, this gives $K_{min}\sim15$.

We find that the singularities appearing in the solutions are of the 
same type as above in the case of $1+1$-dimensional NLS equations, 
i.e., gradient catastrophes in one spatial dimension. Due to the 
symmetry properties of the studied initial data, these coincide with 
the coordinate axes (one component of the gradient blows up). In 
particular we find that
\begin{itemize}
    \item  Solutions to the defocusing variant of the 
    semiclassical DS II equation (\ref{disDSs}) show the 
    same type of break-up as for the corresponding limit of the 
    $1+1$-dimensional NLS equation: the solutions have two break-up 
    points in each spatial direction (not necessarily on the 
    coordinate axes and at the same 
    time) which are generically of cubic type as for generic 
    solutions to the Hopf 
    equation.

    \item  Solutions of the focusing variant of the semiclassical DS 
    II equation (\ref{disDSs}) have 
    in general two break-up points of the same type as solutions of 
    the focusing $1+1$-dimensional NLS equation, a square root cusp 
    for each spatial direction. For initial data with a symmetry with 
    respect to an interchange of the spatial coordinates, these cusps 
    appear at the same time and location. 

\end{itemize}

\subsubsection{Defocusing case}

\begin{figure}[htb!]
\centering
 \includegraphics[width=0.49\textwidth]{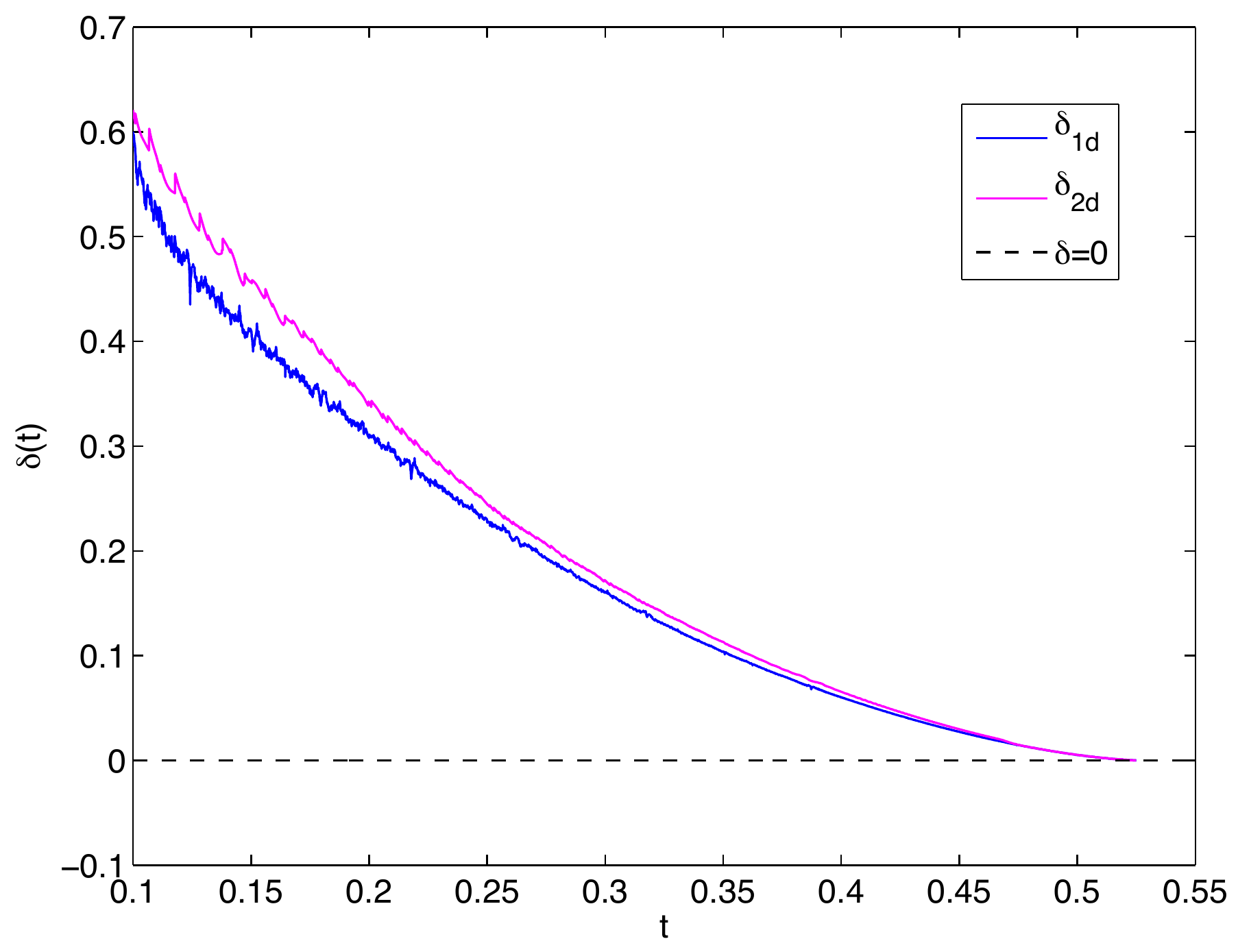} 
 \includegraphics[width=0.49\textwidth]{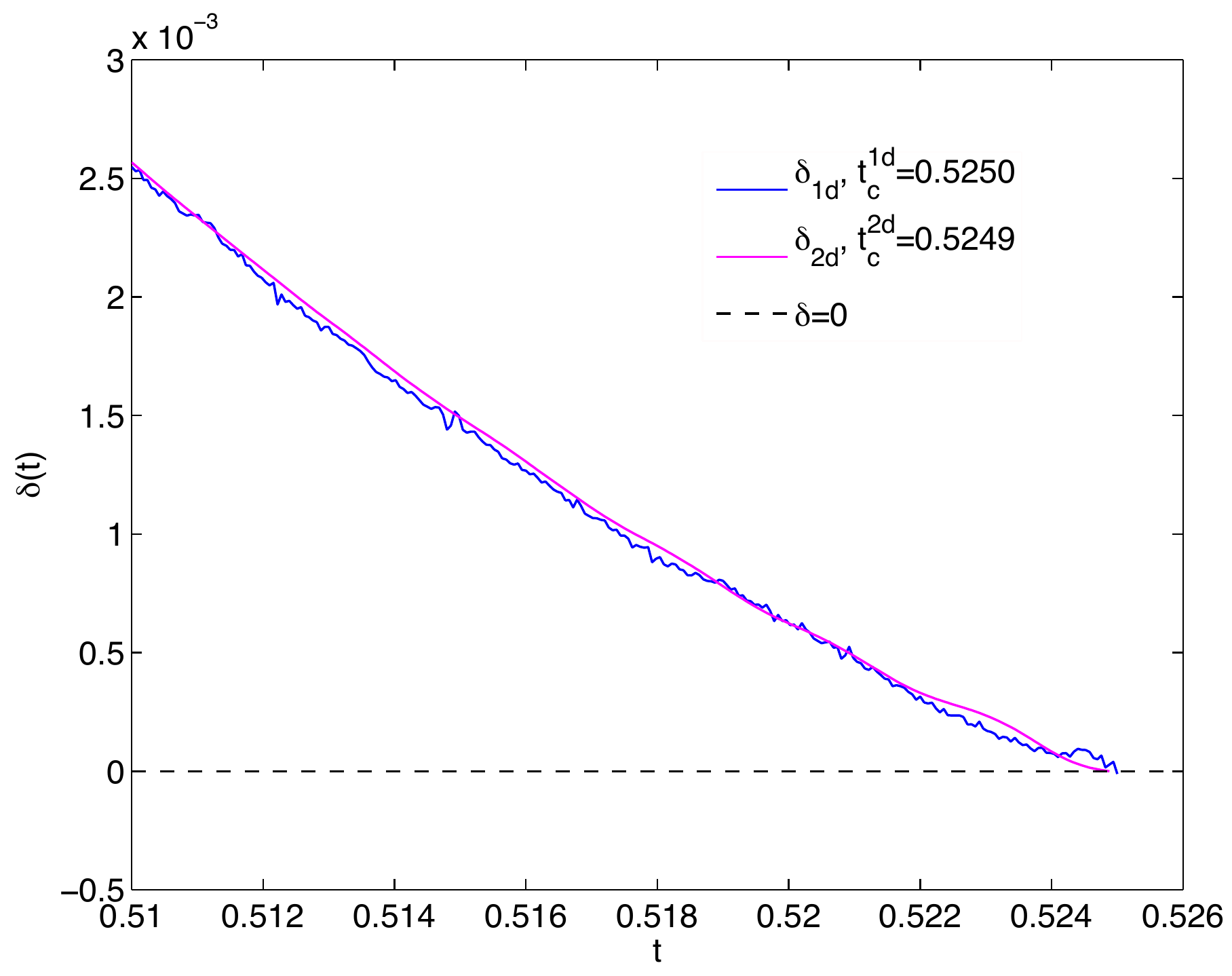} 
\caption{Time dependence of the fitting parameters $\delta_{1d}$ in (\ref{abd}) and $\delta_{2d}$ in (\ref{abd2})  on the 
left, and for $t$ close to $t_c$ on the right, for the numerical 
solution to the defocusing semiclassical DS II system (\ref{disDSs}) with initial data 
(\ref{adinisym})}.
 \label{defocdels}
\end{figure}

We first consider the defocusing system (\ref{disDSs}) $(\rho=1)$ for initial data of the form 
\begin{equation}
u(x,y,0) = e^{-2R^{2}}, \,\,\mbox{with}\,\, R = \sqrt{ x^{2}+y^{2} }, \,\, \mbox{and} \,\, S(x,y,0)=0,
\label{adinisym}
\end{equation}
which thus correspond to Gaussian initial data for the defocusing DS II equation. 
The time step is chosen as $\Delta_t=6*10^{-5}$.
The vanishing of $\delta_{1d}$ in (\ref{abd}) and $\delta_{2d}$ in (\ref{abd2}) occur at
 the same time $t_{c}\sim 0.525$, see Fig. \ref{defocdels}.

\begin{figure}[htb!]
\centering
\includegraphics[width=0.49\textwidth]{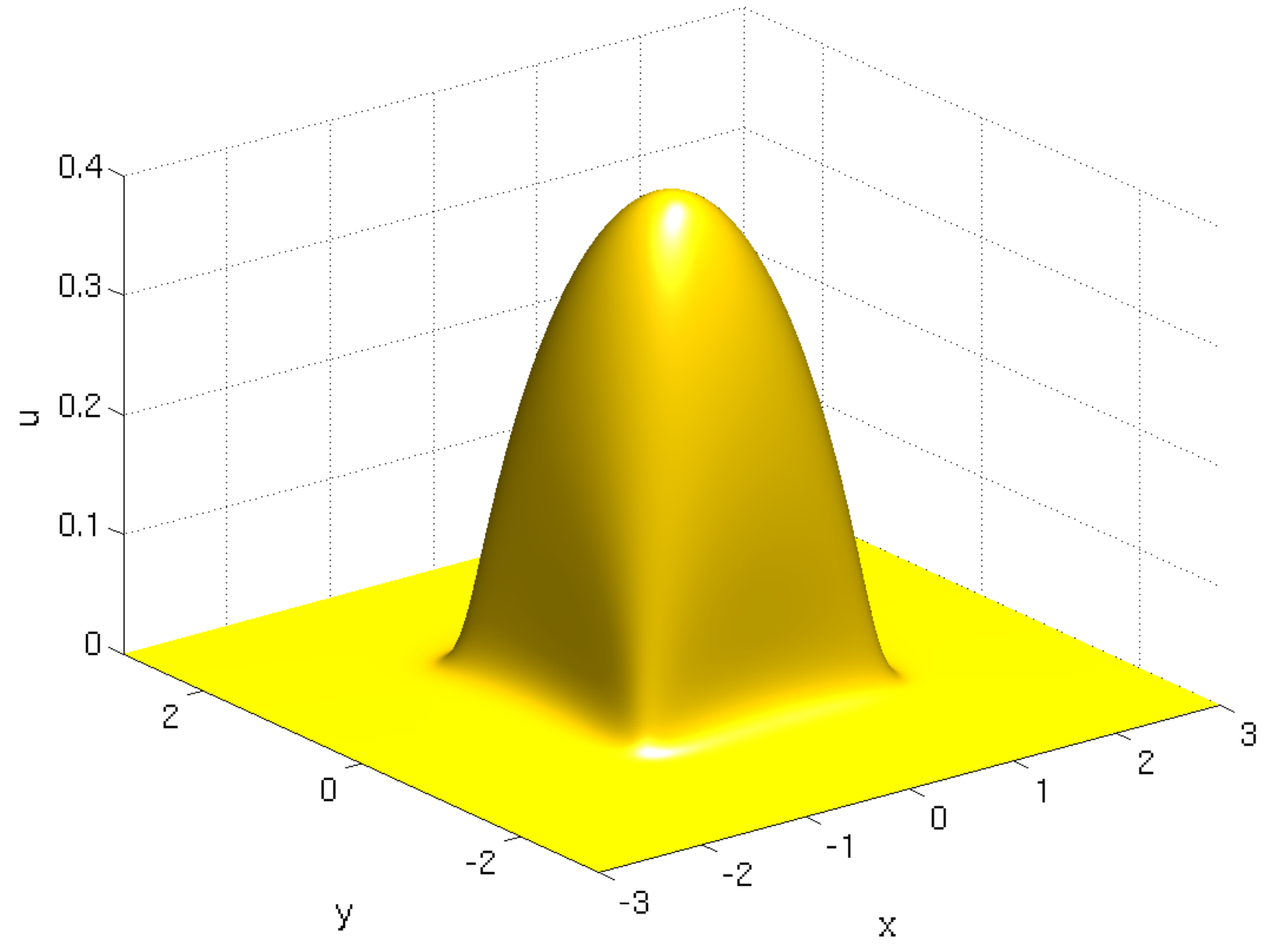}   
\includegraphics[width=0.49\textwidth]{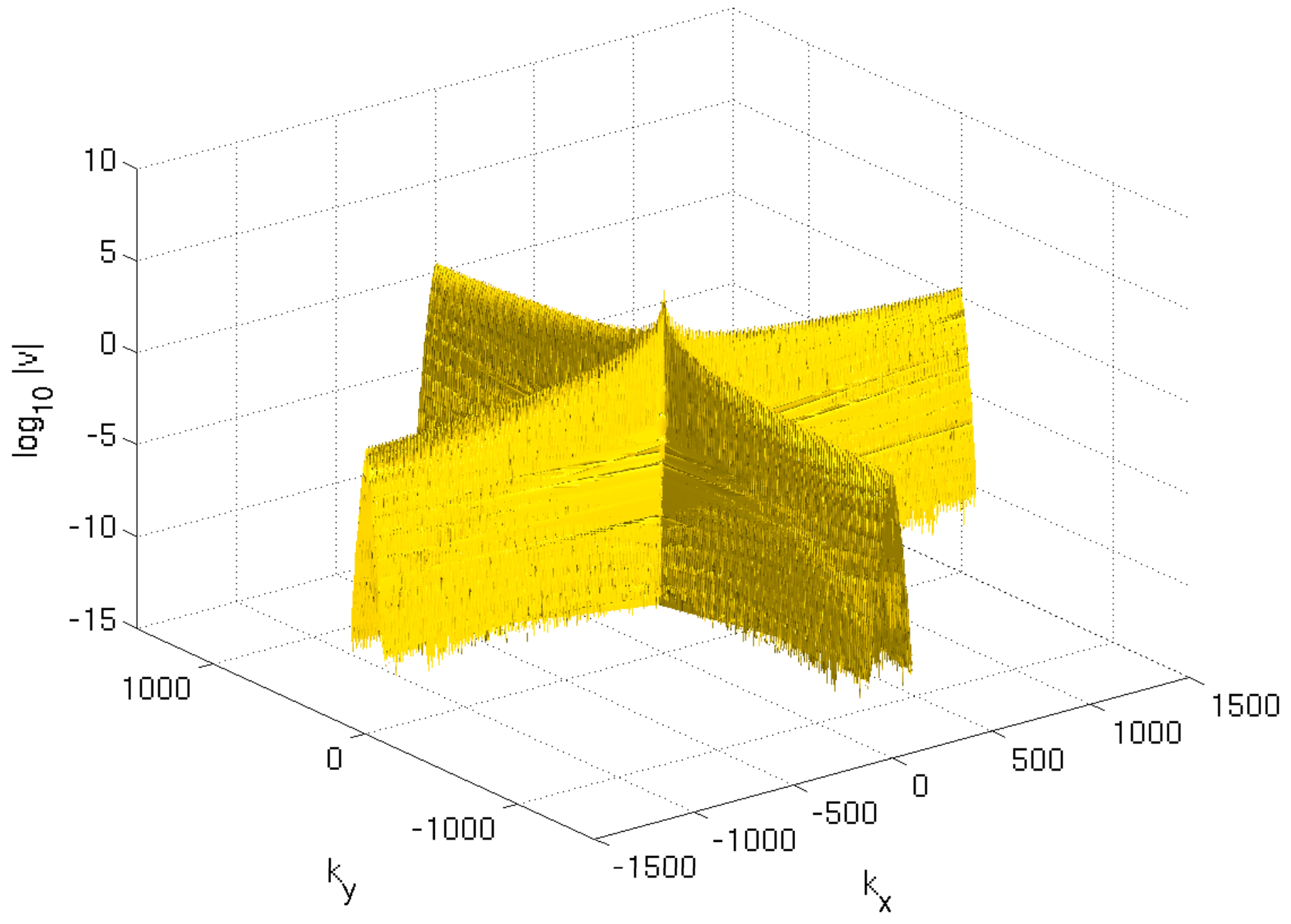} \\    
\includegraphics[width=0.49\textwidth]{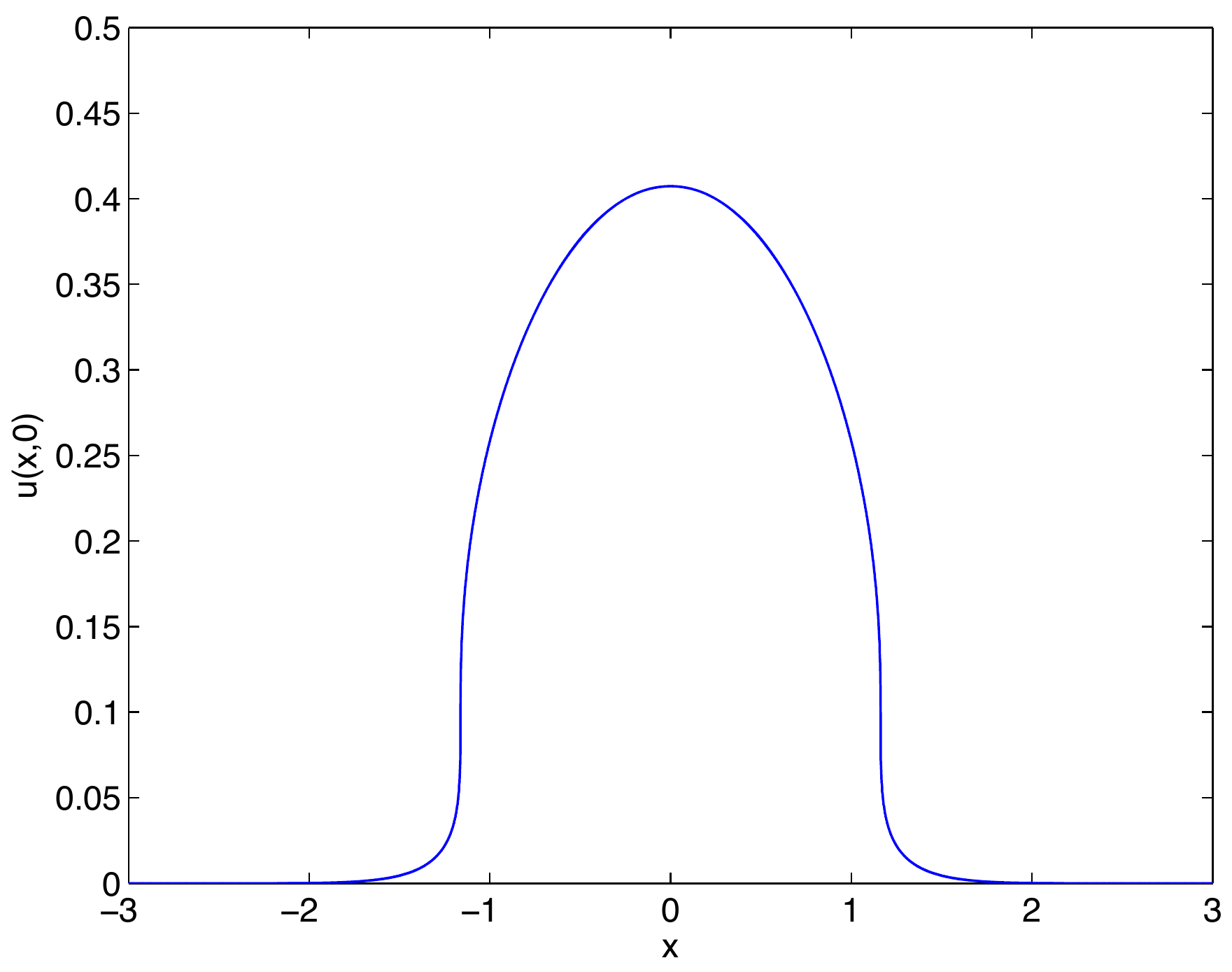}   
\includegraphics[width=0.49\textwidth]{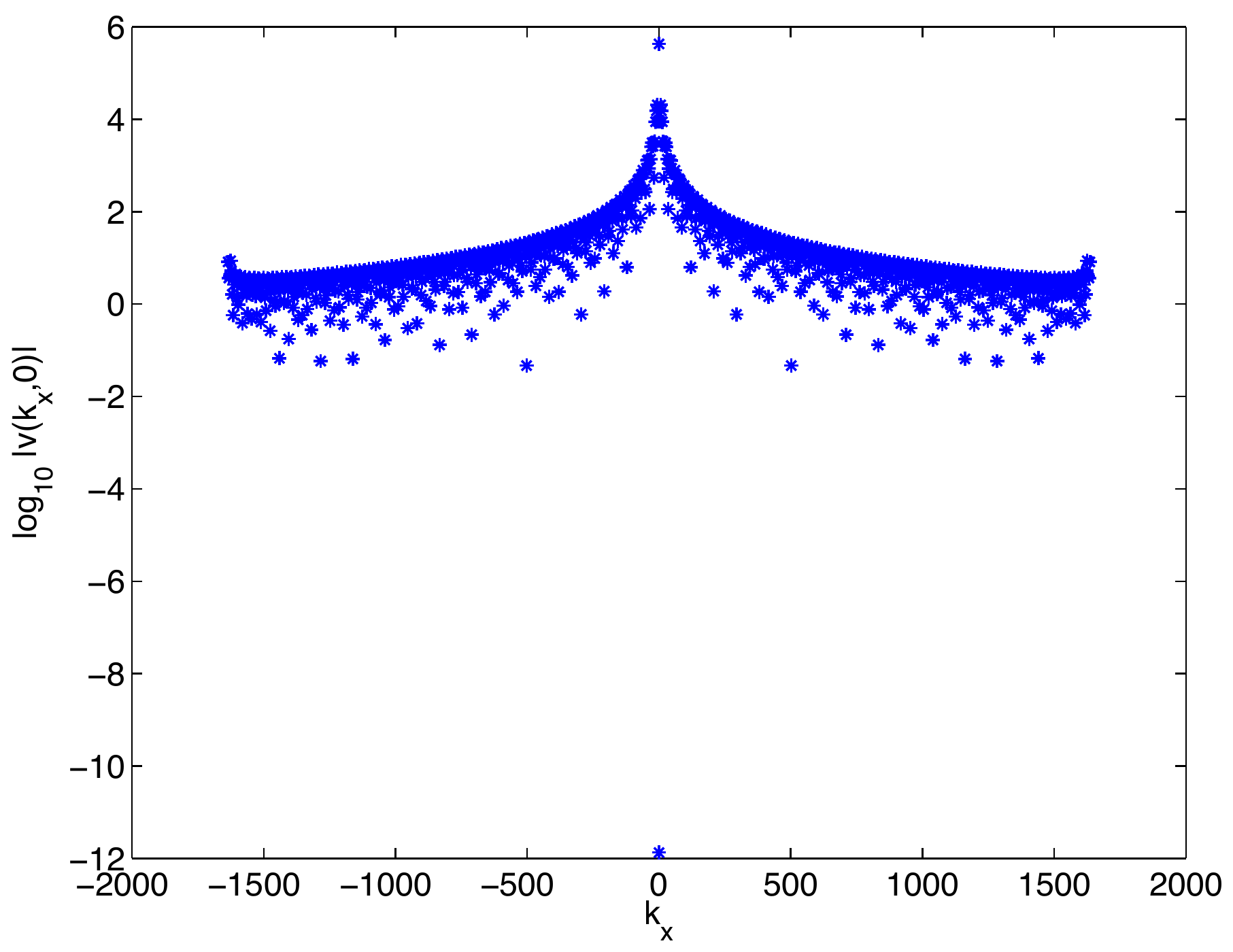} \\    
\caption{Numerical solution to the defocusing semiclassical DS II system 
(\ref{disDSs})  for initial 
data (\ref{adinisym}) at $t=0.525$, and its Fourier coefficients; in 
the lower row the corresponding figures on the $x$- respectively 
$k_{x}$-axis are given.} 
\label{defocuv2d}
\end{figure}

We show the 
solution to the defocusing semiclassical DS II system (\ref{disDSs}) at this time 
$t=0.525$ and its Fourier coefficients in Fig.~\ref{defocuv2d}.
It can be seen that the solution becomes steep on the 4 sides 
parallel to the coordinate axes, and the $x$-derivative of $u$, 
respectively the $y$-derivative, 
become big in two points on the $x$-axis, respectively $y$-axis, 
namely in $(x_c^{\pm}=\pm1.162, 0)$, respectively. $(0, y_c^{\pm}=\pm1.162)$. 
The other parameters attain the following values at $t=0.525$: 
$B_{1d}$ (\ref{abd}) reaches $B_{1d}=1.35$, $B_{2d}$ in (\ref{abd2}) $\sim2.99$, and the numerically computed 
energy (\ref{delE}) $\Delta_E=4.5*10^{-14}$. The 
situation is visibly similar to the $1+1$-dimensional example 
shown in Fig.~\ref{dsNLSu}, just that now 
four singularities form at the same time for symmetry reasons. This 
implies as in Fig.~\ref{dsNLSfourier} strong oscillations in the 
Fourier coefficients, now both in $k_{x}$ and $k_{y}$ direction. A 
direct consequence of this is that  the fitting errors cannot be used 
as an indicator of the quality of the fitting (one has $\Delta_{1d}$ 
of the order of $\sim5$ and $\Delta_{2d} \sim 1$).  But the fitting 
appears to be very reliable as in the $1+1$-dimensional case which is 
also confirmed by the value of $B_{1d} \sim \frac{4}{3}$ which again indicates a cubic 
singularity. 
\begin{figure}[htb!]
\centering
\includegraphics[width=0.49\textwidth]{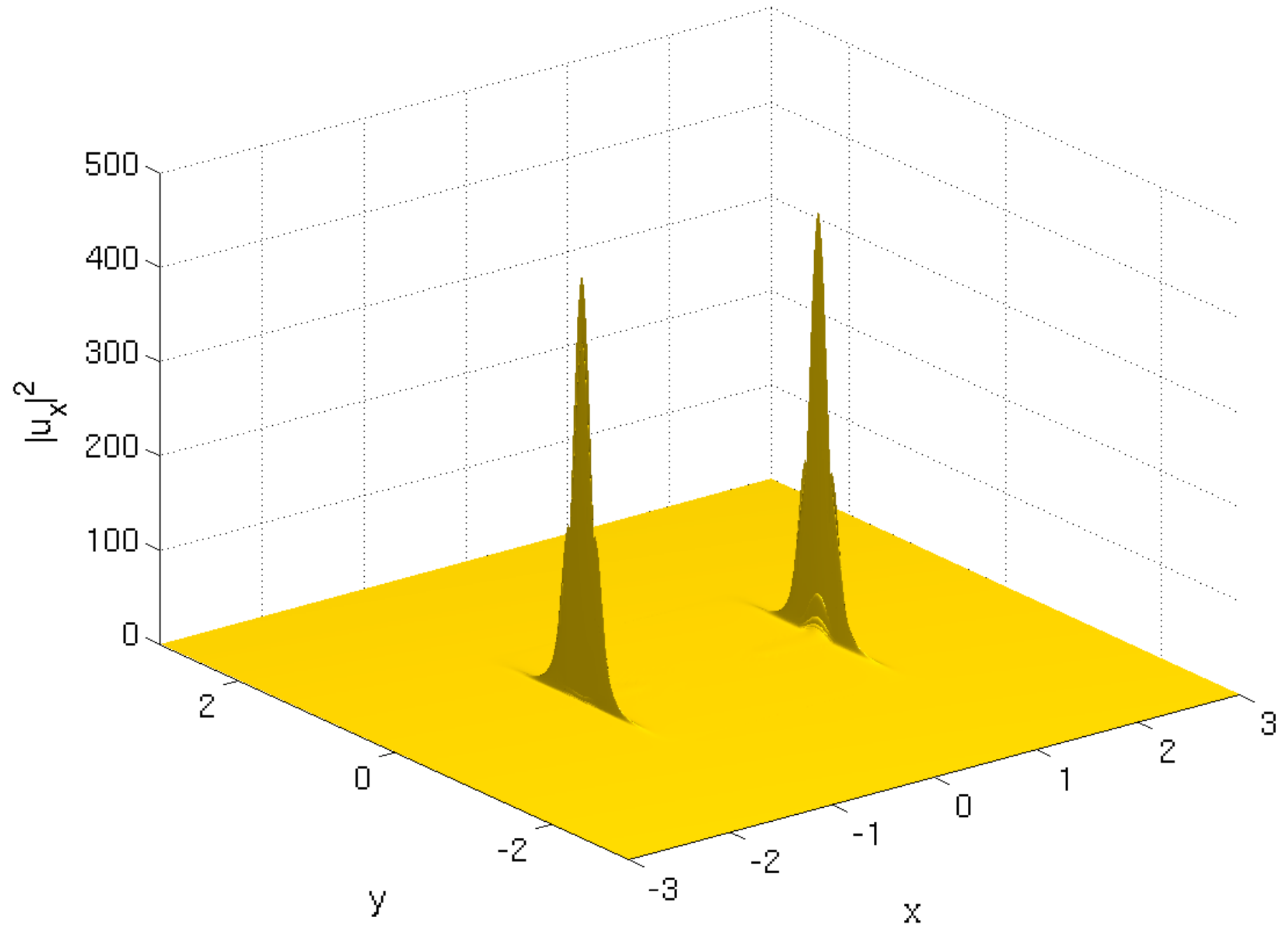}   
\includegraphics[width=0.49\textwidth]{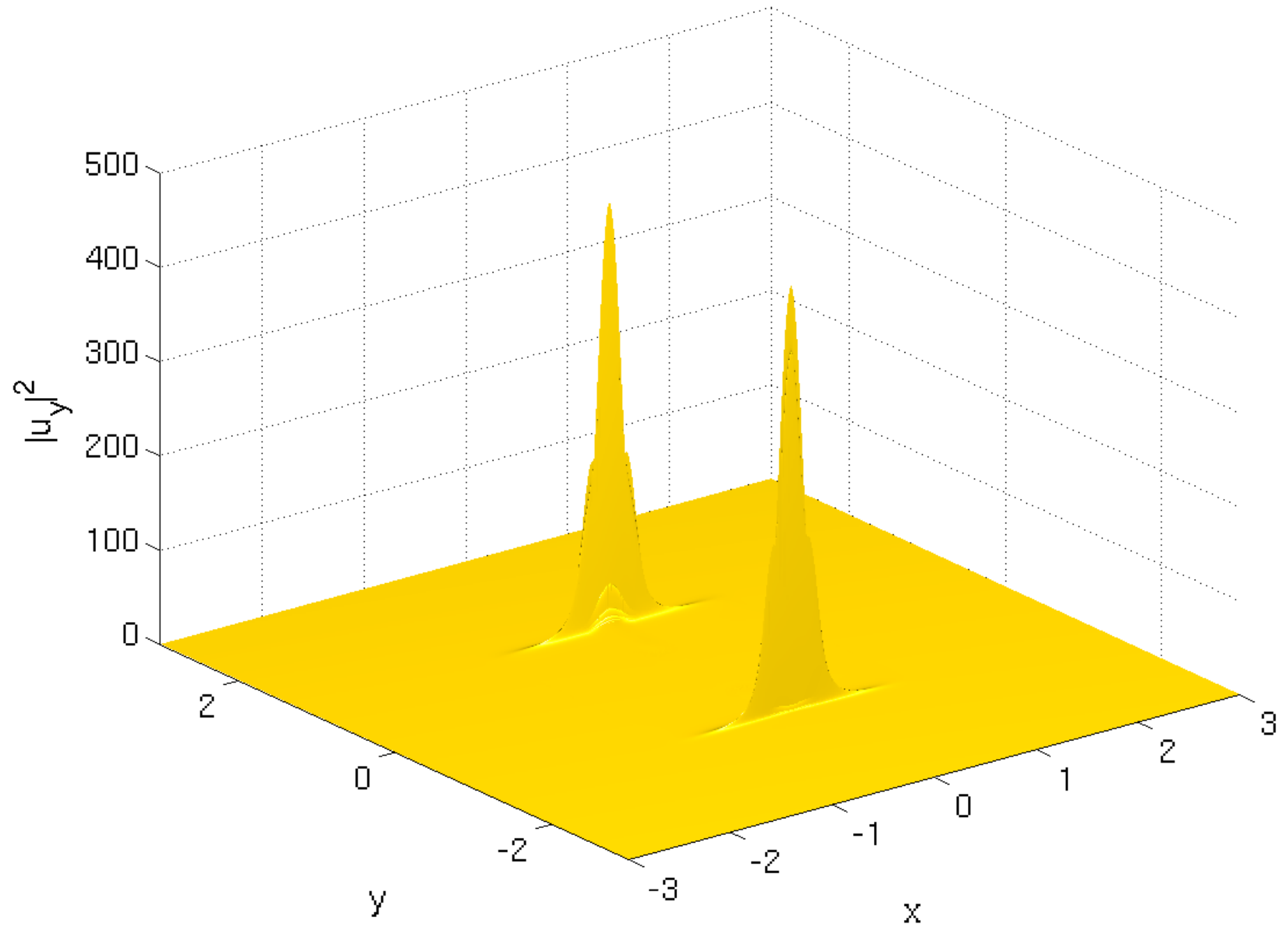}   
\caption{Derivatives of the numerical solution to the defocusing  semiclassical DS II 
system (\ref{disDSs}) with initial data (\ref{adinisym});  $|u_x|^2$ (left) and $|u_y|^2$ (right) at $t=0.525$. } 
\label{defocuxy2d}
\end{figure}

The results of this subsection can be summarized in the following 
\begin{conjecture}
Solutions to the defocusing semiclassical DS II system (\ref{disDSs}) 
for generic rapidly decreasing initial data with a single hump develop 
four points of gradient catastrophe in finite time. At each of these 
points, only one component of the gradient becomes infinite. The 
singularity is thus one dimensional, the solution at these points 
$x_{c}$ and $y_{c}$ respectively behaves as  $(x-x_{c})^{1/3}$ 
respectively $(y-y_{c})^{1/3}$.
\end{conjecture}

\subsubsection{Focusing case}

\begin{figure}[htb!]
\centering
 \includegraphics[width=0.5\textwidth]{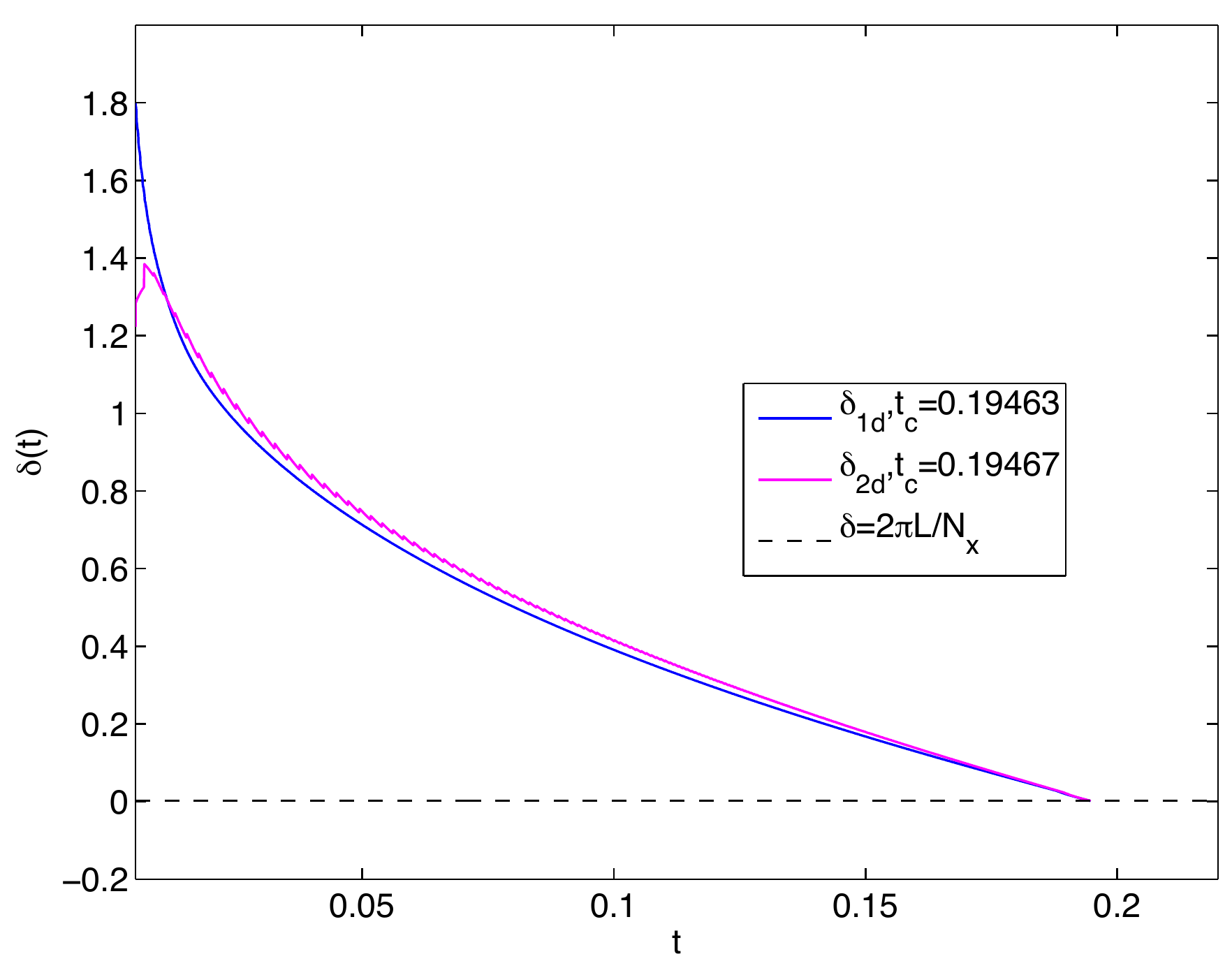} 
\caption{Time dependence of the fitting parameter $\delta$ 
% on the 
%left, and for $t$ close to $t_c$ on the right, 
for the numerical solution to the focusing semiclassical DS II system with initial data (\ref{adinifoc})}
 \label{delsfoc01}
\end{figure}

\begin{figure}[htb!]
\centering
\includegraphics[width=0.49\textwidth]{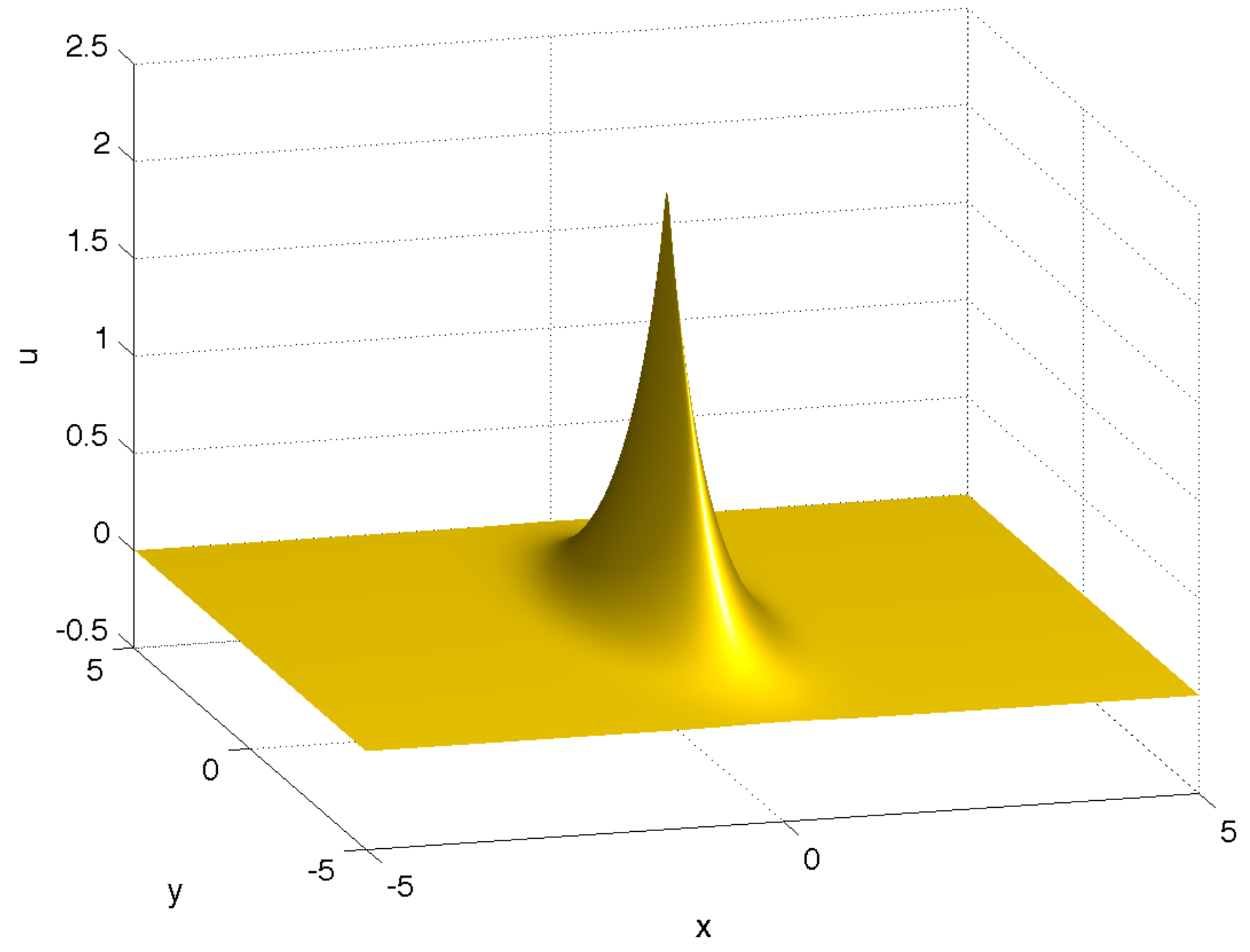}   
\includegraphics[width=0.49\textwidth]{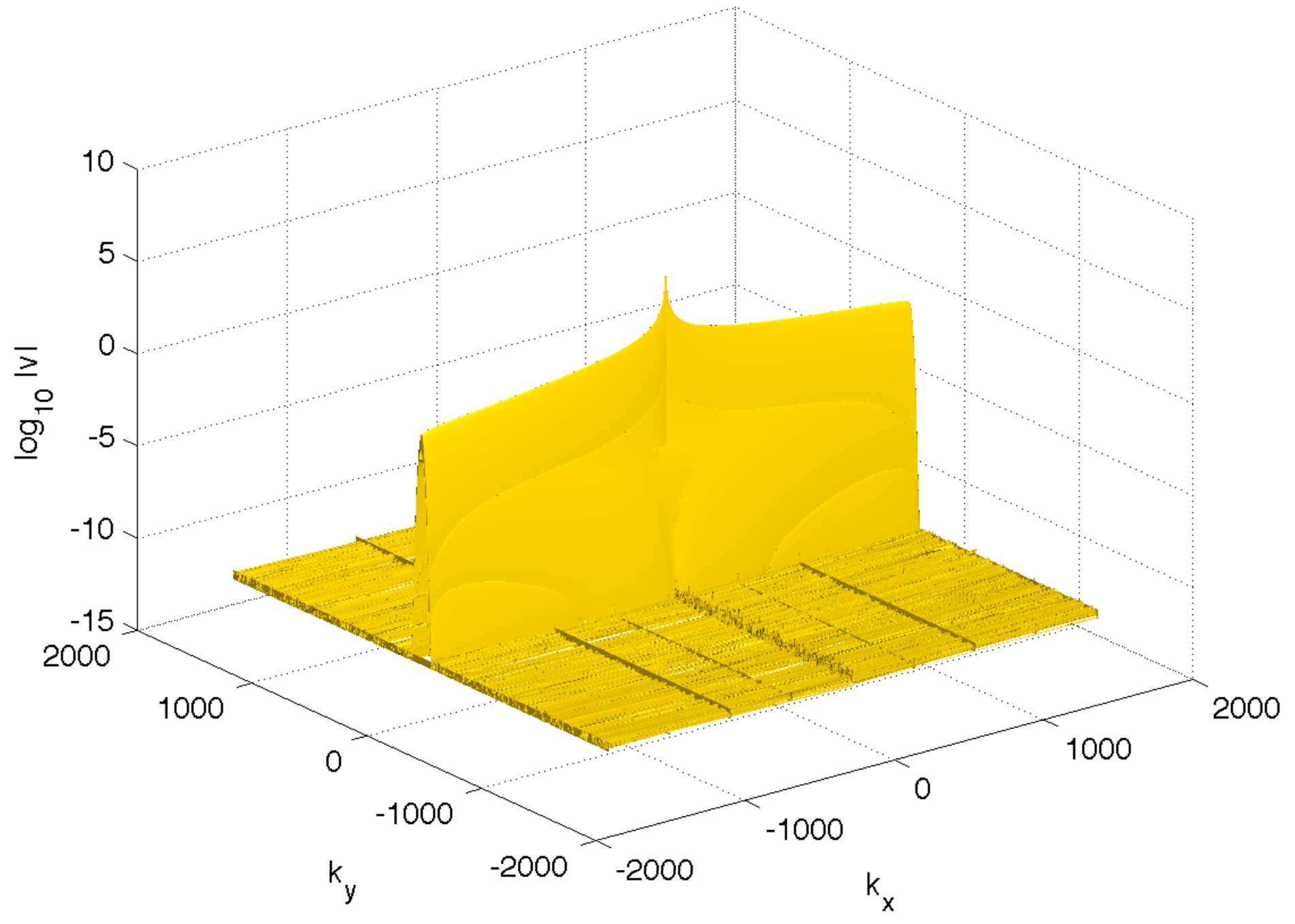} \\    
\caption{Numerical solution to the focusing semiclassical DS II system  
(\ref{disDSs}) with 
initial data (\ref{adinifoc}) at $t=0.1946$ on the left, and its Fourier 
coefficients on the right.} %(top), %
%plotted on the $x$-axis (resp. $k_x$-axis) on the bottom}
 \label{foc01uv2d}
\end{figure}

A similar study as above is presented for the focusing ($\rho=-1$) 
semiclassical DS II system (\ref{disDSs}), first with non-symmetric initial data
\begin{equation}
u(x,y,0) = e^{-2R^{2}}, \,\,\mbox{with}\,\, R = \sqrt{ x^{2}+ 0.1 y^{2} }, \,\, \mbox{and} \,\, S(x,y,0)=0.
\label{adinifoc}
\end{equation}
The time step  is chosen as $\Delta_t=3*10^{-5}$.
The vanishing of $\delta_{1d}$ in (\ref{abd}) and $\delta_{2d}$ in (\ref{abd2}) occur in this case 
roughly at the same time, $t_c\sim 0.1946$, see Fig.~\ref{delsfoc01}. 
The
solution to the focusing semiclassical DS II system  (\ref{disDSs}) at $t=0.1946$ and its Fourier 
coefficients can be seen in Fig.~\ref{foc01uv2d}.
Visibly the solution develops a cusp in the $x$-direction at $t_c$, 
which is also reflected by both $\delta_{1d}$ and $\delta_{2d}$ 
vanishing. The parameter $B_{1d}$ in (\ref{abd}) reaches a 
value of $B_{1d}(t_c)\sim 1.24$ and $B_{2d}(t_c) \sim 2.77$ in (\ref{abd2}).  As in 
the $1+1$-dimensional case we thus do not recover a value of $B_{1d}$ 
close to $1.5$, but we get essentially what was observed there. One 
can conclude that the solution at the critical point has a square 
root type cusp. Presumably a second such cusp would form at a later 
time in $y$-direction if the code could be run beyond the first 
critical time.

\begin{figure}[htb!]
\centering
 \includegraphics[width=0.49\textwidth]{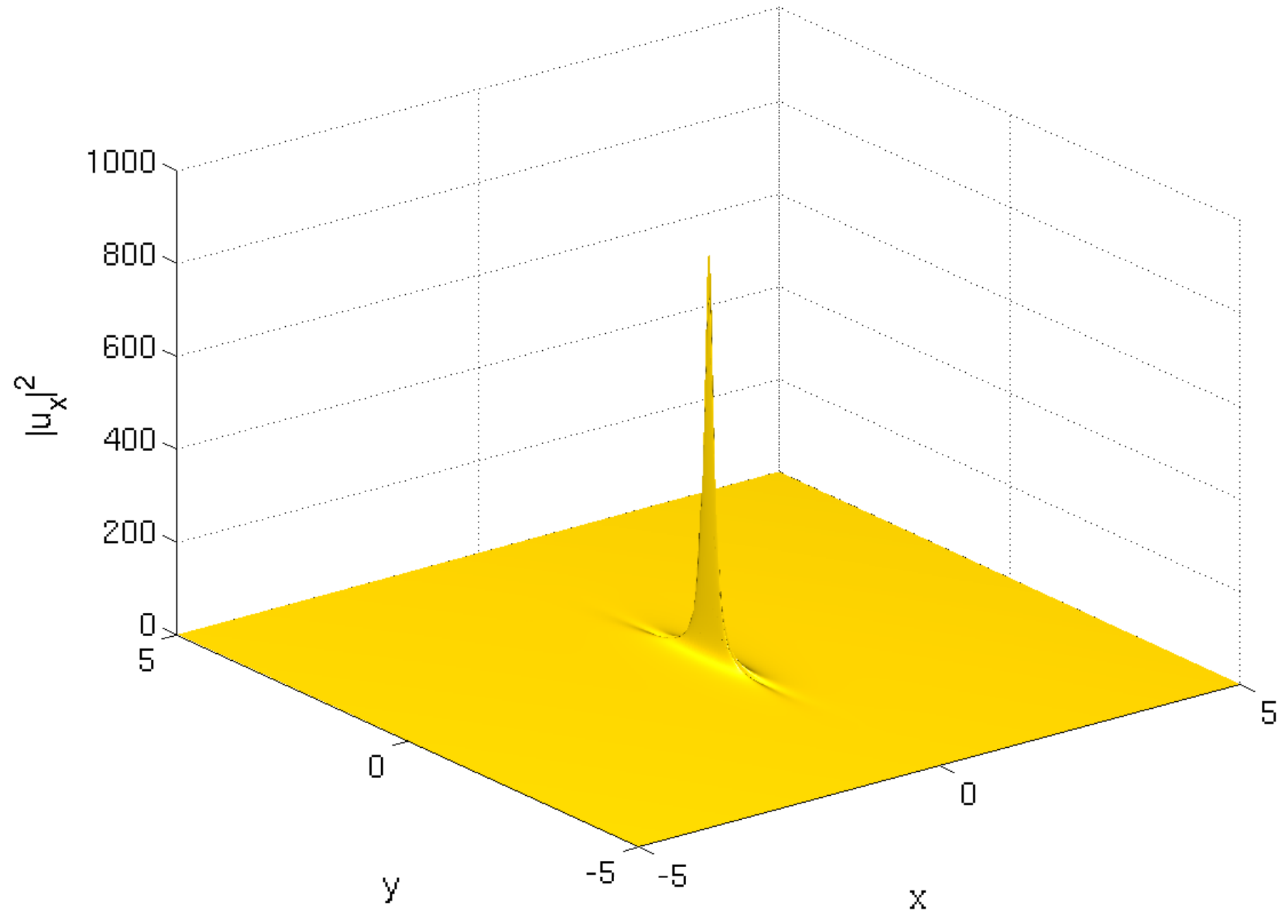} 
 \includegraphics[width=0.49\textwidth]{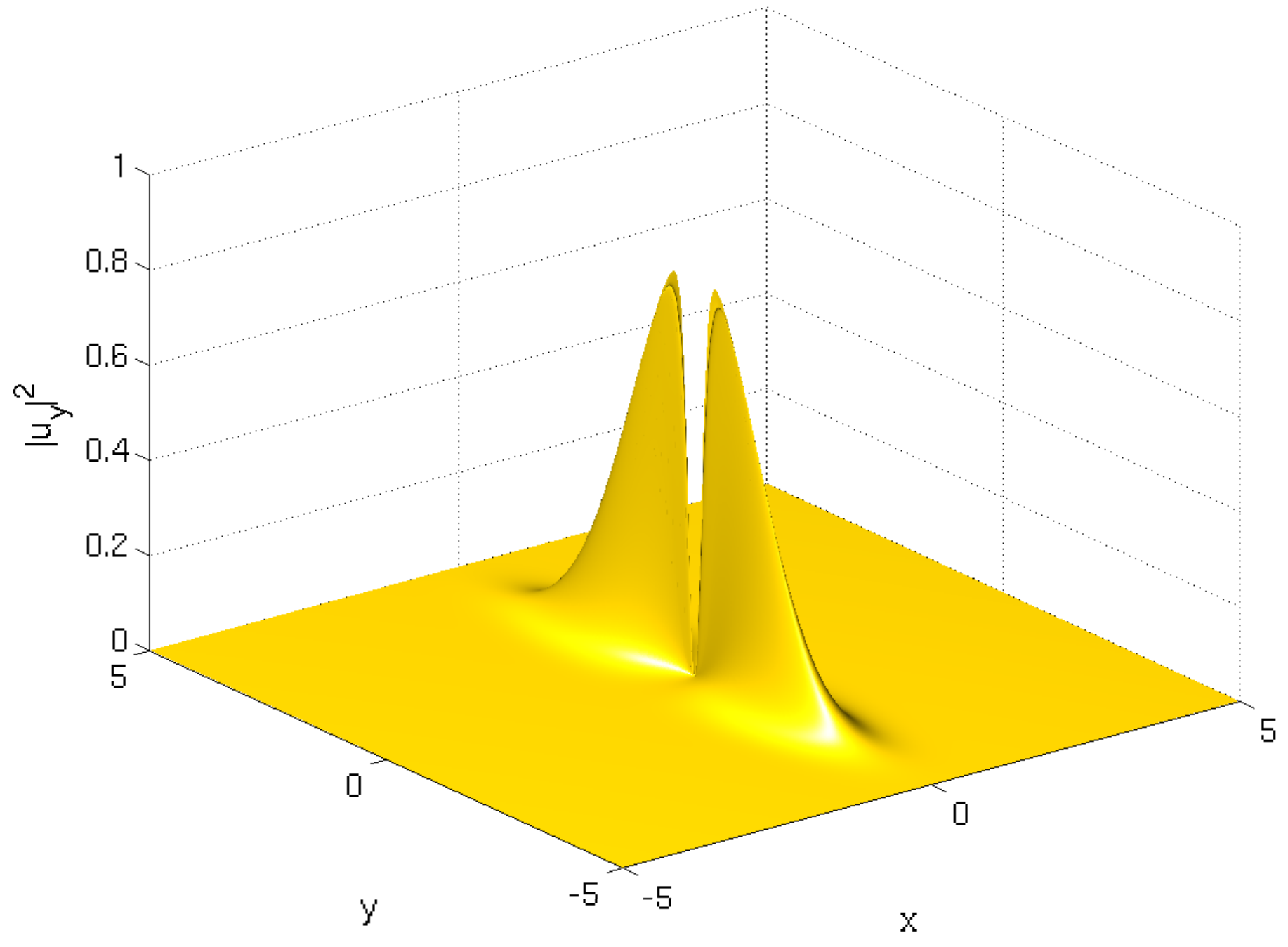} 
\caption{Derivatives of the numerical solution to the focusing semiclassical DS II system (\ref{disDSs}) with initial 
data (\ref{adinifoc}),  $|u_x|^2$ (left) and $|u_y|^2$ (right) at $t=0.1946$. }
 \label{Gradfoc01}
\end{figure}

\begin{figure}[htb!]
\centering
\includegraphics[width=0.49\textwidth]{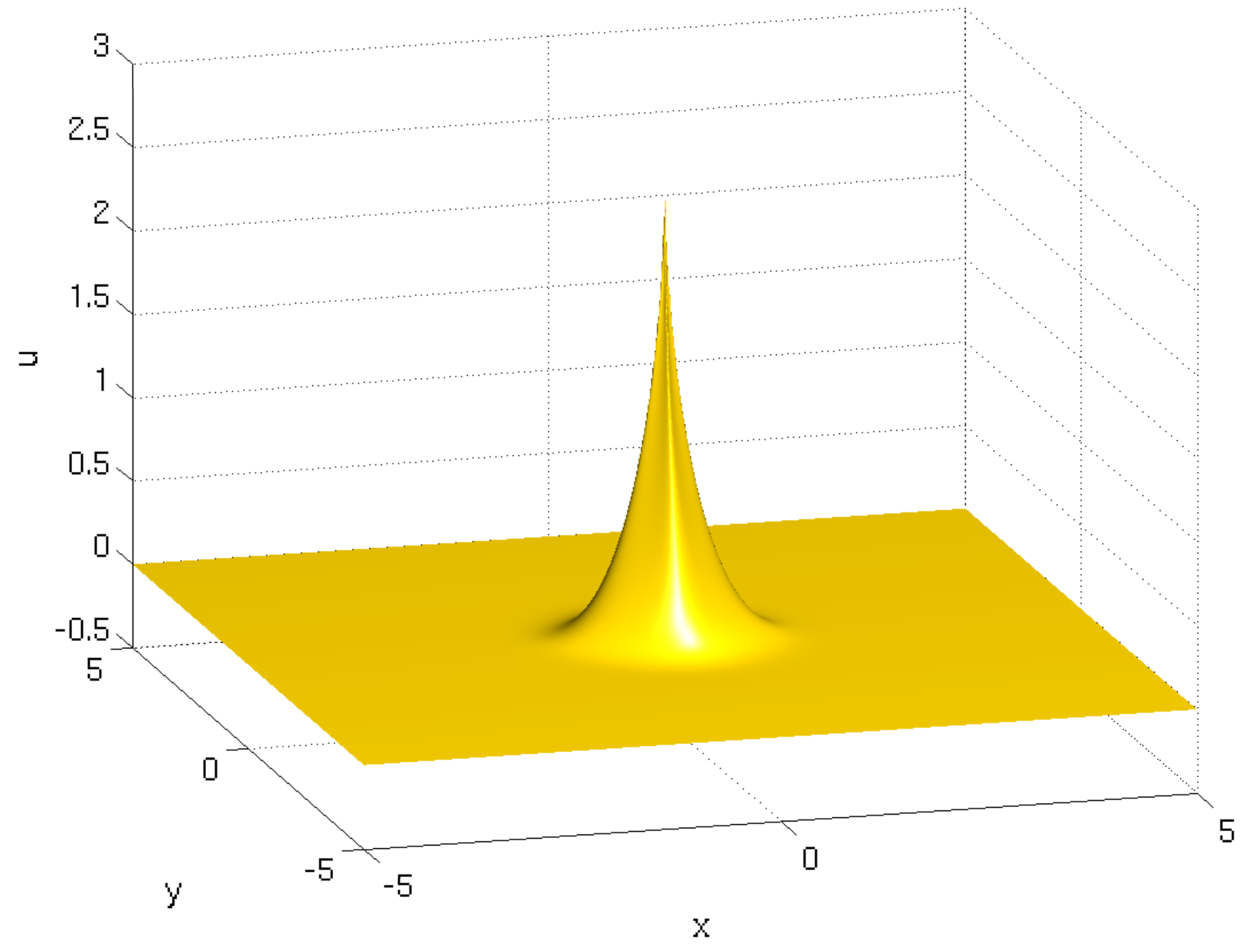}   
\includegraphics[width=0.49\textwidth]{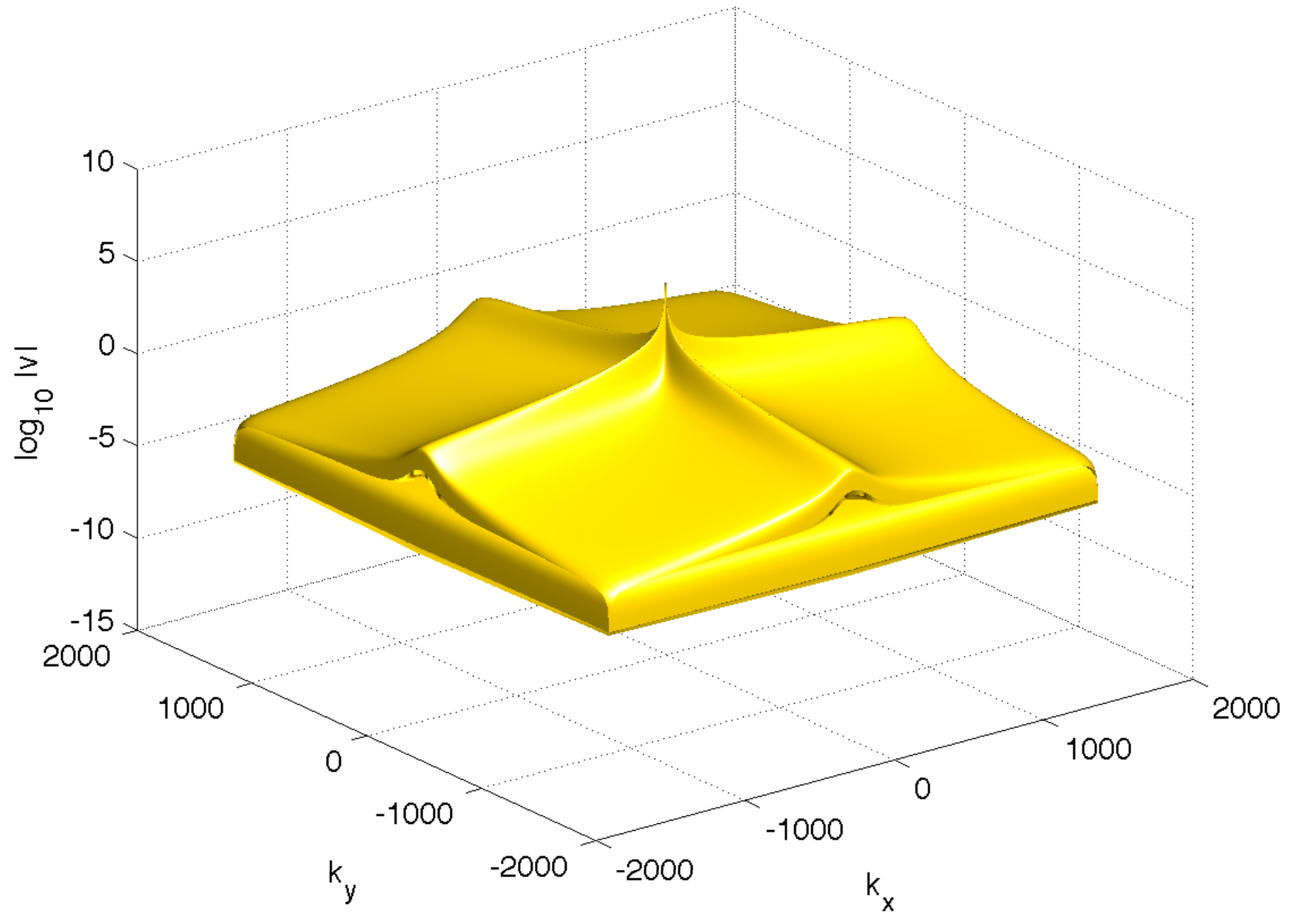} \\    
\caption{Numerical solution to the focusing semiclassical DS II system (\ref{disDSs}) with initial 
data (\ref{syminifoc}) at $t=0.2153$, and its Fourier coefficients.}
 \label{foc1uv2d}
\end{figure}

\begin{figure}[htb!]
\centering
\includegraphics[width=0.55\textwidth]{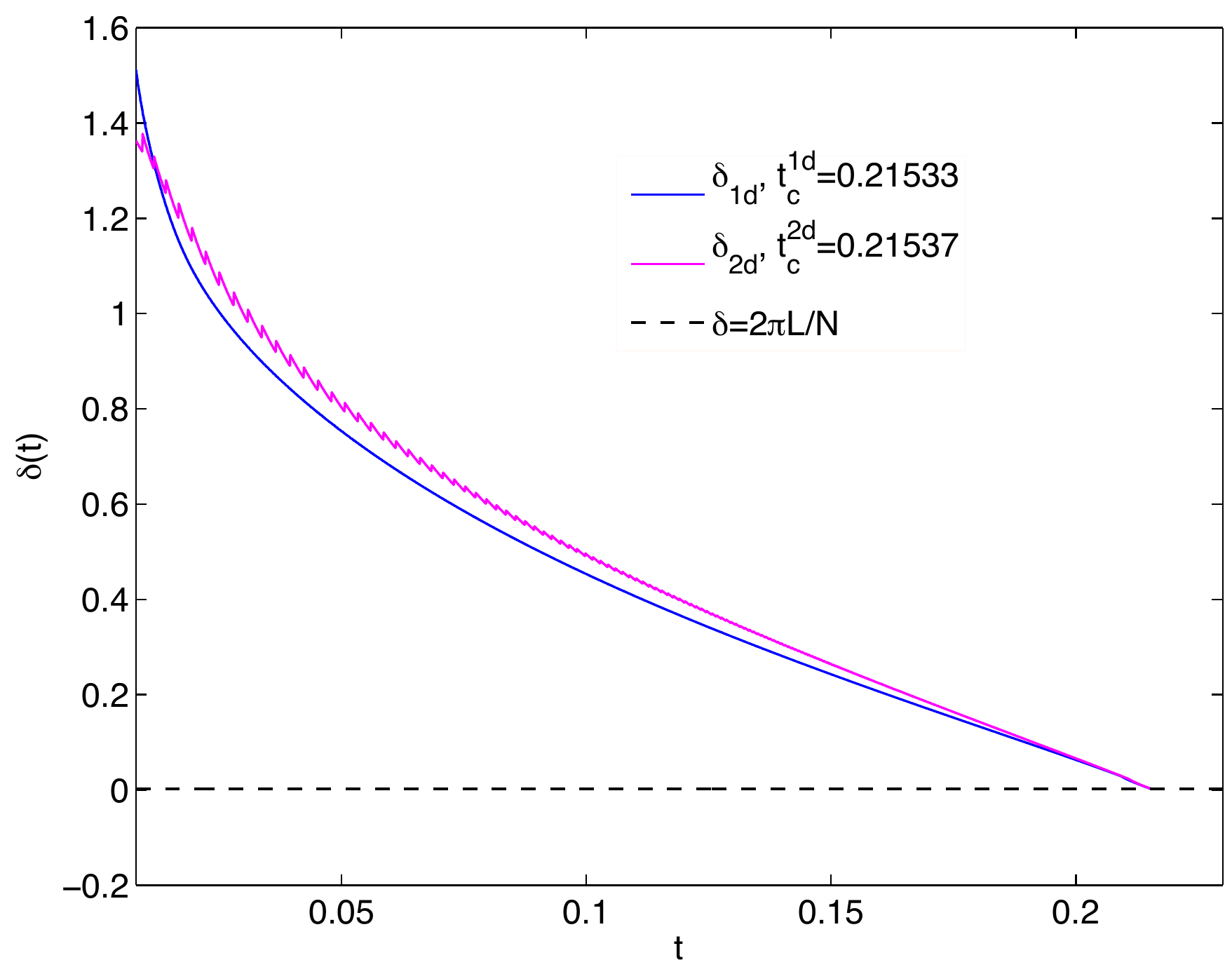} 
\caption{Time dependence of the fitting parameter $\delta$ 
%on the left, and for $t$ close to $t_c$ on the right, 
for the numerical 
solution to the focusing semicalssical DS II system (\ref{disDSs}) with initial data 
(\ref{syminifoc}).}
 \label{delsfoc1}
\end{figure}

\begin{figure}[htb!]
\centering
 \includegraphics[width=0.49\textwidth]{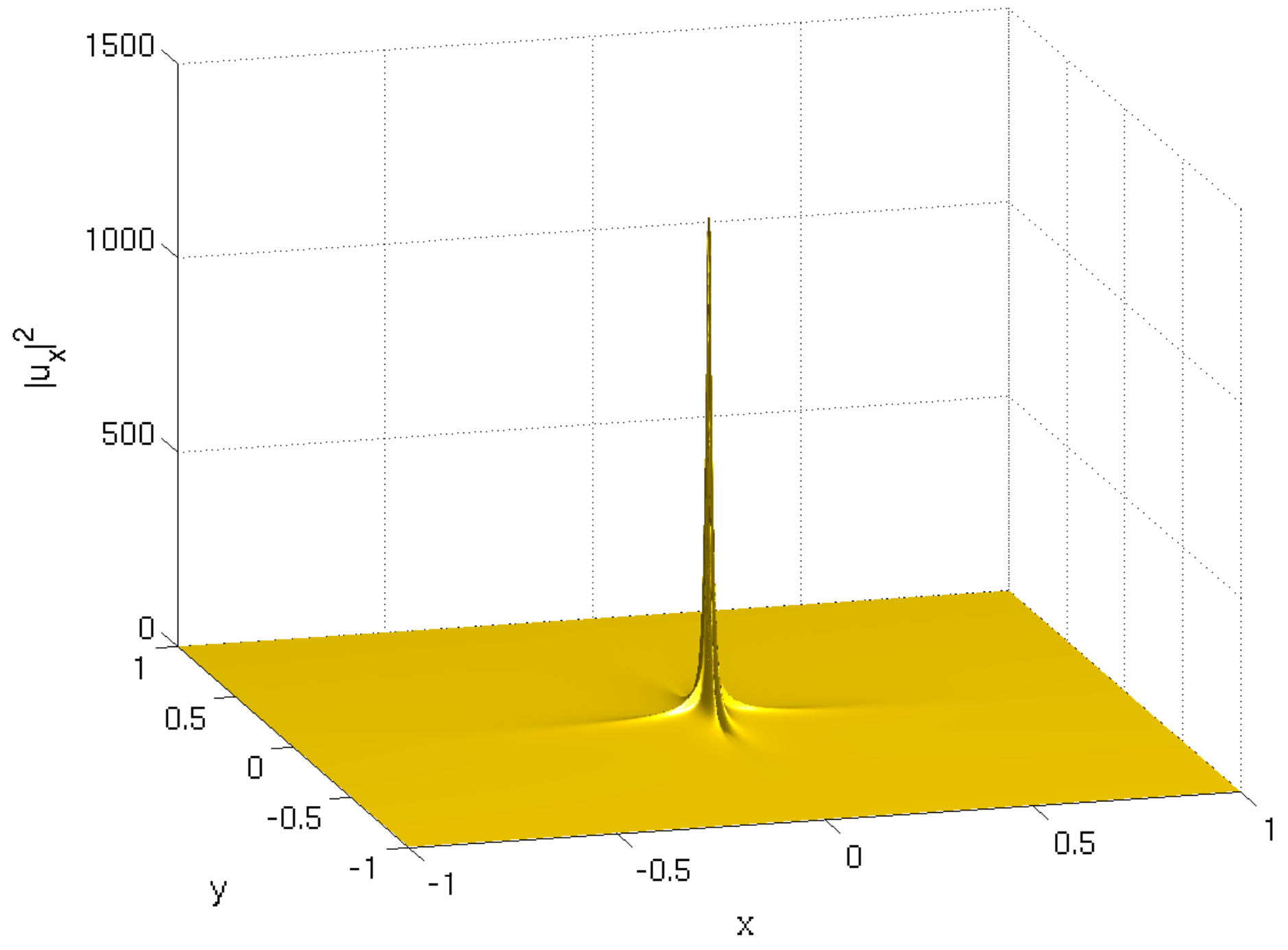} 
 \includegraphics[width=0.49\textwidth]{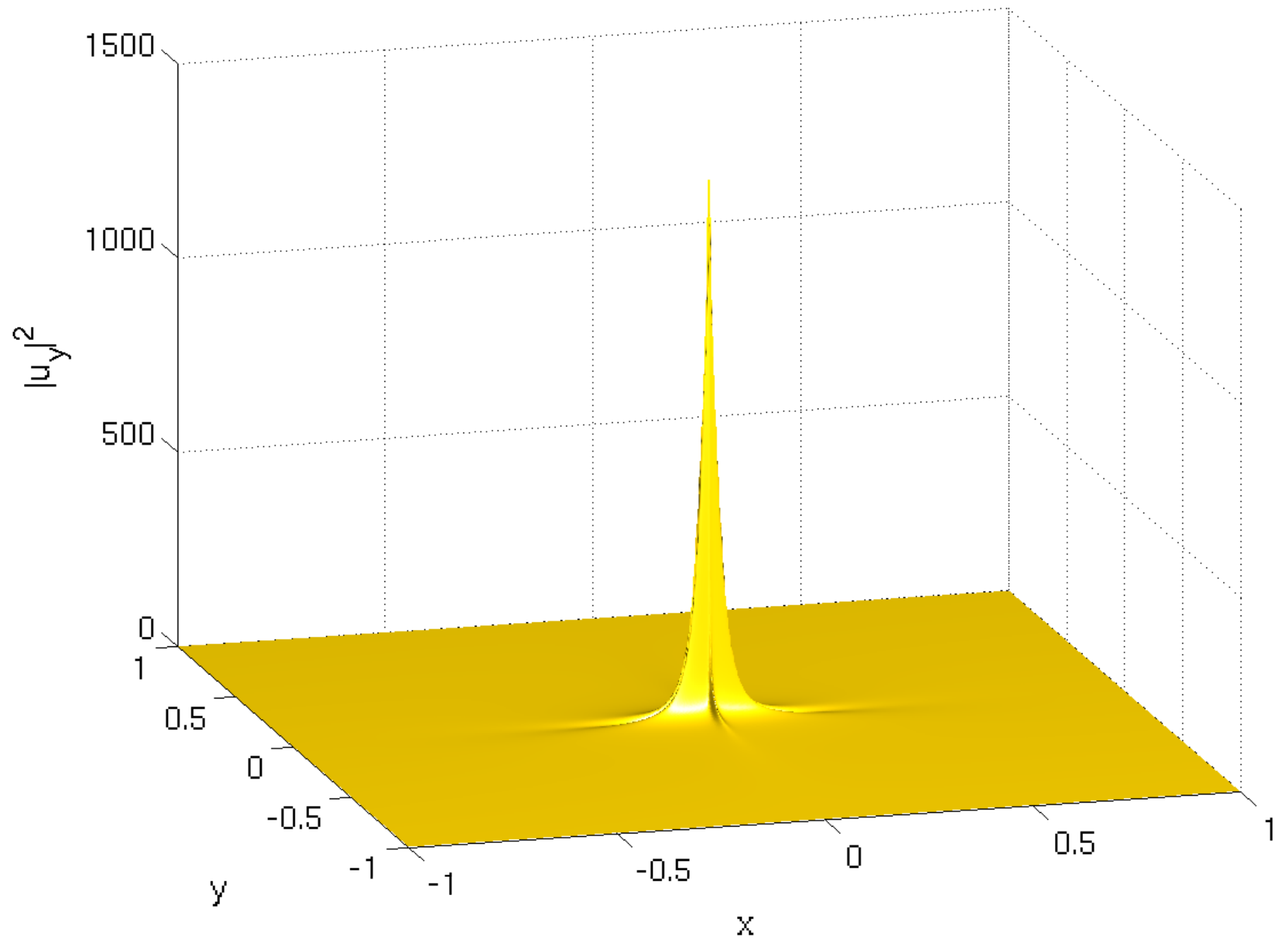} 
\caption{%Time evolution of $\|u_x\|_{\infty}$ on the left, and 
Derivatives of the numerical solution to the focusing semiclassical DS II system with initial data (\ref{syminifoc}), 
$|u_x|^2$ (left) and $|u_y|^2$ (right) at $t=0.2153$.}
 \label{Gradfoc1}
\end{figure}

At $t=t_c$, the $x$-gradient of $u$ begins to explode, with $\| u_x 
\|_{\infty} \sim 30$, see Fig.~\ref{Gradfoc01}, 
where we show $|u_x|^2$ on the left and $|u_y|^2$ on the right. We observe 
 that the gradient catastrophe only appears in one spatial point, 
 $(x_c, y_c)$, here $(0,0)$.  The fitting error $\Delta_{1d}$ is of 
 the order $ 0.2$ 
 which is again similar to what was observed in the $1+1$-dimensional 
 case, and $\Delta_{2d}\sim 2$ at $t_c$. 
\\
\\
\\
The situation is quite different if we consider symmetric initial data, 
\begin{equation}
u(x,y,0) = e^{-2R^{2}}, \,\,\mbox{with}\,\, R = \sqrt{ x^{2}+ y^{2} 
}, \,\, \mbox{and} \,\, S(x,y,0)=0.
 \label{syminifoc}
 \end{equation}
In this  case a cusp occurs in both spatial directions at the same 
time, as  can be seen in Fig. \ref{foc1uv2d}. There we show the 
solution to the focusing semiclassical DS II system  (\ref{disDSs}) with initial data (\ref{syminifoc}) at 
$t=t_c$, the latter being determined, as before, by using a fitting 
for the asymptotic behavior of the Fourier coefficients.
The bounds for the fitting are chosen as 
%$k_{min}=10, k_{max}=max(k_x)/2$, 
previously
and yield a vanishing of $\delta_{1d}$ in (\ref{abd}) and $\delta_{2d}$ in (\ref{abd2}) at roughly the same time $t_c\sim0.2153$, see Fig. \ref{delsfoc1}. 
The parameter $B_{1d}$ (\ref{abd}) reaches a value of $B_{1d}\sim 1.21$ at this time, and $B_{2d}\sim2.55$ (\ref{abd2}). 
The fitting errors are roughly of the same order as in the previous case, 
one gets $\Delta_{1d} \sim 0.3$ and $\Delta_{2d} \sim 0.7$. The $L_{\infty}$-norm of $u_x$ 
clearly explodes with $\| u_x \|_{\infty} \sim 35$, as well as the $L_{\infty}$-norm of $u_y$ which reaches 
the same value. 
%We show in Fig. \ref{Gradfoc1} the . 
%This time the same is the case for $u_y$, as we can see in Fig. \ref{Gradfoc1}, where 
We show in Fig. \ref{Gradfoc1} the $x$- and $y$-derivatives of $u$.

The results of this subsection can be summarized in the following 
\begin{conjecture}
Solutions to the focusing semiclassical DS II system (\ref{disDSs}) 
for generic rapidly decreasing initial data with a single hump develop 
two points of gradient catastrophe in finite time. At each of these 
points only one component of the gradient is unbounded. The 
singularity is thus one dimensional, the solution at these points 
$x_{c}$ and $y_{c}$ respectively behaves as  $(x-x_{c})^{1/2}$ 
respectively $(y-y_{c})^{1/2}$. If the initial data are invariant 
under an exchange of $x$ and $y$, these two points coincide. 
\end{conjecture}

\section{Semiclassical Limit of Davey-Stewartson II solutions}
In this section, we numerically study solutions to the DS II equation 
for the initial data of the previous section for several values of 
$\epsilon \ll 1$. We then investigate the \emph{scaling laws}, i.e, 
the dependence on $\epsilon$ of the difference between the DS II and semiclassical DS II solutions.
In the previous section we had shown that the 
singularities of the semiclassical DS II system are as in the corresponding 
$1+1$ dimensional situations. The same is observed for the dispersive 
regularizations near the singularity here, i.e., for the difference 
of semiclassical DS II and DS II solutions for finite small 
$\epsilon$, both for
the same initial data: we find the same scalings in $\epsilon$ as in 
the $1+1$ dimensional case, $\epsilon^{2/7}$ for the defocusing case 
and $\epsilon^{2/5}$ for the focusing case.

\subsection{Defocusing case}

%(strong $\mathcal{O}(1)$ cubic defocusing nonlinearity, zero initial phase data)\\
% In \cite{KR}, we identified the Driscoll's method \cite{Dris} as a good choice for the time integration of the DS II equation. Consequently, 
% experiments described in the following have been done with this time-integration scheme, as explained in section 2.
We consider zero initial phase data of the form
\begin{equation}
\Psi_0(x,y)=e^{-R^2}, \,\, R=\sqrt{x^2+y^2}
\label{uinigauss},
\end{equation}
which correspond to the initial data (\ref{adinisym}) studied before for the 
defocusing semiclassical DS II system (\ref{disDSs}), now for defocusing DS II ((\ref{DSnonlocal}) with $\rho=1$).
The computations are carried out with $N_{x}=N_{y}=2^{14}$ Fourier 
modes, 
$L_{x}=L_{y}=5$, and $\Delta_t=2*10^{-4}$ for different values of 
$\epsilon$, until $t_{max}=1$, almost twice the critical time of 
the corresponding semiclassical system identified in Sec. 3.2.1.%  \sim 2* t_{c}^{dds}$.

\begin{figure}[htb!]
\begin{center}
\includegraphics[width=\textwidth]{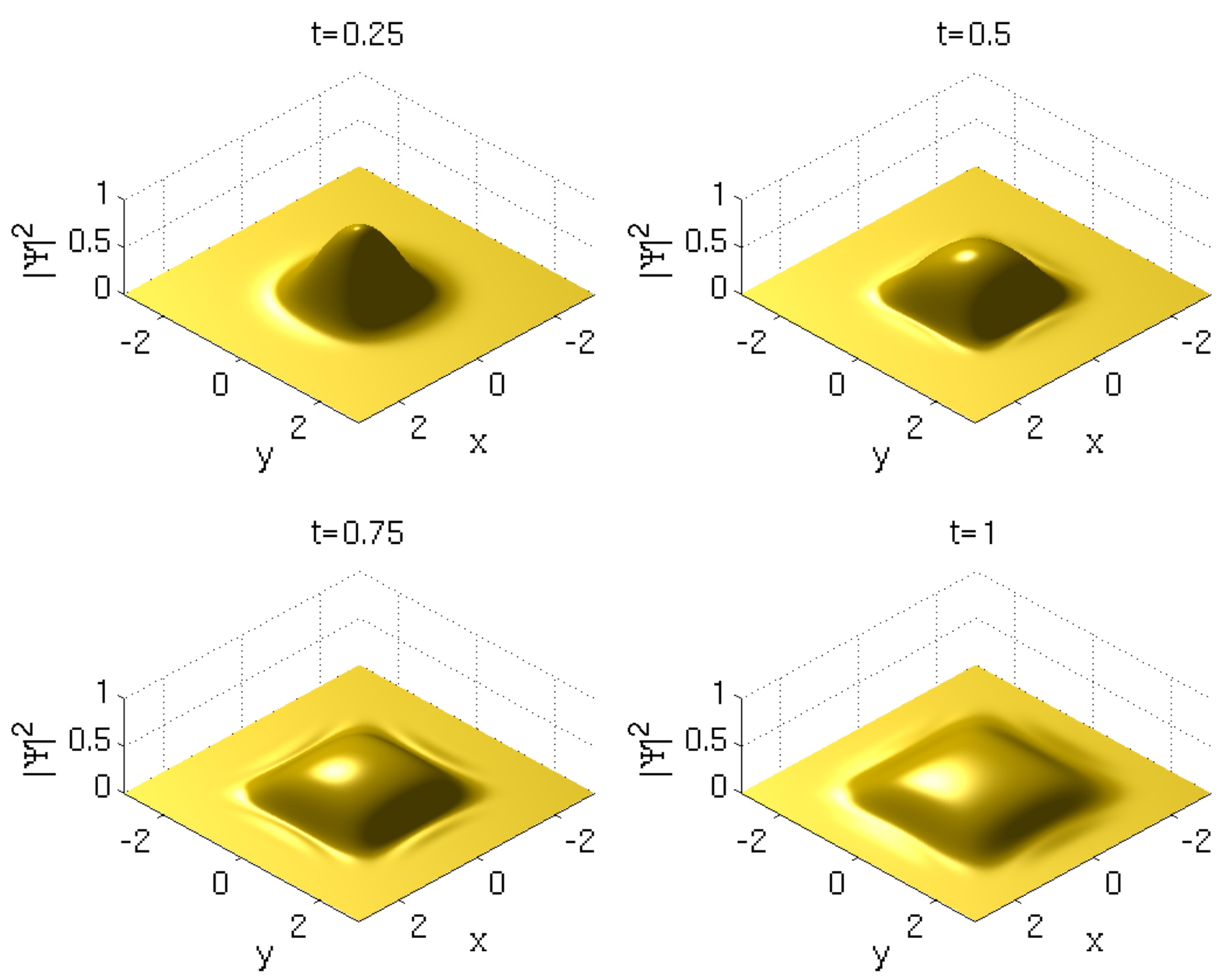}
\caption{Solution to the defocusing DS II equation for the initial 
 data (\ref{uinigauss}) and $\epsilon=0.1$ for several values of $t$.}
\label{figdefoc1}
\end{center}
\end{figure}

The defocusing effect of the equation for these initial data can be 
seen in Fig. \ref{figdefoc1}, where $|\Psi|^2$ is shown for several 
values of $t$. The compression of the initial pulse into some almost 
pyramidal shape leads to a steepening on the 4 sides orthogonal to the 
coordinate axes and to oscillations at the bottom edges of the 
`pyramid', see also \cite{KR}. 

\begin{figure}[htb!]
\begin{center}
\includegraphics[width=0.7\textwidth]{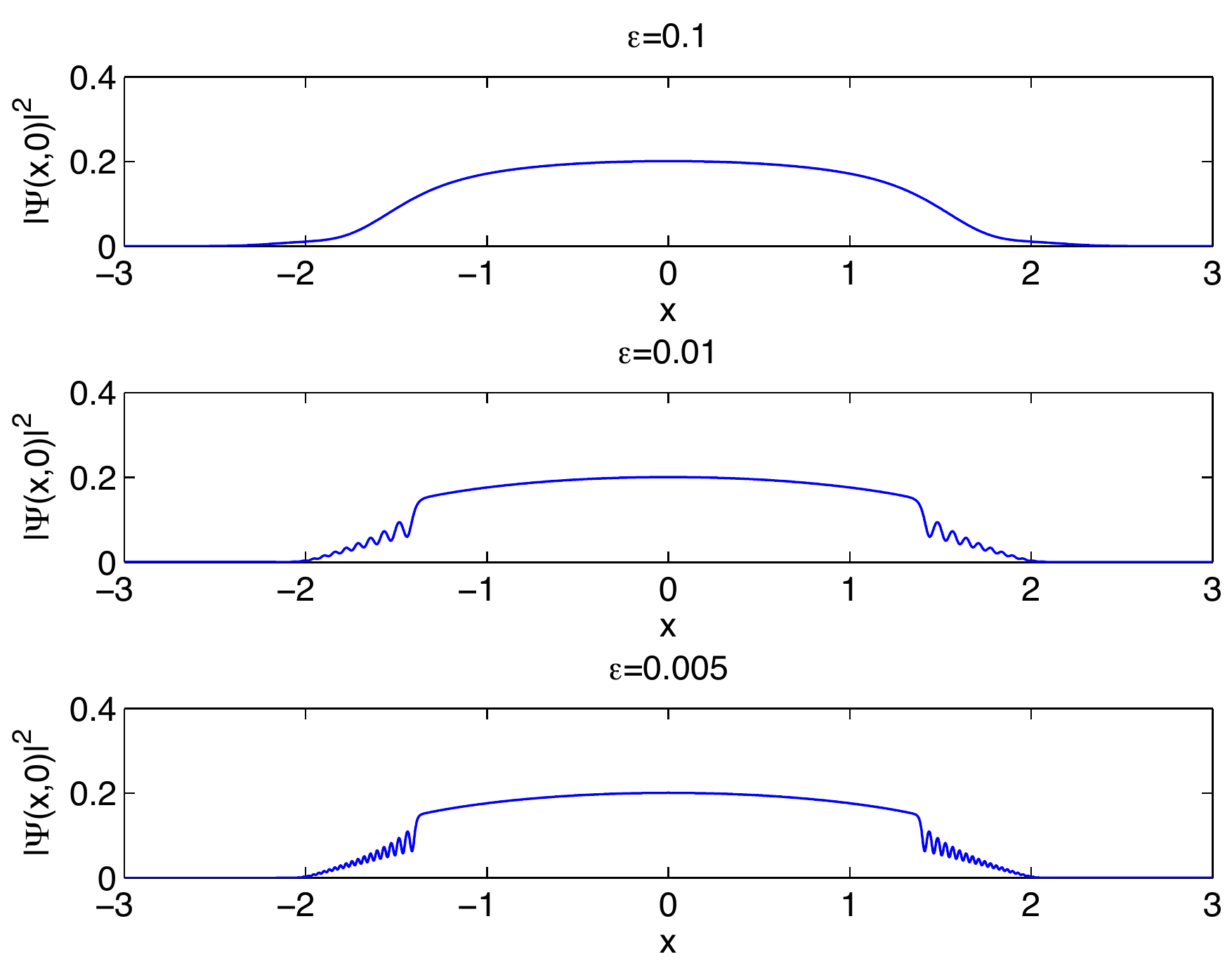} 
\caption{Absolute value of the numerical solutions of the defocusing DS II for the 
initial data (\ref{uinigauss}) for $y=0$ and $t=1$ for different values of $\epsilon$.}
\label{difepsu}
\end{center}
\end{figure}

\begin{figure}[htb!]
\begin{center}
\includegraphics[width=0.32\textwidth]{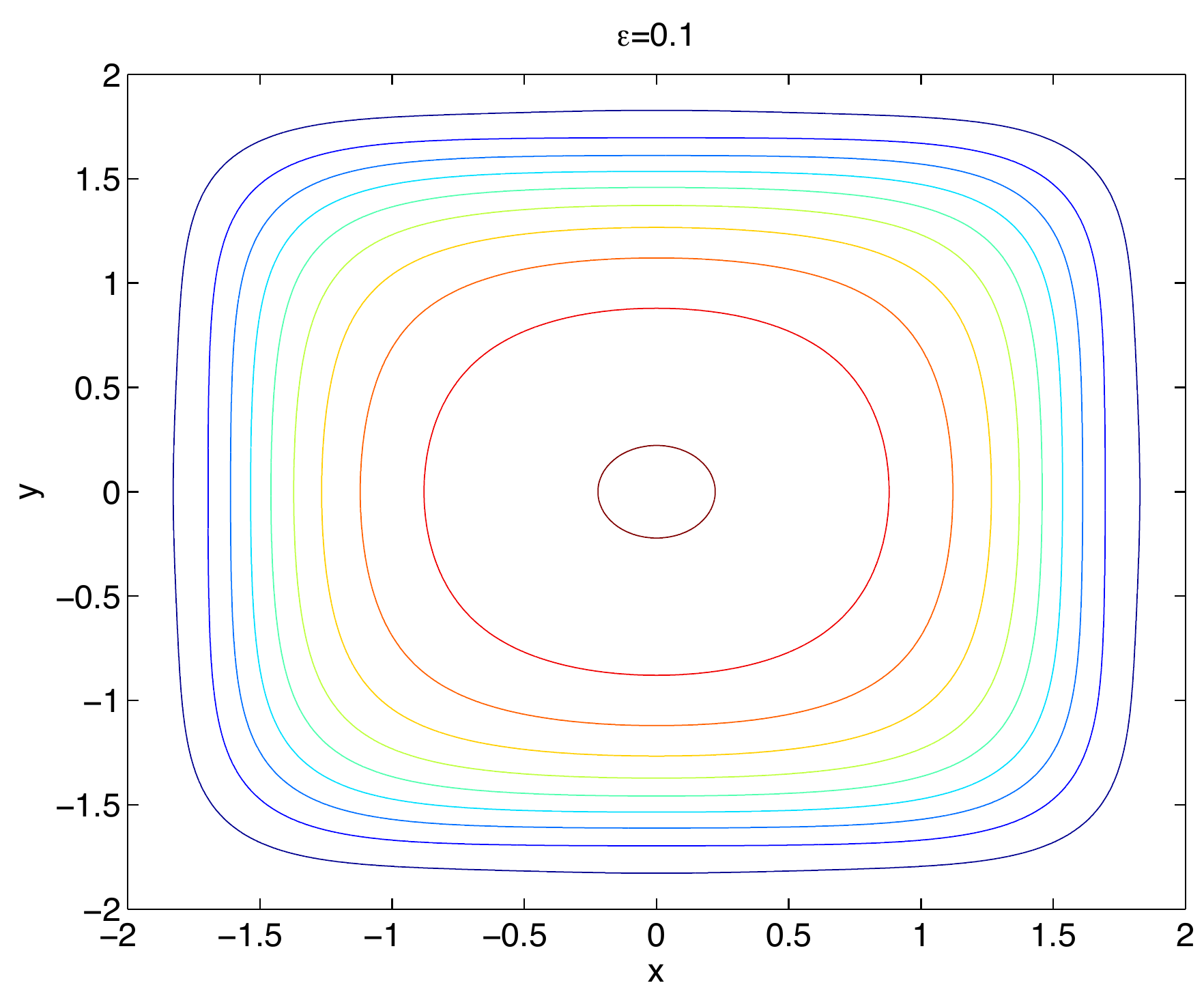} 
\includegraphics[width=0.32\textwidth]{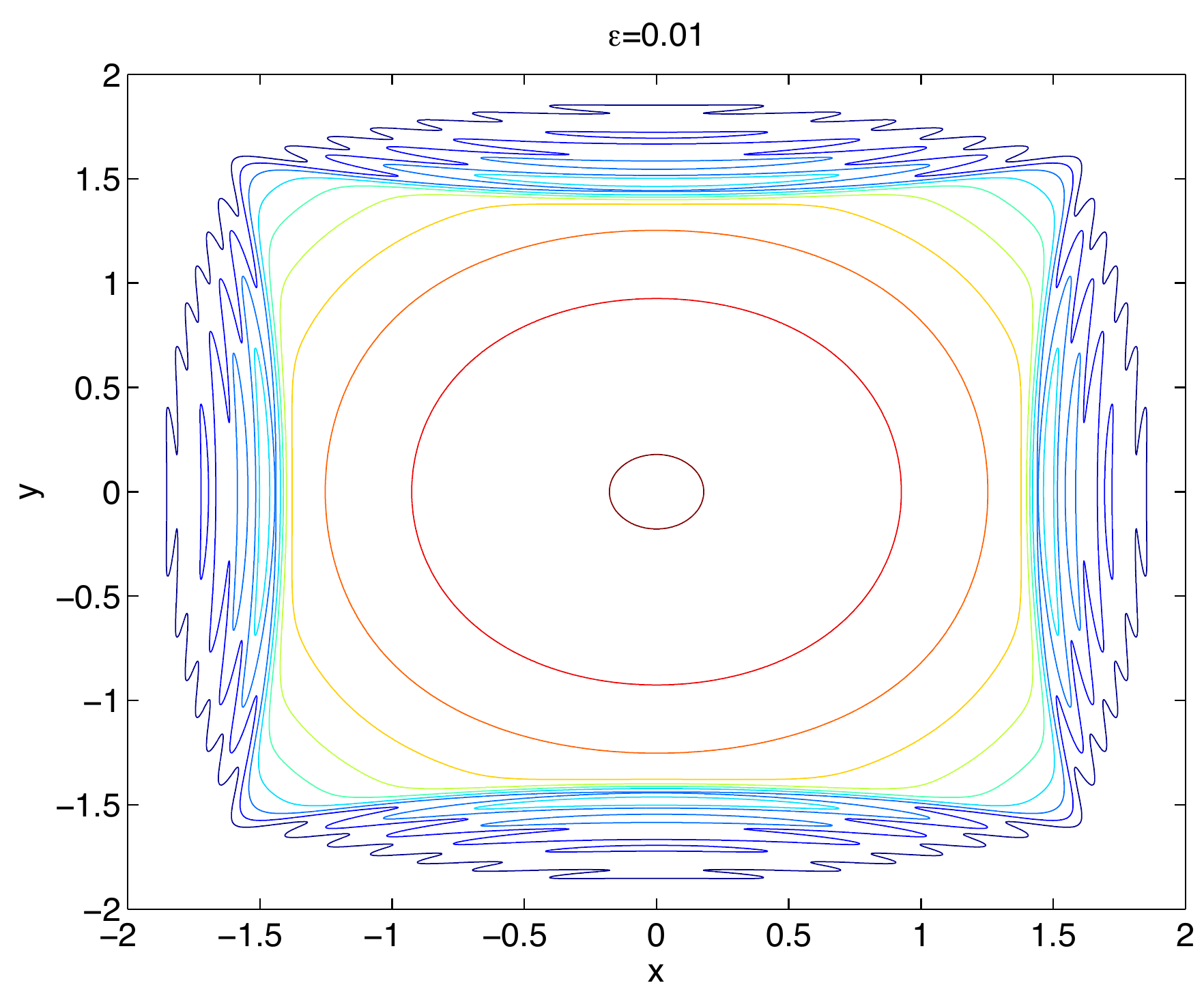} 
\includegraphics[width=0.32\textwidth]{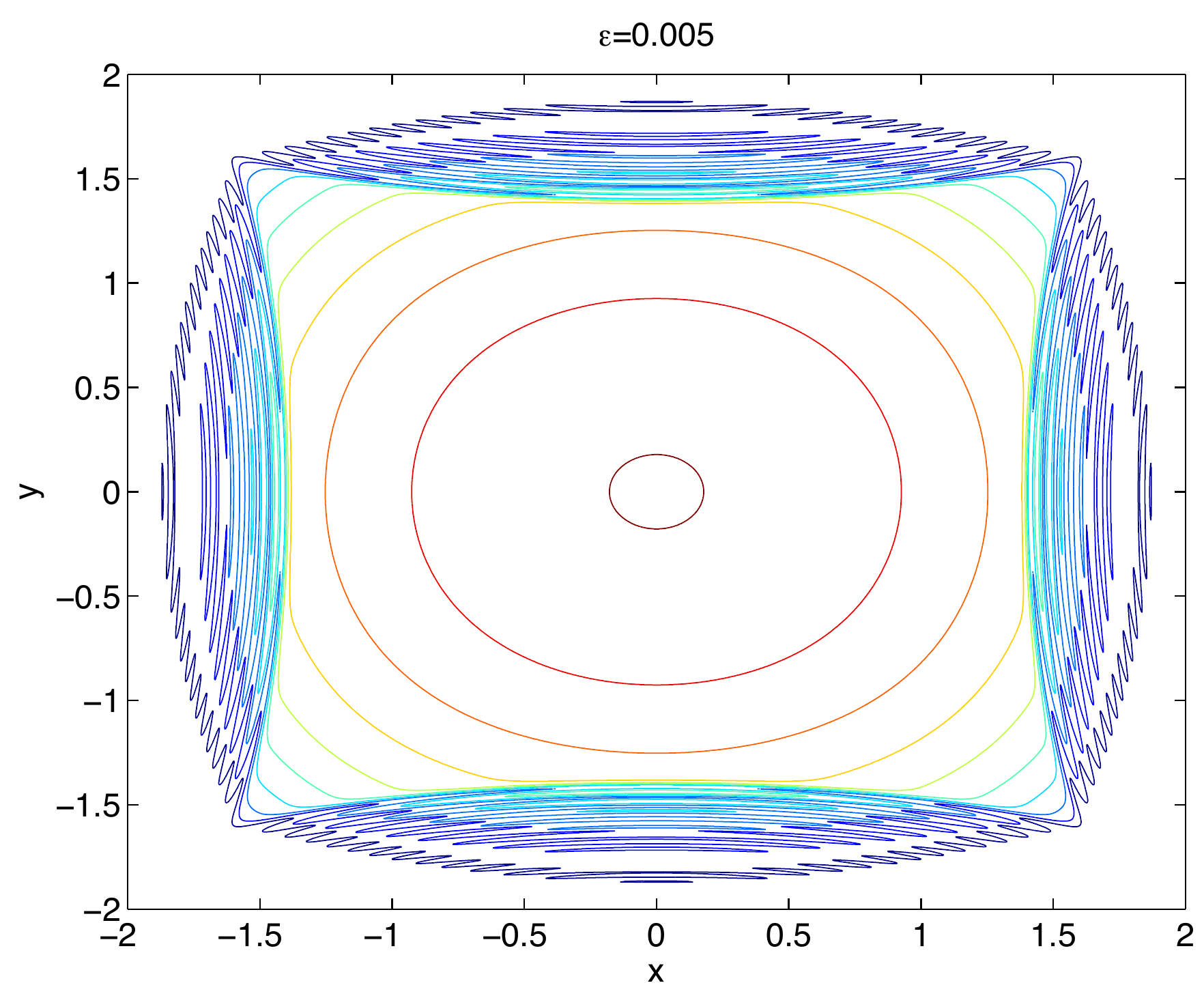} 
\caption{Contour plot of $|\Psi|^{2}$ for solutions of the defocusing DS II 
equation for the initial data 
(\ref{uinigauss}) at $t=1$ for different values of $\epsilon$.}
\label{contdefoc}
\end{center}
\end{figure}

The situation is similar for smaller values of $\epsilon \in [0.005, 
0.1]$. The oscillations become more rapid and more confined to a zone 
the smaller $\epsilon$ is, as can be seen in Fig. \ref{difepsu}, where 
we show the square of the absolute value of $\Psi(x,0,t)$ in dependence 
of $x$ at $t_{max}=1$ for different values of $\epsilon$. It can be 
seen that a lens shaped zone forms in the vicinity of each of the 
shocks of the semiclassical DS II system which should delimit for 
$\epsilon\to0$ the oscillations.  
%ici uy0 3 eps et contours
A contour plot for the modulus squared of these solutions can be 
found in Fig.~\ref{contdefoc}.

\begin{figure}[htb!]
\begin{center}
\includegraphics[width=0.40\textwidth]{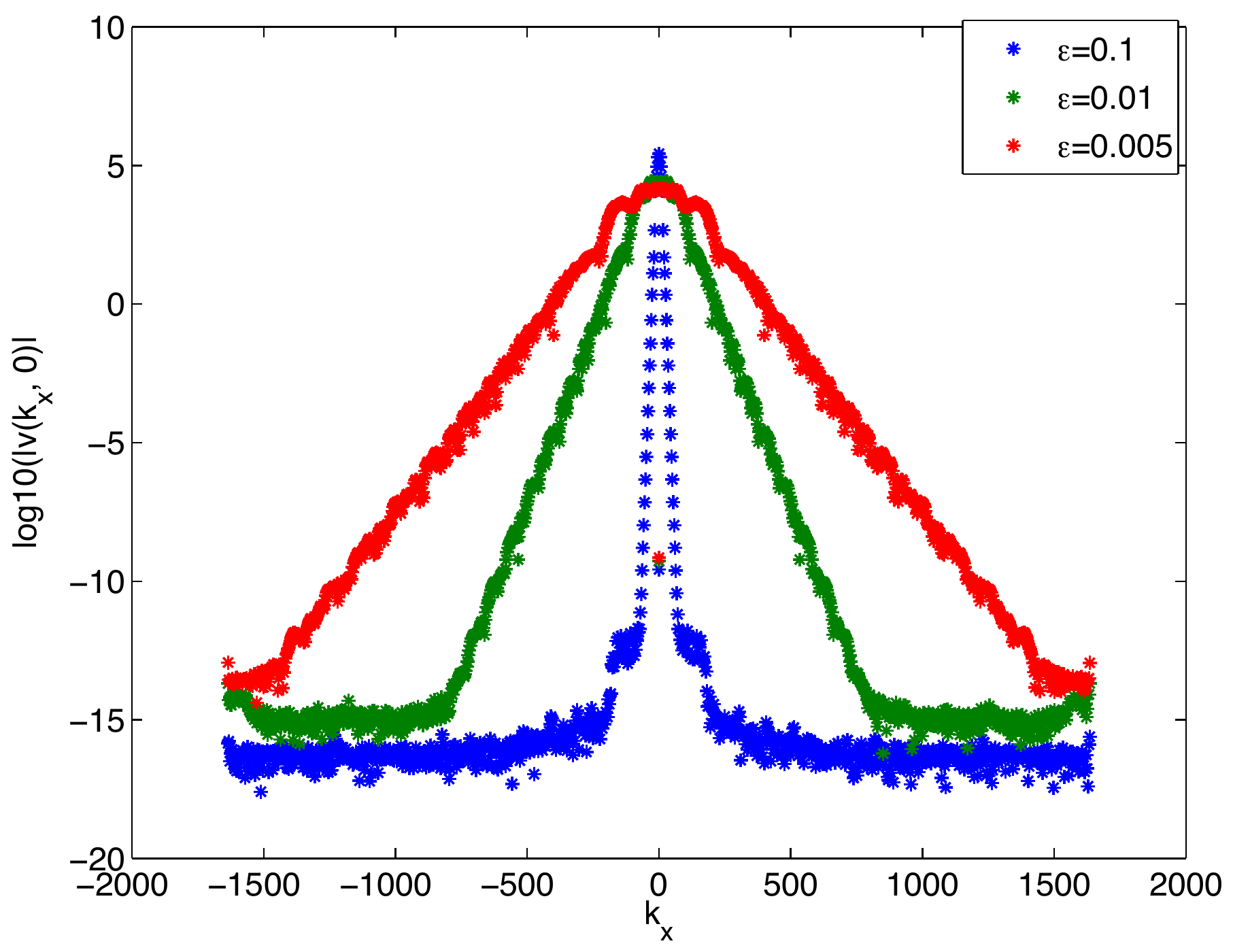}
\caption{Modulus of the Fourier coefficients for the solutions in 
Fig.~\ref{difepsu}, plotted on the $k_x$-axis at the maximal time of 
computation, $t_{max}=1$ for different values of $\epsilon$.}
% (left) and time evolution of the quantity 
%$\Delta_E$ (\ref{delE}) (right) for different values of $\epsilon$.}
\label{coefdefoc}
\end{center}
\end{figure}

We ensure that the system is well resolved numerically by checking both the decay of the Fourier coefficients, which decrease here
to machine precision, see Fig. \ref{coefdefoc}, and also the time 
evolution of the quantity $\Delta_E$ (\ref{delE}).
The numerically computed energy $E[\Psi](t)$, which is  a conserved 
quantity of DS (\ref{hamil}) for the exact solution, is here 
evaluated on $\mathbb{T}^2$. 
The quantity $\Delta_{E}$ increases  due to unavoidable numerical errors, but stays 
below $10^{-8}$ until the end of the computation for all studied cases. 
This is of the same order as the results for the 
semiclassical limit of the defocusing NLS equation in \cite{BJM,JSLDMD,ckkdvnls}.

\begin{figure}[htb!]
\centering
 \includegraphics[width=0.4\textwidth]{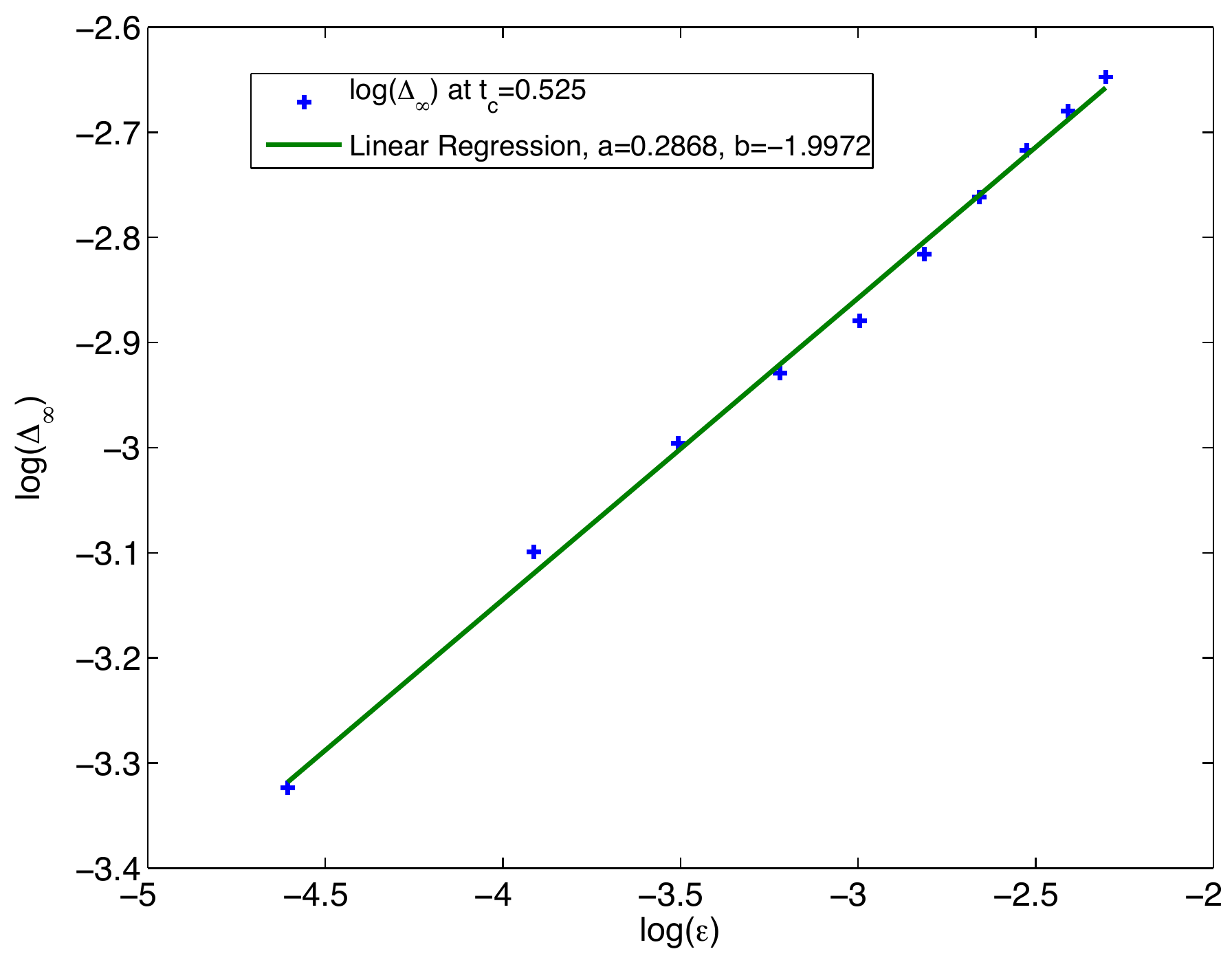} 
 \caption{$L_{\infty}$ norm  $\Delta_{\infty}$ of the difference 
 between defocusing DS II solutions and the solutions to the 
 defocusing system (\ref{disDSs}) for the 
 initial data (\ref{uinigauss}), respectively (\ref{adinisym}) in dependence 
 of $\epsilon$  at $t_c=0.525$. }
 \label{scaltcdefoc}
\end{figure}

An important question is the scaling with $\epsilon$ of the 
$L_{\infty}$ norm of the difference between semiclassical DS II
and DS II solutions for the same initial data. 
The $L_{\infty}$ norm of this difference is shown in 
Fig.~\ref{scaltcdefoc} at the critical time $t_c=0.525$ 
in dependence of $\epsilon$ for  $0.01\leq  \epsilon \leq 0.1$. 
A linear regression analysis ($\log_{10} \Delta_{\infty} = a \log_{10} \epsilon + b$) shows that $\Delta_{\infty}$ decreases as 
\begin{align}
\mathcal{O} \left( \epsilon^{0.28} \right)  \sim \mathcal{O} \left( \epsilon^{2/7} \right) \,\, \mbox{at} \,\, t=t_c=0.525, \,\, \mbox{with}\,\, a=0.2868  \,\,\mbox{and} \,\,  b= -1.9972.
\end{align}
The correlation coefficient is $r = 0.998$. 

This means that the same 
scaling is found as in the defocusing NLS case for which an 
asymptotic description at break-up was conjectured in \cite{DGK13}. 
Thus it appears that the essentially one-dimensional character of the 
singularity of the solution of the semiclassical DS II system (\ref{disDSs})
implies that the regularization effect of the dispersion in the full 
DS II system is as in the $1+1$-dimensional case. It has to be 
checked whether the special PI2 solution appearing in the asymptotic 
description of the NLS solution near the critical point plays a role 
also in the $2+1$-dimensional case.

\subsection{Focusing case}

\begin{figure}[htb!]
\begin{center}
\includegraphics[width=\textwidth]{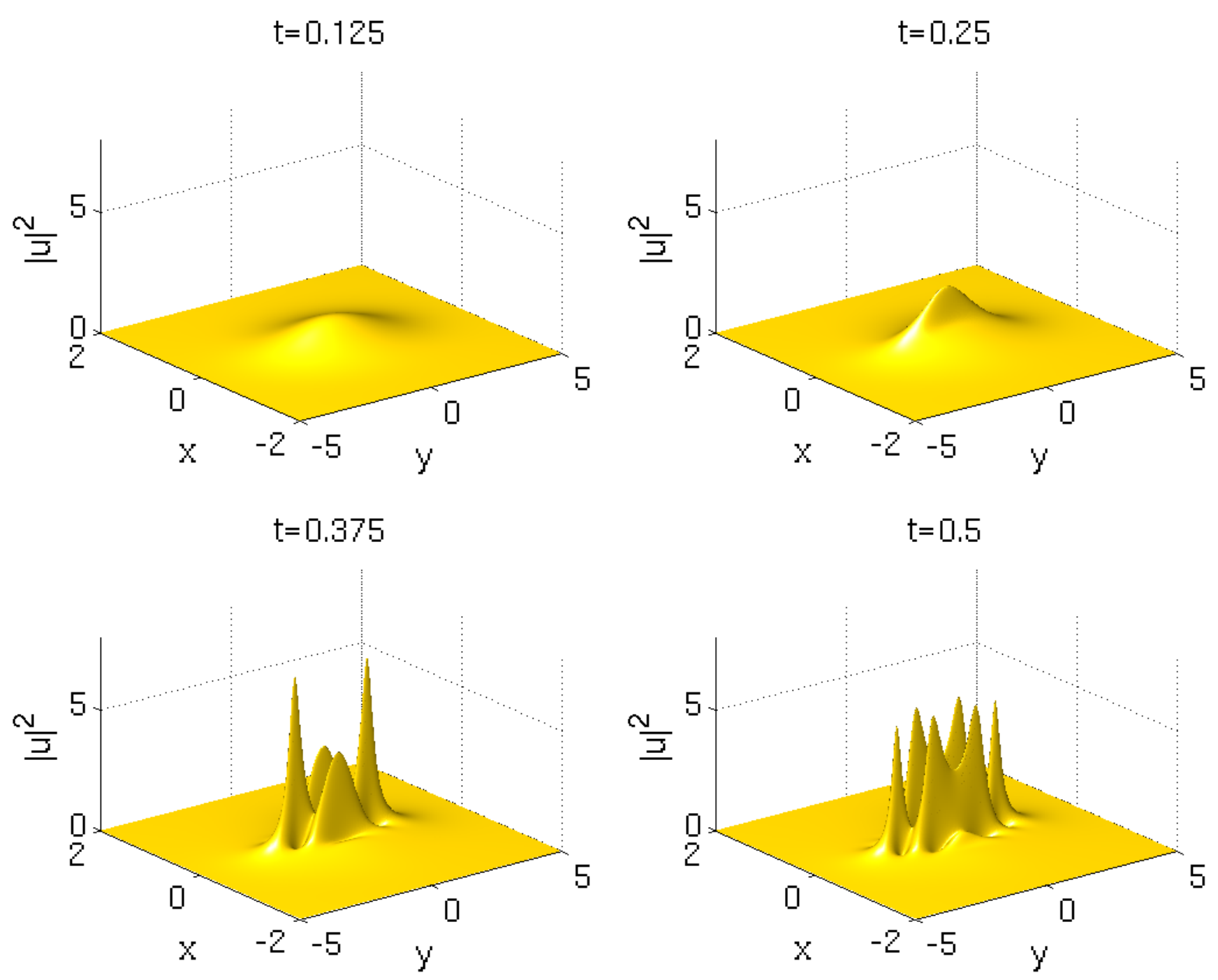} 
\caption{Modulus of the solution to the focusing DS II equation with initial data (\ref{adini}) and $\epsilon=0.1$ at several times.} 
\label{Ade01ut}
\end{center}
\end{figure}

We consider now solutions of the focusing DS II equation for small $\epsilon$. 
The initial condition corresponding to (\ref{adinifoc})
is
\begin{equation}
\Psi_0(x,y) = e^{-R^{2}}, \,\,\mbox{with}\,\, R = \sqrt{ x^{2}+0.1 y^{2} }.
\label{adini}
\end{equation}
\\
For $\epsilon=0.1$, the computation is carried out with 
$2^{14} \times 2^{14}$ points 
for 
$ x \times y \, \in \, [-5\pi,5\pi]\times[-5\pi,5\pi]$, and $\Delta_t=8*10^{-5}$.
For smaller values of $\epsilon$, we take $N_x=N_y=2^{15}$ to ensure 
sufficient resolution in Fourier space up to the maximal time of computation $t_{max}$.
The latter is chosen to be $t_{max}=0.5$, almost twice 
the break-up time of 
the corresponding focusing semiclassical DS II system found in sect. 3.2.2. 
\\ 
\\
For $\epsilon=0.1$, the initial peak grows until its maximal height (here $\sim 3.24$) at 
$t\sim0.275$. At later times it breaks up into smaller humps, see Fig. \ref{Ade01ut} as in the case
of the one-dimensional cubic NLS equation in the semiclassical limit, 
see for instance \cite{ckkdvnls,DGK}.

\begin{figure}[htb!]
\begin{center}
\includegraphics[width=0.49\textwidth]{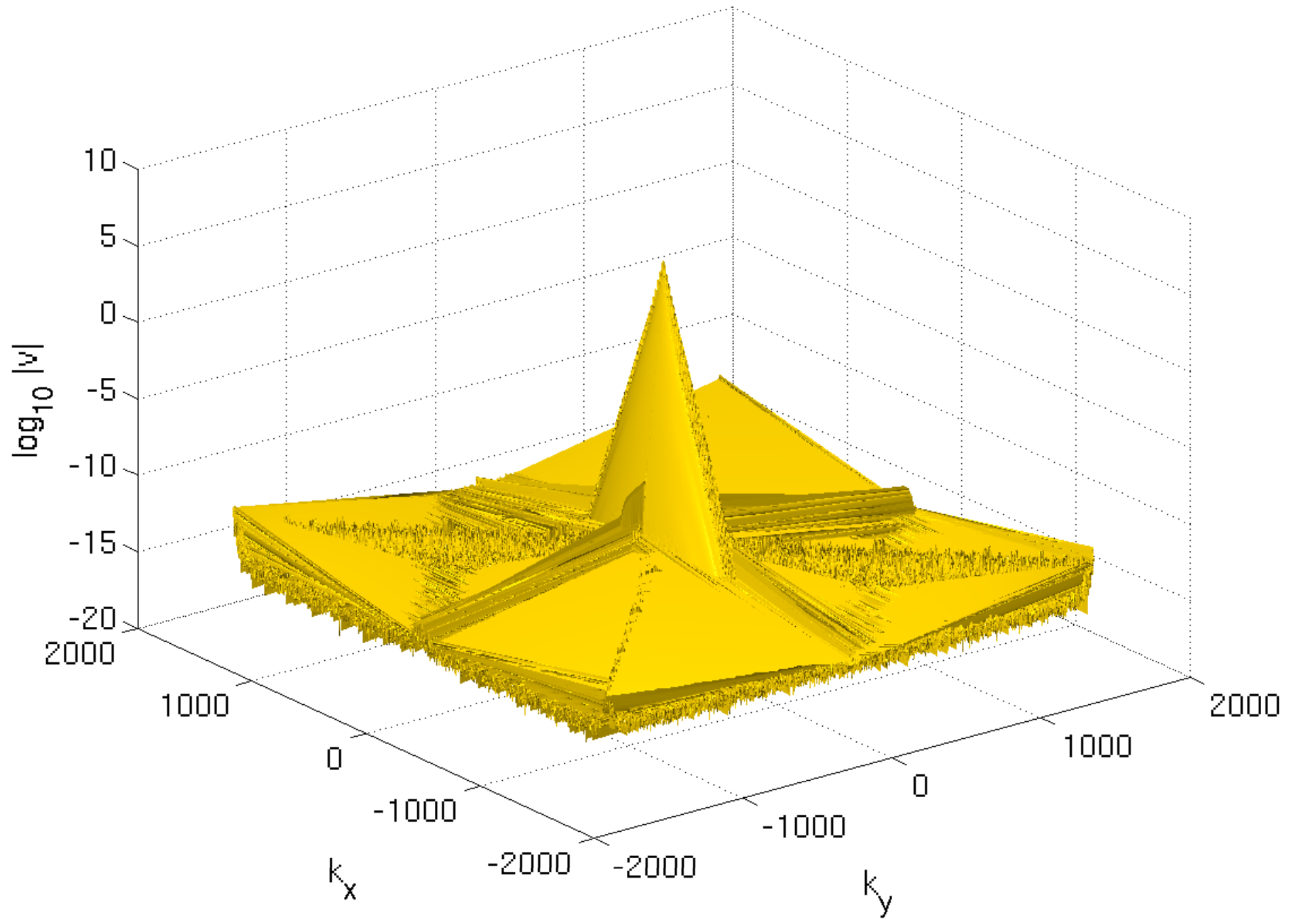} 
\includegraphics[width=0.49\textwidth]{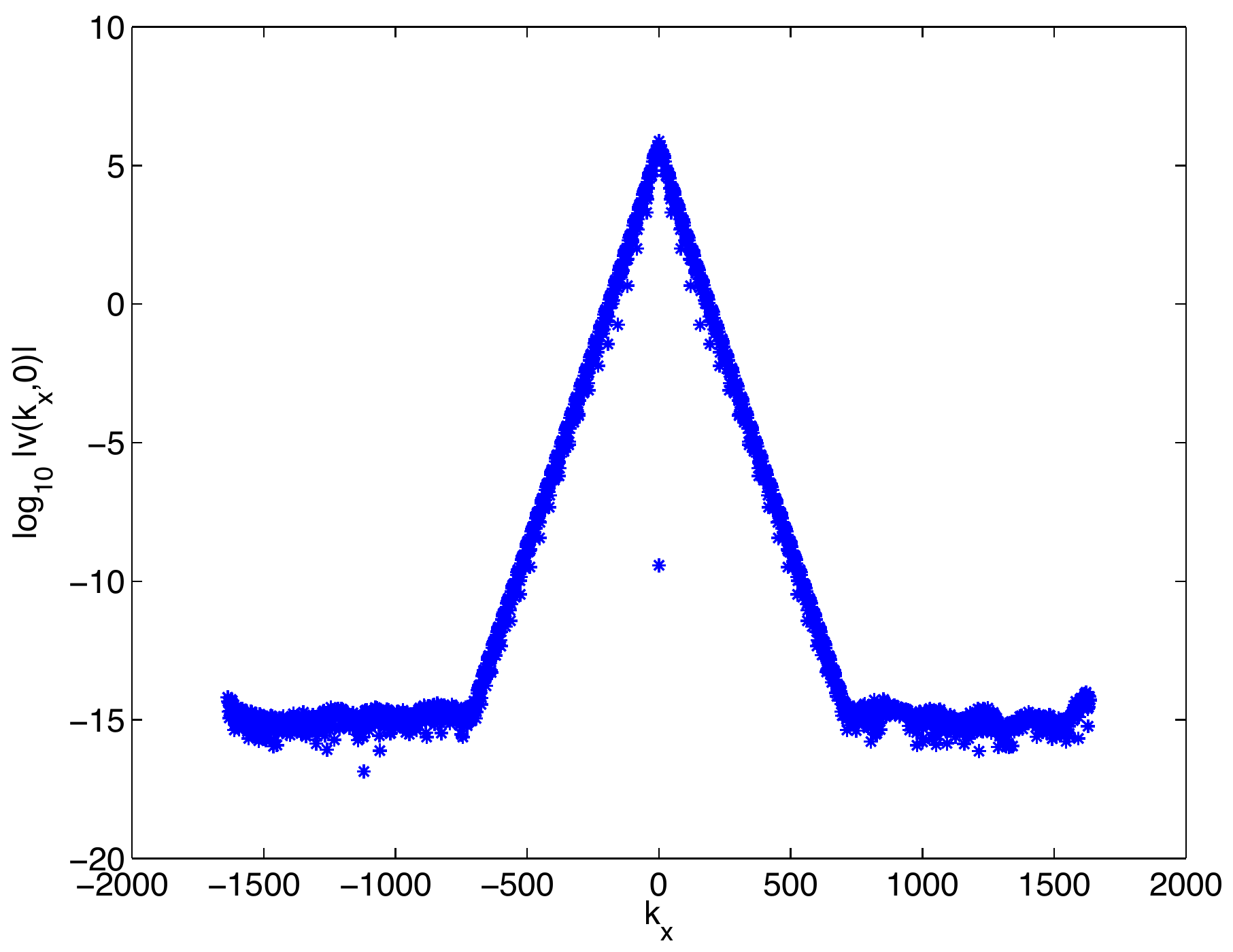}
\caption{Modulus of the Fourier coefficients of the solution shown in Fig. 
\ref{Ade01ut} at $t=0.5$ on the left, and plotted on the $k_x$-axis on the right.} 
\label{Ade01coef}
\end{center}
\end{figure}

The Fourier coefficients decrease to machine precision at 
$t_{max}=0.5$ as can be seen in Fig. \ref{Ade01coef}, and the 
numerically computed energy is of the order 
$\Delta_E \sim 1.5 *10^{-11}$ at this time.
%
% 
%Denoting by $t^{\epsilon}_{m}$, the time where the initial peak reaches its maximal height,
%we observe in Fig. \ref{amplnreps}, where we show the time evolution of both  $\underset{(x,y)}{\max} \left(|u(x,y,t)|^2\right)$ and $\Delta_E(t)$ for several values of $\epsilon$,     
%that as $\epsilon$ decreases,  $t^{\epsilon}_{m}$ decreases as well, and approaches 
%the corresponding break-up time of the dispersionless DS system $t_{b}^{dds, \,0.1}=0.1952$. 
%Indeed, it as been shown in \cite{KRM} that when a function admits a
%strong maximum, $\Delta_E$ suffers from an increase, but stays at least under $10^{-3}$.
%
%$t_{b}^{dds,\, \nu}$ for the studied values of $\nu$ are given in Table \ref{tbnu}.

\begin{figure}[htb!]
\begin{center}
\includegraphics[width=0.3\textwidth]{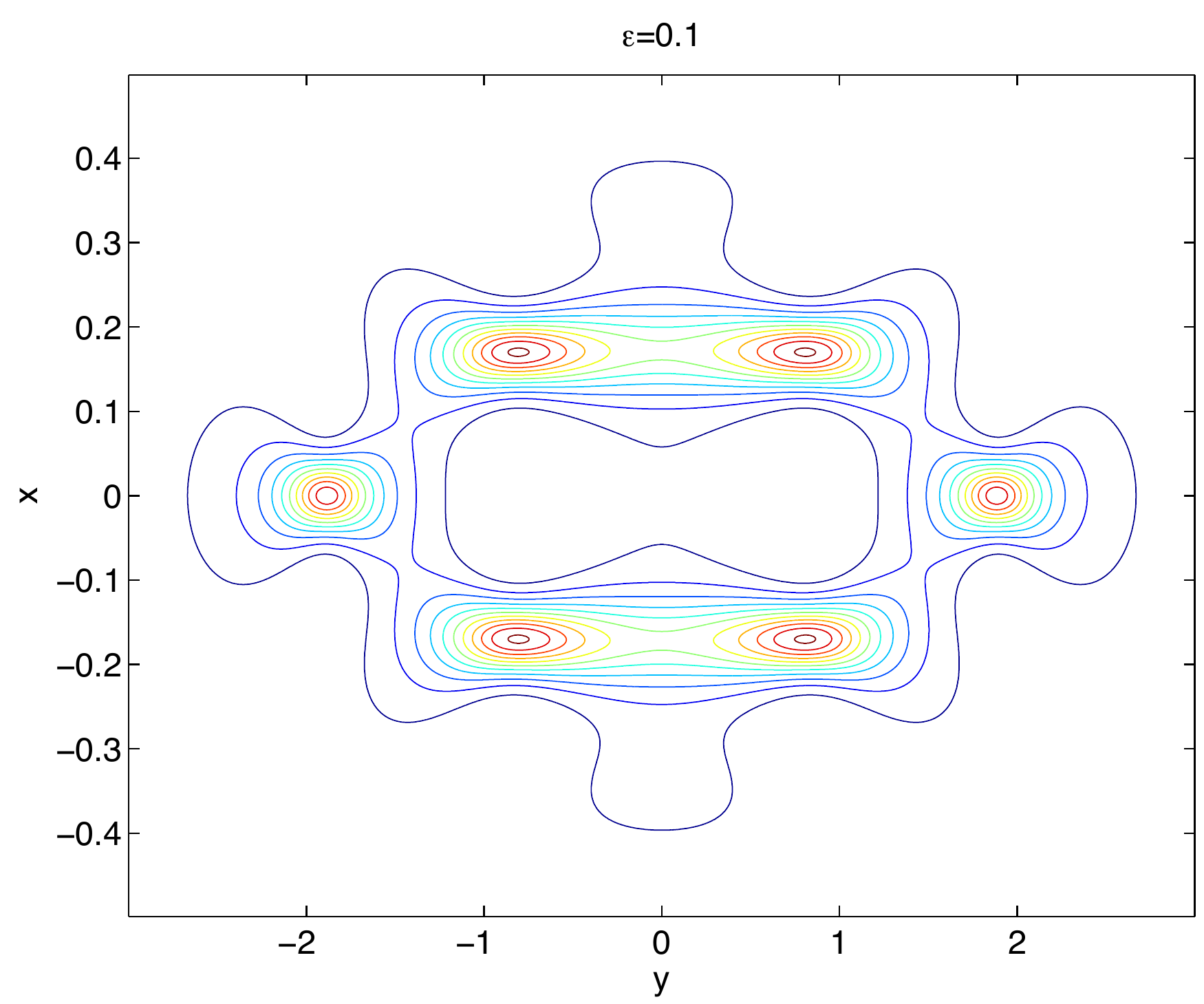}
\includegraphics[width=0.3\textwidth]{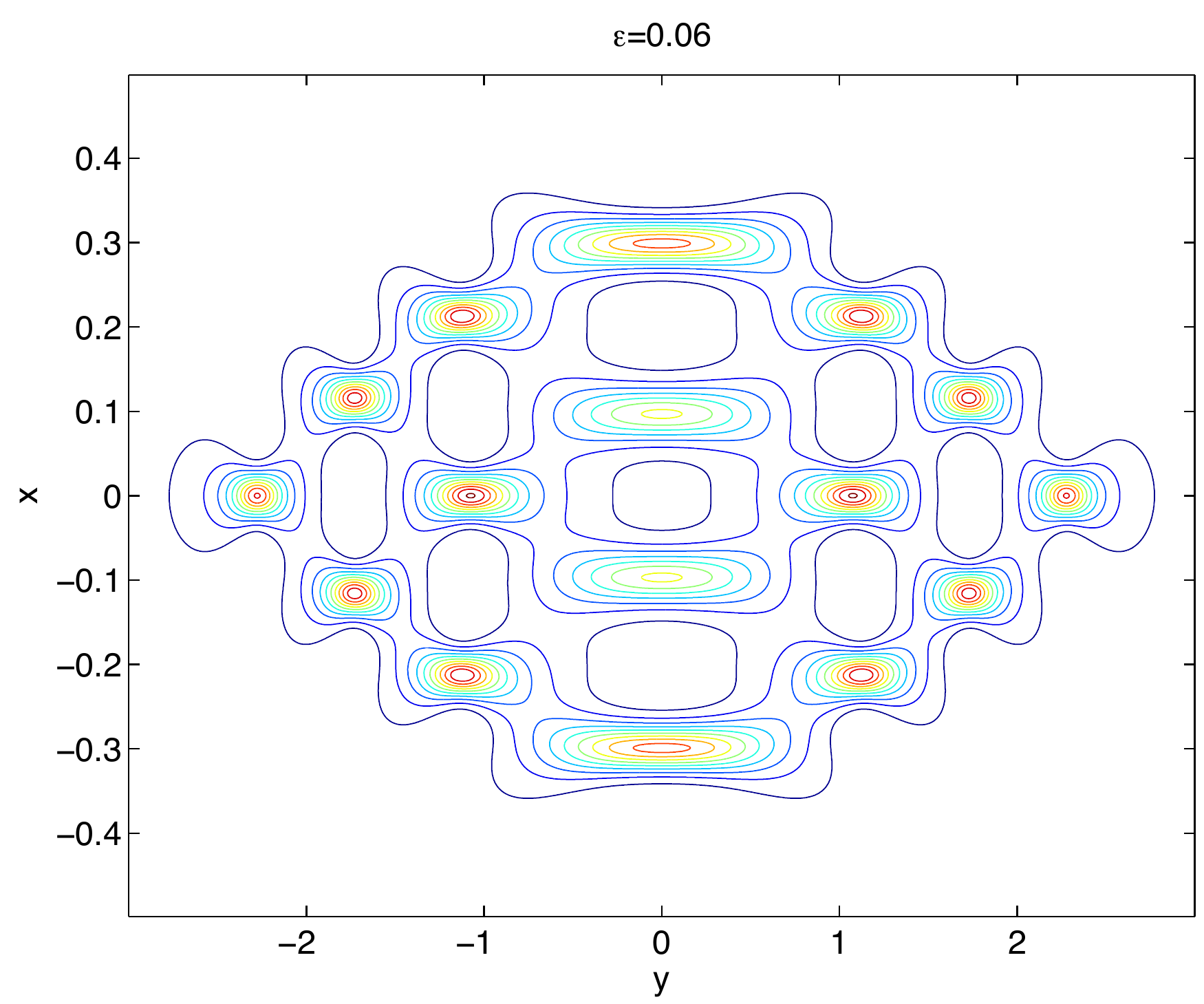}
\includegraphics[width=0.3\textwidth]{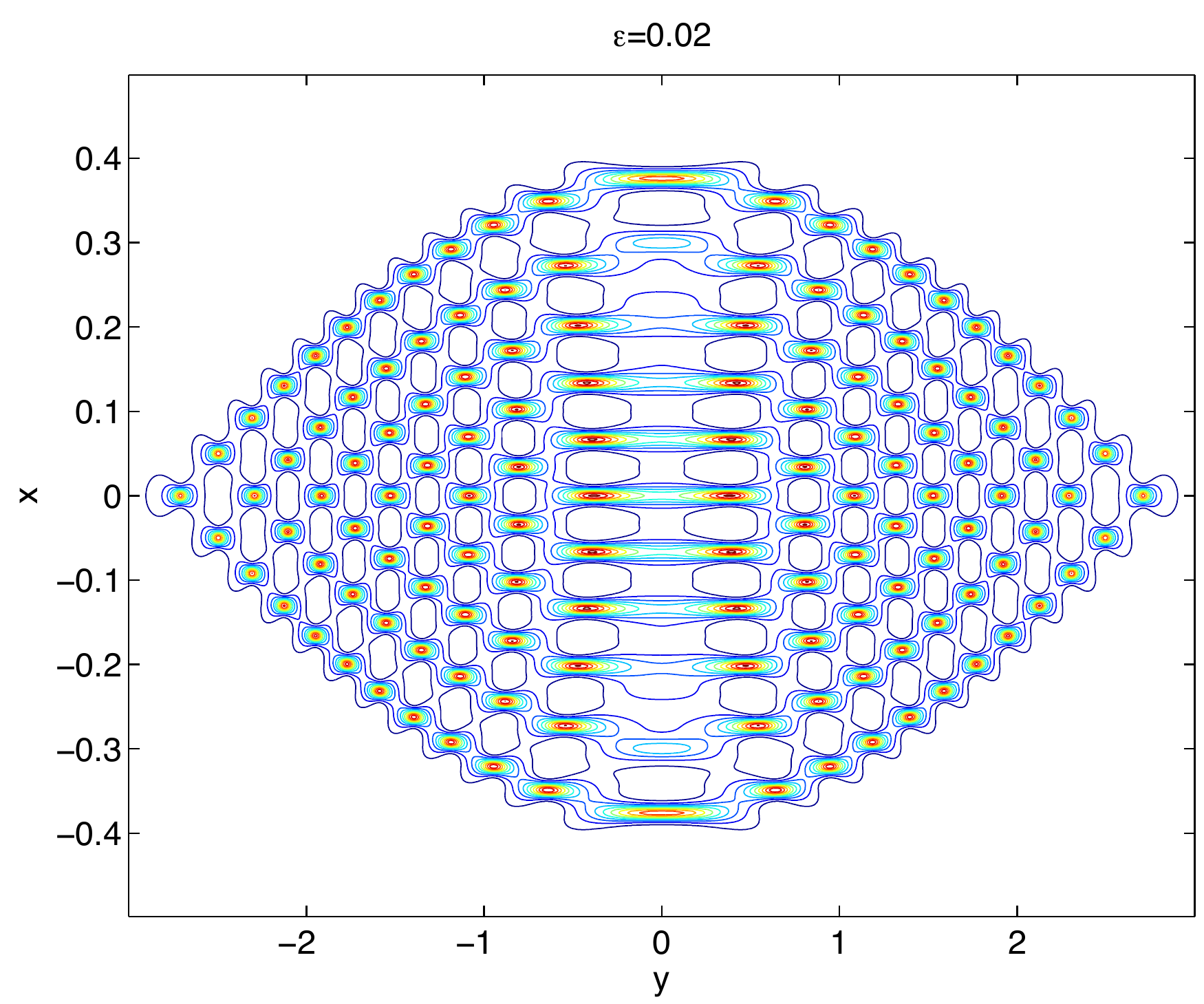}\\
\caption{Contour plots of the numerical solutions of the focusing DS II equation at $t=t_{max}=0.5$ 
with initial data (\ref{adini}) for different values of $\epsilon$.}
\label{dsfoce01cont}
\end{center}
\end{figure}

The situation is similar for smaller values of $\epsilon$, we observe 
as expected an increase of the number of oscillations in the 
dispersive shock as can be seen in Fig. \ref{dsfoce01cont}, where we show the contour plots of the solutions of the focusing DS II equation at $t=t_{max}=0.5$ 
with initial data (\ref{adini}) for different values of $\epsilon$. 
Again the oscillations of appear to be more and more confined for 
smaller $\epsilon$ to a lense shaped zone. 

\begin{figure}[htb!]
\begin{center}
\includegraphics[width=0.35\textwidth]{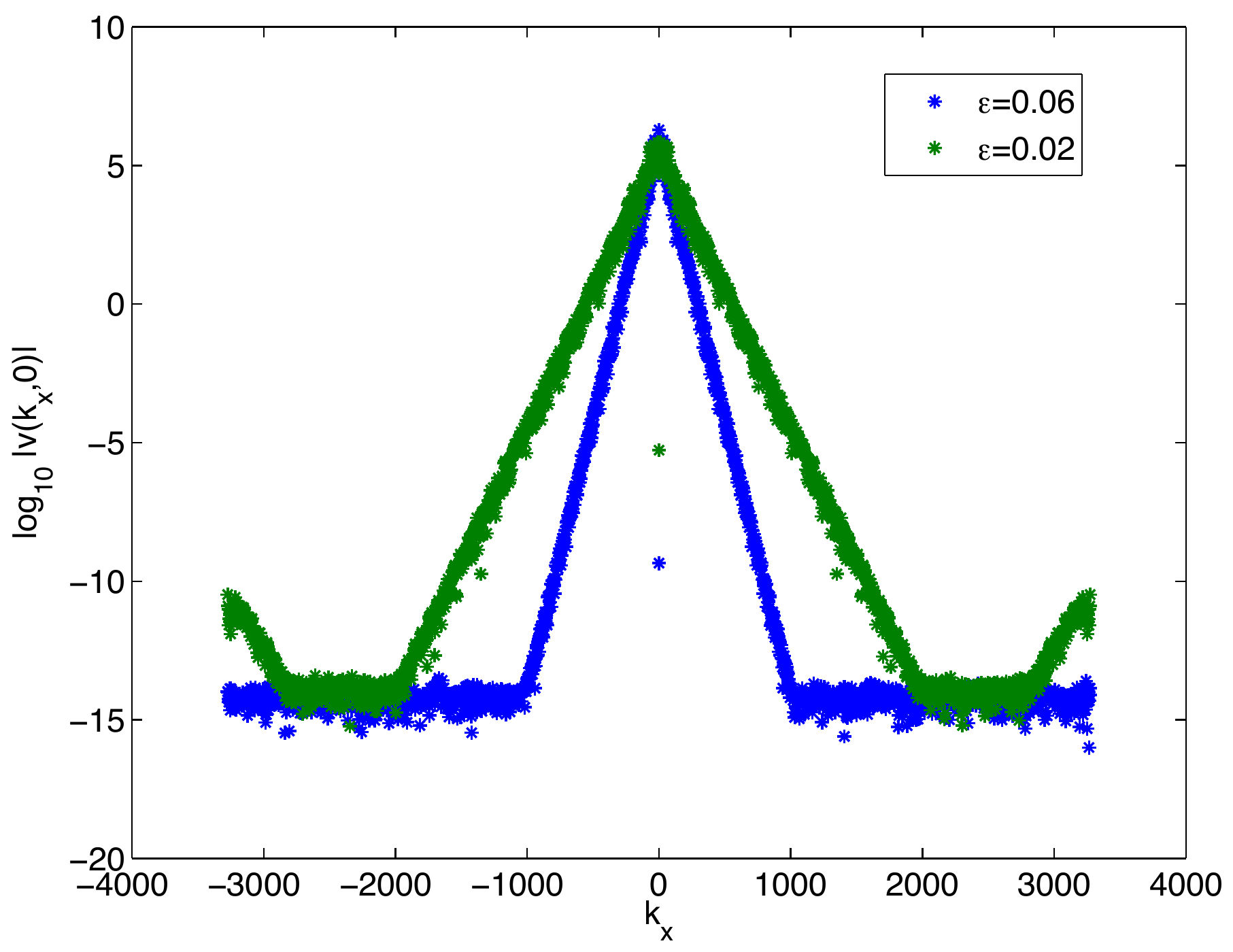}
\caption{Fourier coefficients of the solutions shown in Fig. \ref{dsfoce01cont} at $t=0.5$ plotted on the $k_x$-axis for $\epsilon=0.06$ and $\epsilon=0.02$.} 
\label{coefepss}
\end{center}
\end{figure}

For all situations studied, we check the  decay of the Fourier 
coefficients up to $t_{max}$, and the precision indicated by $\Delta_E$. 
We can see in Fig. \ref{coefepss}, that for the situations shown in 
Fig. \ref{dsfoce01cont}, the Fourier coefficients decrease to machine 
precision for $\epsilon=0.06$. For $\epsilon=0.02$, the phenomenon of 
modulational instability leads to a slight increase of the latter for 
high wave numbers, but they still decrease to $10^{-10}$, 
which is more than satisfactory here. This is of course due to the high 
spatial resolution used in the simulations and shows why such a 
resolution is needed here. 
At $t=t_{max}$, the numerically computed energy is of the order of 
$\Delta_E \sim 5.7*10^{-11}$ for the case $\epsilon=0.06$ and 
$\Delta_E \sim 1.5*10^{-7}$ for the case $\epsilon=0.02$. This 
indicates a numerical error well below plotting accuracy. 

In \cite{BertTob}, the semiclassical limit of the $1+1$-dimensional 
focusing cubic NLS equation is studied. It  
is shown that near the point of gradient catastrophe $(x_c, t_c)$, 
each spike of the NLS solutions is asymptotically described by the 
Peregrine breather, an exact rational solution to NLS, and has the height $3|u^{\epsilon}(x_c, 
t_c)|$. The authors illustrated numerically this relation for $\epsilon=\frac{1}{33}$.
% and 
%uniform shape of the rational breather solution to the NLS, scaled to the size $\mathcal{O} (\epsilon)$.\\
Here, we have determined in Section 3.2.2 the break-up time 
numerically with some potential small error, and it is not clear 
whether a similar relation holds also for DS II. Nevertheless, we compare in Table \ref{Bert} 
the values of $|\Psi^{\epsilon}(x_c, t_c)|$ and  
$|\Psi^{\epsilon}(x_c, t_{peak})|$, where $t_{peak}$ corresponds to the time, where the first 
spike appears in the numerical solution, 
before the appearance of oscillations. 
We find that as $\epsilon$ decreases, the ratio  
$\frac{|\Psi^{\epsilon}(x_c, t_{peak})|}{|\Psi^{\epsilon}(x_c, t_c)|}$ tends also here to $3$.

\begin{table}[htb!]
\centering
\begin{tabular}{|c|c|c|c|c|}
\hline 
$\epsilon$ & $0.1$ & $0.06$ & $0.04$ & $0.02$\\
\hline 
&&&&\\
$\frac{|\Psi^{\epsilon}(x_c, t_{peak})|}{|\Psi^{\epsilon}(x_c, t_c)|}$ & $2.5807$ & $2.7772$ & $2.8792$ & $2.9811$ \\
&&&&\\
\hline 
\end{tabular}
\caption{ Ratio $\frac{|\Psi^{\epsilon}(x_c, t_{peak})|}{|\Psi^{\epsilon}(x_c, t_c)|}$ for several values of $\epsilon$ for initial data of the form (\ref{adinifoc}). }
\label{Bert}
\end{table}

\begin{remark}
If one considers initial data of the form $\Psi_0(x,y) = e^{-R^{2}}, \,\,\mbox{with}\,\, R = \sqrt{ x^{2}+\nu y^{2} },$ and $\nu<1$,
the situation is similar. We observe the appearance of dispersive shocks, and as 
$\epsilon$ decreases, the number of oscillations increases. The ratio 
$\frac{|\Psi^{\epsilon}(x_c, t_{peak})|}{|\Psi^{\epsilon}(x_c, t_c)|}$ tends also to $3$ as $\epsilon$ tends to $0$ for other values of $\nu<1$.
\end{remark}
\\

%Scaling laws
%
%
We now study the scaling of the difference of the DS II and the semiclassical 
DS II solution for the initial data (\ref{adinisym}), 
at $t=t_c=0.1946$, for different values of $\epsilon$.
We consider values of $\epsilon$ between $0.1$ and $0.01$, 
with $N_x = N_y=2^{14}$. Note that we use here less resolution since 
the maximal time of computation is $t_c\sim 0.1946$, and that the 
modulational instability and other typical numerical problems are due 
to the formation of dispersive shocks. Thus we do not need the high 
spatial resolution we used before for the study in the semiclassical limit. 

\begin{figure}[htb!]
\centering
 \includegraphics[width=0.45\textwidth]{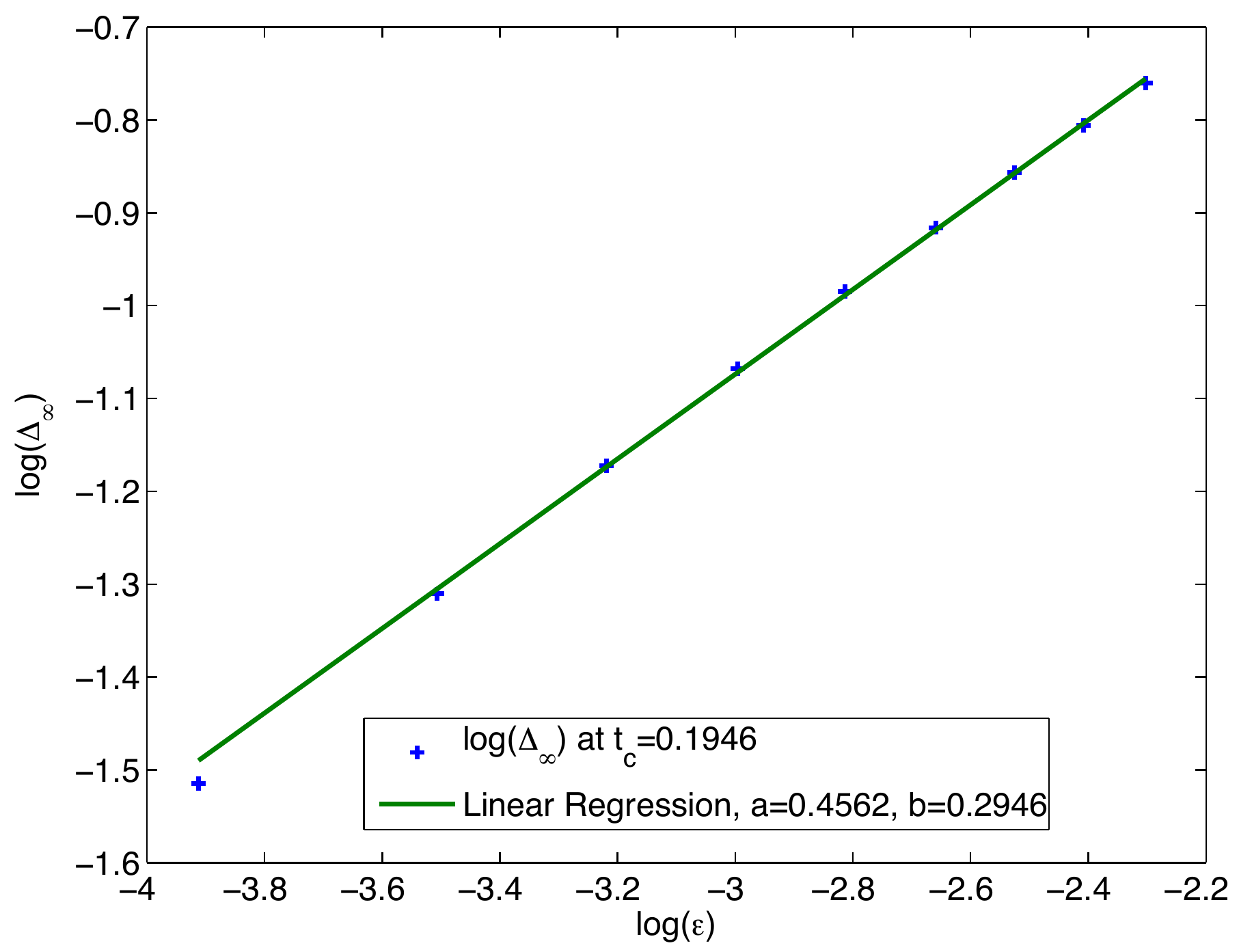} 
 \caption{$L_{\infty}$ norm  $\Delta_{\infty}$ of the difference 
 between focusing DS II solution and the corresponding solution to the 
 focusing semiclassical DS II system (\ref{disDSs}) for the initial 
 data (\ref{adinifoc}) respectively (\ref{adinisym}) in dependence of 
 $\epsilon$  at $t_c=0.1946$. }
 \label{scaltcfoce01}
\end{figure}

At $t_c\sim 0.1946$,  the 
$L_{\infty}$ norm of the difference between semiclassical DS II
and DS II solutions for the same initial data roughly decreases as
$\mathcal{O} \left( \epsilon^{0.45} \right)$. Indeed, by doing a linear regression analysis ($\log_{10} \Delta_{\infty} = a \log_{10} \epsilon + b$), we find 
$a=0.4562$, $b=0.2946$ and $r=0.999$, ($r$ being the correlation 
coefficient) as can be seen in Fig. \ref{scaltcfoce01}.

\begin{figure}[htb!]
\centering
\includegraphics[width=0.5\textwidth]{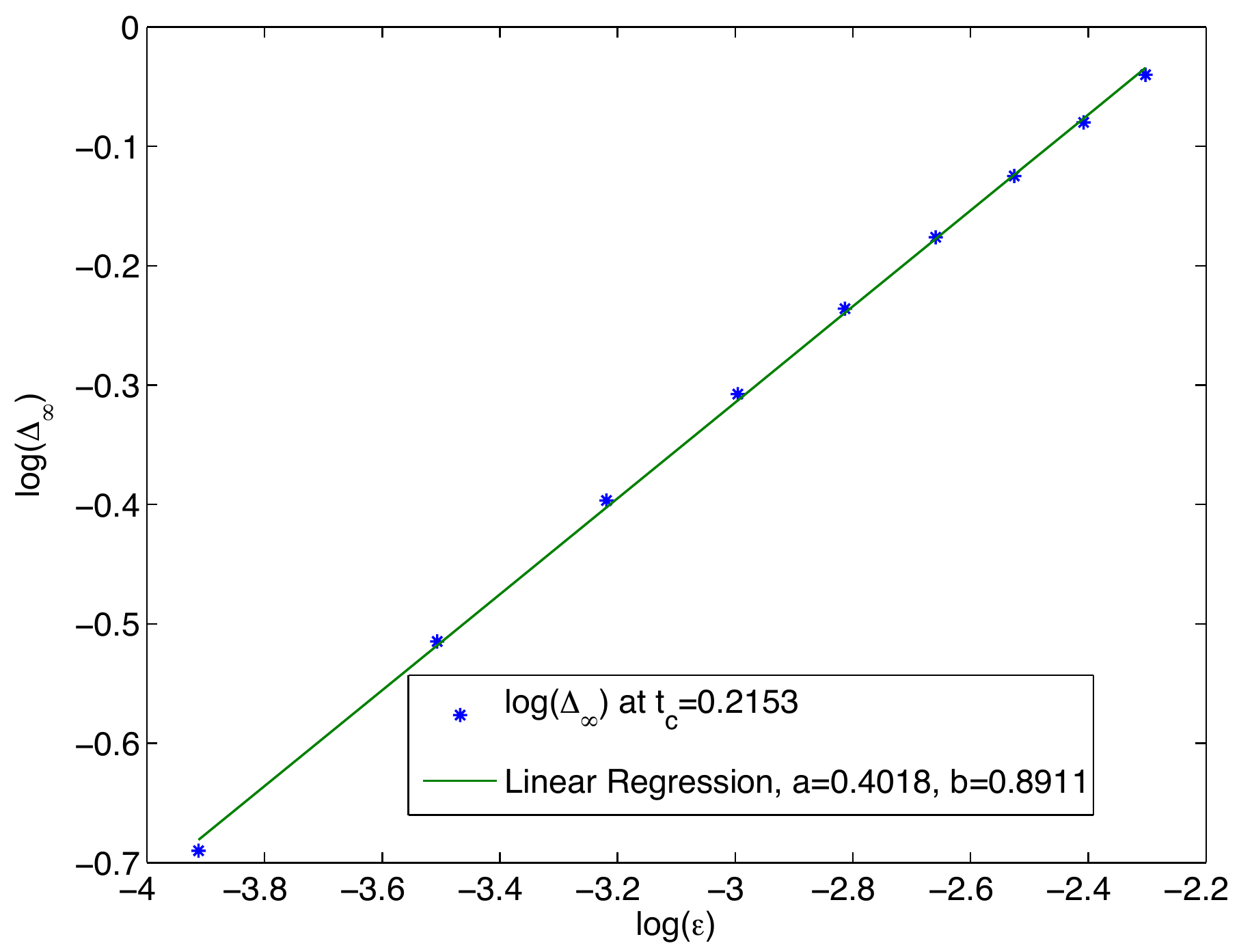} 
 \caption{ $L_{\infty}$-norm of the difference between the solutions 
 to the focusing  DS II solution and the corresponding solution to the 
 focusing semiclassical DS II system for symmetric initial data 
 at $t_c=0.2153$ in dependence of $\epsilon$.}
 \label{scalepsse1b}
\end{figure}

The scaling at the break up time is similar for a symmetric initial data of the form
\begin{equation}
\Psi_0(x,y) = e^{-R^{2}}, \,\,\mbox{with}\,\, R = \sqrt{ x^{2}+y^{2} },
\end{equation}
i.e., the second case studied in Sec. 3.2.2. For $0.1 \leq \epsilon 
\leq 0.02$, we find that the $L_{\infty}$-norm of the difference 
between the solutions to the focusing DS II 
and the corresponding system (\ref{disDSs}) at $t_c=0.2153$ 
scales as 
$\epsilon^{0.4}$. The situation is shown in Fig. \ref{scalepsse1b}, for which a linear regression 
($\log(\Delta_{\infty})= a \log(\epsilon) + b$) gives 
$a=0.4018$, $b=0.8911$ and $r=0.999$. Note that this is the same 
scaling conjectured in the focusing NLS case, see \cite{DGK13} for an 
asymptotic description at break-up. In the asymptotic description of 
the critical behavior of $1+1$-dimensional focusing NLS solutions, 
the \emph{tritronqu\'ee} solution of the PI equation appeared. It 
remains to be checked whether this solution also plays a role in the 
context of the focusing DS II equation. 

Since in \cite{Roid1}, preliminary results in this context suggest 
that for symmetric initial data, the solutions of the DS II equation 
in the semiclassical limit blow up for larger $t$, we study this special case in the following section.

\section{Blow-up in solutions to the Davey-Stewartson II equations in 
the semiclassical limit}
In this section we study numerically the possibility of a blow-up in 
solutions to the focusing DS II system in the small dispersion limit. Since 
the formation of a dispersive shock implies that the initial peak 
decomposes into smaller ones, a blow-up appears only to be possible 
if the initial hump continues to grow without limits. In this sense 
blow-up and dispersive shocks appear to be competing phenomena. In 
the examples studied in the previous section, there is clearly no 
blow-up. 
To identify numerically a potential blow-up, we will use again the 
asymptotic behavior of the Fourier coefficients to indicate as for the 
semiclassical systems the appearance of this 
singularity by the vanishing of $\delta$, i.e., the disappearance of 
the exponential decay of the coefficients. We first test this 
approach for the $1+1$-dimensional focusing quintic NLS equation, for 
which previous studies of blow-up exist, see for instance \cite{SS} 
for a review of the topic and references. Then we investigate this 
phenomenon for the semiclassical DS II system for initial data with 
a symmetry with respect to an exchange of $x$ and $y$. We find that 
there is indeed a blow-up in this case, and that the difference 
between break-up and blow-up time scales roughly as $\epsilon$ (the 
corresponding scaling in the $1+1$-dimensional 
case is $\epsilon^{4/5}$).  

\subsection{Focusing Quintic NLS Equation} 
We first consider the quintic NLS equation to check the efficiency of 
our methods to detect blow-up phenomena.
It is well known that solutions to focusing NLS equations of the form (\ref{NLSeqgen})
 can have blow-up, if $\sigma d \geq 2 $.
Thus the simplest case to investigate blow-up phenomena for 
1+1-dimensional focusing 
NLS equations is  $\sigma = 2$, i.e., the focusing quintic NLS equation, 
%\\
%The latter is given by 
\begin{equation}
    i\epsilon  \Psi_{t}+\frac{\epsilon^2}{2} \Psi_{xx}+ \frac{1}{2} |\Psi|^{4} 
    \Psi=0.
\label{quintNLS}
\end{equation}
It is well known (see \cite{MR}), that its solutions 
can blow up in finite time  for  initial data with negative energy $E[\Psi]$,
\begin{equation}
    E[\Psi] = 
    \int_{\mathbb{R}}^{}\left( \frac{\epsilon^2}{2} |\partial_{x}\Psi|^{2}-\frac{|\Psi|^{6}}{6}\right)dx
    \label{quintener}.
\end{equation}
In the semiclassical regime ($\epsilon \to 0$), this condition is 
obviously met for arbitrary non trivial  initial data in $L_{2}$ for sufficiently small $\epsilon$.\\

\begin{figure}[htb!]
\centering
 \includegraphics[width=0.5\textwidth]{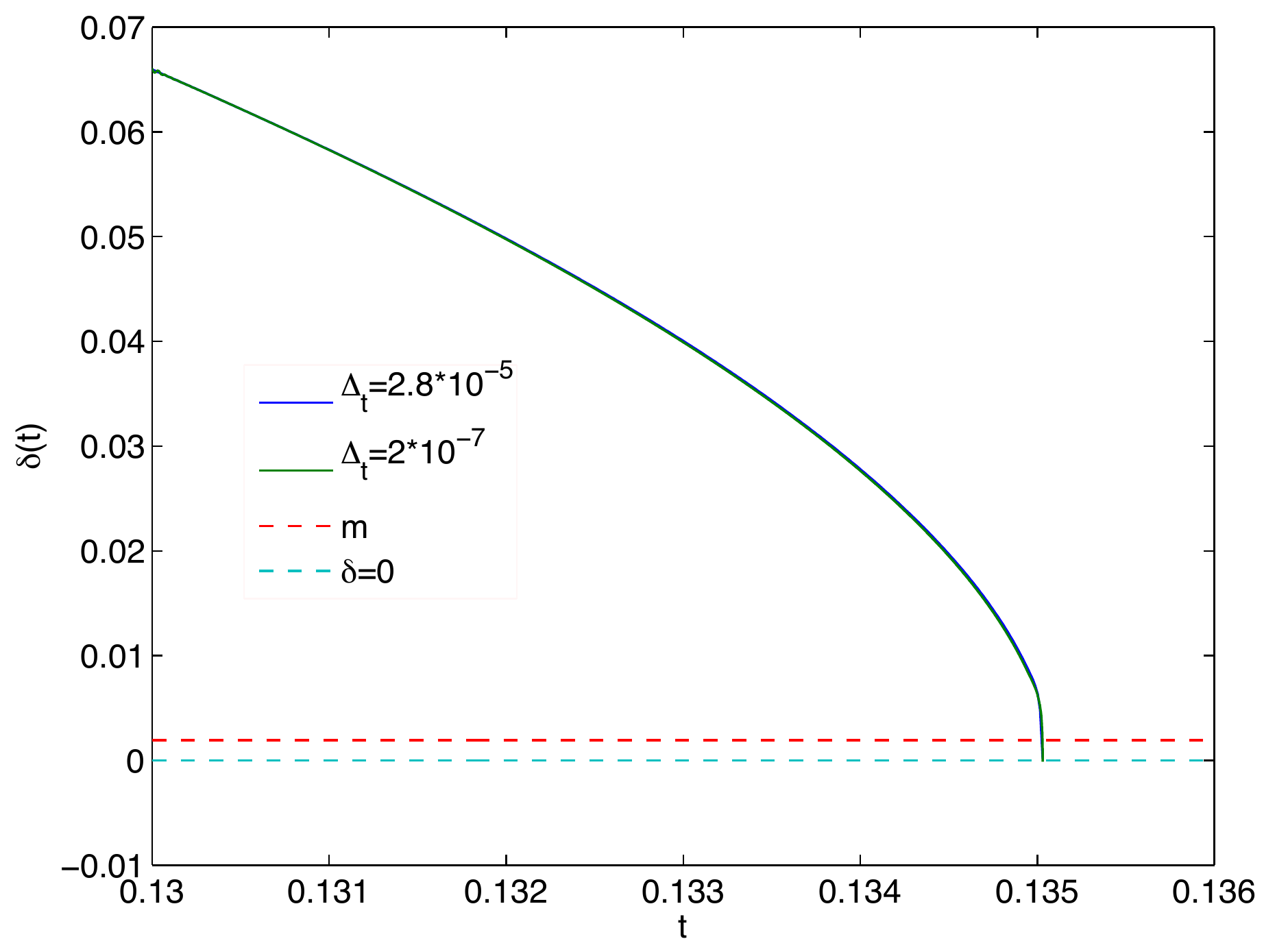} 
 \caption{Time dependence of the fitting parameter $\delta$ (\ref{abd})
 for $t$ close to the blow-up time $t^*$ for 
 the numerical solution to the focusing $1+1$-dimensional quintic NLS equation with $\epsilon=1$ and initial data $\Psi_{0} (x) =1.8i\exp(-x^{2})$.
The fitting  is done for  $5<k<\max(k)/2$. }
 \label{qexpdel}
\end{figure}

We study here two situations where a blow-up occurs in the solutions to this equation. First, we look at an example for $\epsilon=1$ studied in
 \cite{Sti} and in \cite{KRM} for the initial data 
$\Psi_{0}(x)=1.8i\exp(-x^{2})$ having  negative energy. 
Since we aim to study the semiclassical limit of the 
Davey-Stewartson system, where blow-up is expected to appear for 
special classes of initial data, see \cite{Roid1}, we also consider a 
typical example in the semiclassical limit to the quintic NLS 
equation with initial data of the form $\Psi(x,0)= \mbox{sech} \, x$ 
and $\epsilon=0.1$. 
\\
\\
In the first experiment, ($\Psi_{0}(x)=1.8i\exp(-x^{2})$, $\epsilon=1$),
the computation is carried out with $2^{14}$ Fourier modes for $x \in [-5\pi, 5\pi]$, and the fitting of the Fourier coefficients is done for $5<k<\max(k)/2$. 
 This case has been studied in \cite{Sti} and the blow-up time $t^{*}$ has been identified.
 Note however, that in this paper, they considered the following form of the quintic NLS equation, 
 \begin{equation}
     i\epsilon  \Psi_{t}+\epsilon^2 \Psi_{xx}+ |\Psi|^{4} 
    \Psi=0,
 \label{nlssti}
  \end{equation}
  i.e., the change of $t\to t/2$ in (\ref{quintNLS}). To compare our result with the ones of \cite{Sti}, we thus consider this form of the equation for the first experiment.  
With this time scale, the blow-up time $t^{*}$ has been identified in \cite{Sti} to be $t^* \sim 0.135$. 
  \\
  \\
We recover exactly this value from the fitting of the Fourier coefficients, see Fig. \ref{qexpdel},
where we show the time dependence of the fitting parameter $\delta$ (\ref{abd}).

\begin{figure}[htb!]
\centering
 \includegraphics[width=0.49\textwidth]{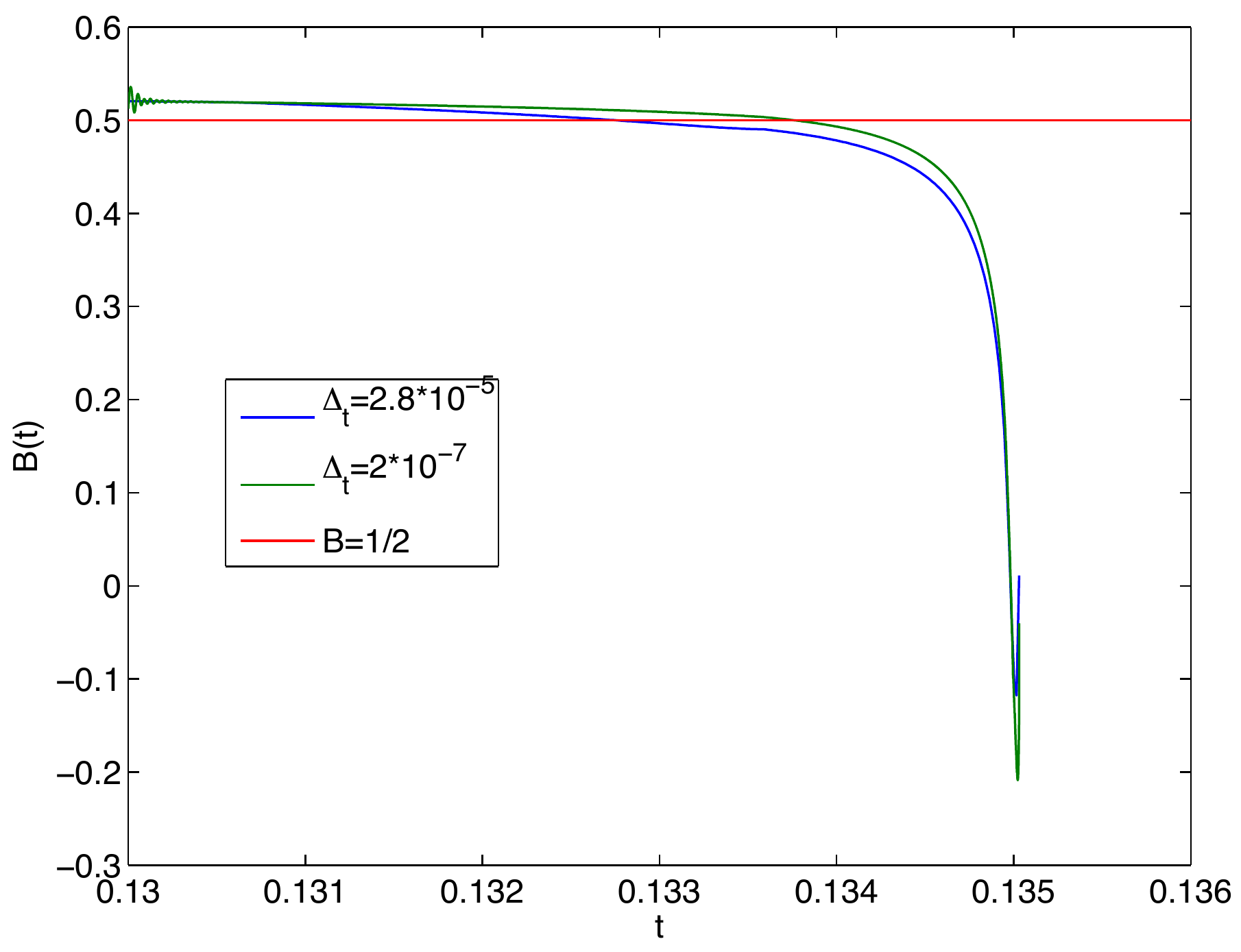} 
 \includegraphics[width=0.48\textwidth]{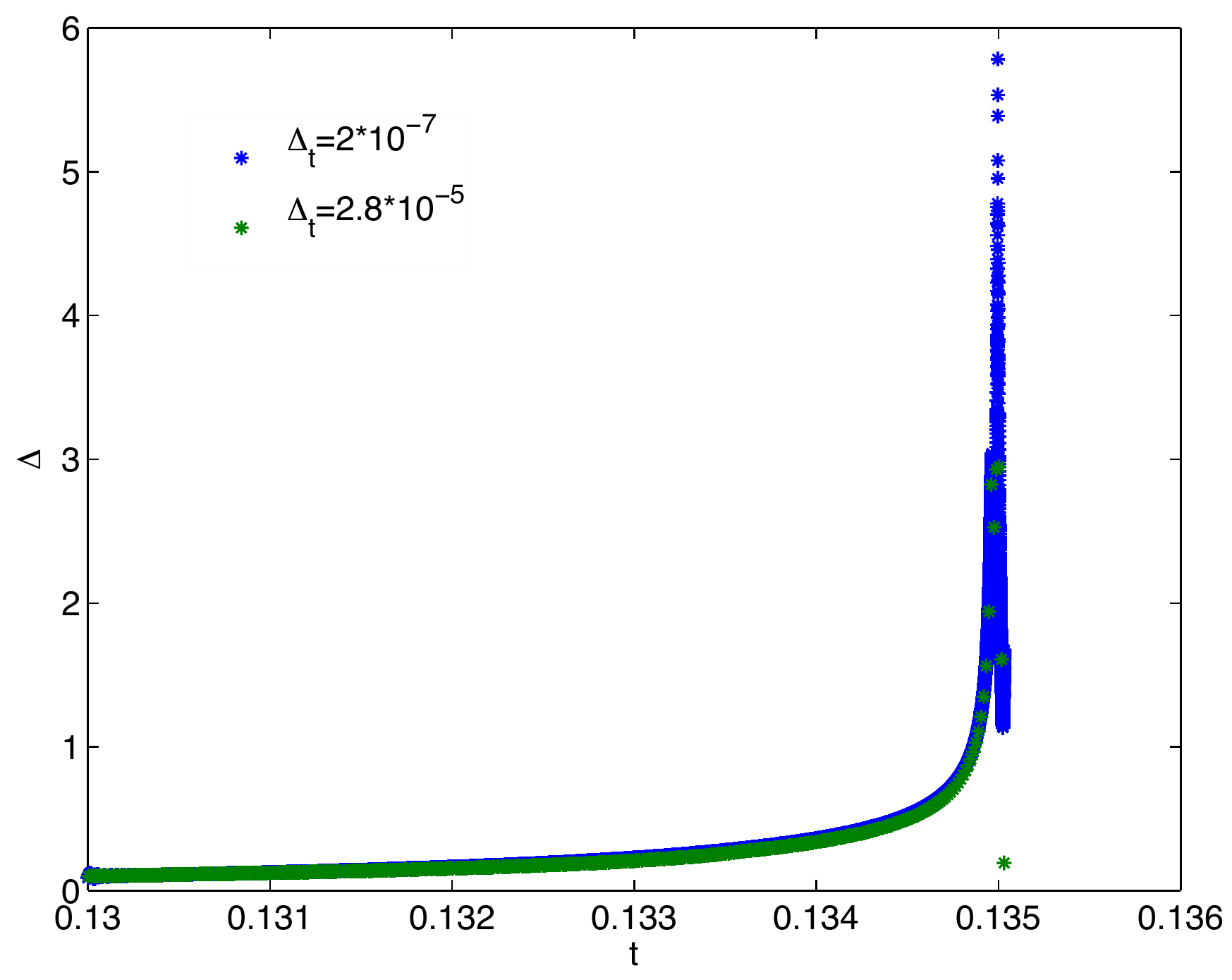} 
 \caption{Time dependence of the fitting parameter $B$ (\ref{abd}) and of the 
 fitting error $\Delta$ (\ref{abd}) for the numerical solution to the focusing 
 $1+1$-dimensional quintic NLS equation with $\epsilon=1$ and initial data $\Psi_{0} (x) =1.8i\exp(-x^{2})$.}
 \label{qexpalp}
\end{figure}

We can see that $\delta$ (\ref{abd}) decreases rapidly  as expected and vanishes 
at $t^{*}\sim0.135$.
In addition, we can roughly determine via the fitting parameter $B$ (\ref{abd})
that the singularity corresponds to a blow-up here.  
Indeed for $t$ approaching the blow-up time, the parameter $B$ stays close 
to $0.5$, corresponding to a $1/\sqrt{x}$ singularity as actually 
expected, before decreasing rapidly, see Fig. \ref{qexpalp}. The 
reason for this behavior is obviously the blow-up  which completely 
destroys the Fourier coefficients. This can be also seen from the 
fitting error which increases at the same time, see also Fig. \ref{qexpalp}. 
A mesh refinement close to the blow up time does not reduce this 
behavior, as we can infer from the previous pictures,
% in Fig. \ref{qexpmeshes}, 
 where we show the 
time evolution of $\delta$, $B$ (\ref{abd}) and of $\Delta$ for $t\sim t^{*}$ for 
two different mesh sizes. In the first case, $\Delta_t= 2.8*10^{-5}$, and in the second case, $\Delta_t= 2*10^{-7}$. In the latter case, 
the loss of precision in $\Delta$ is clearer, and the value of $B$ 
stays longer close to $0.5$. But it appears not worth to use such 
a high resolution, which is computationally expensive,  since it does 
not have an influence on the quantities as $t^{*}$ we are interested 
in, and this is even more true for the second experiment below. The fact 
that the results essentially do not change if higher resolution is used clearly 
confirms that blow-up occurs. But close to $t^{*}$, phenomena  as the 
rapid decrease of $B$ and the loss of precision of $\Delta$ cannot be avoided.
%\begin{figure}[htb!]
%\centering
% \includegraphics[width=0.32\textwidth]{qNLS/del12meshqexp.pdf} 
% \includegraphics[width=0.32\textwidth]{qNLS/alp12meshqexp.pdf} 
% \includegraphics[width=0.32\textwidth]{qNLS/erfit2meshqexp.pdf} 
% \caption{Time dependence of the fitting parameters $\delta$ and $B$ 
% and of $\Delta$ for $t$ close to $t^*$ for the numerical solution to 
% the focusing $1+1$-dimensional quintic NLS equation with $\epsilon=1$ and initial data $\Psi_{0} (x) =1.8i\exp(-x^{2})$. }
% \label{qexpmeshes}
%\end{figure}

\begin{figure}[htb!]
\centering
 \includegraphics[width=0.32\textwidth]{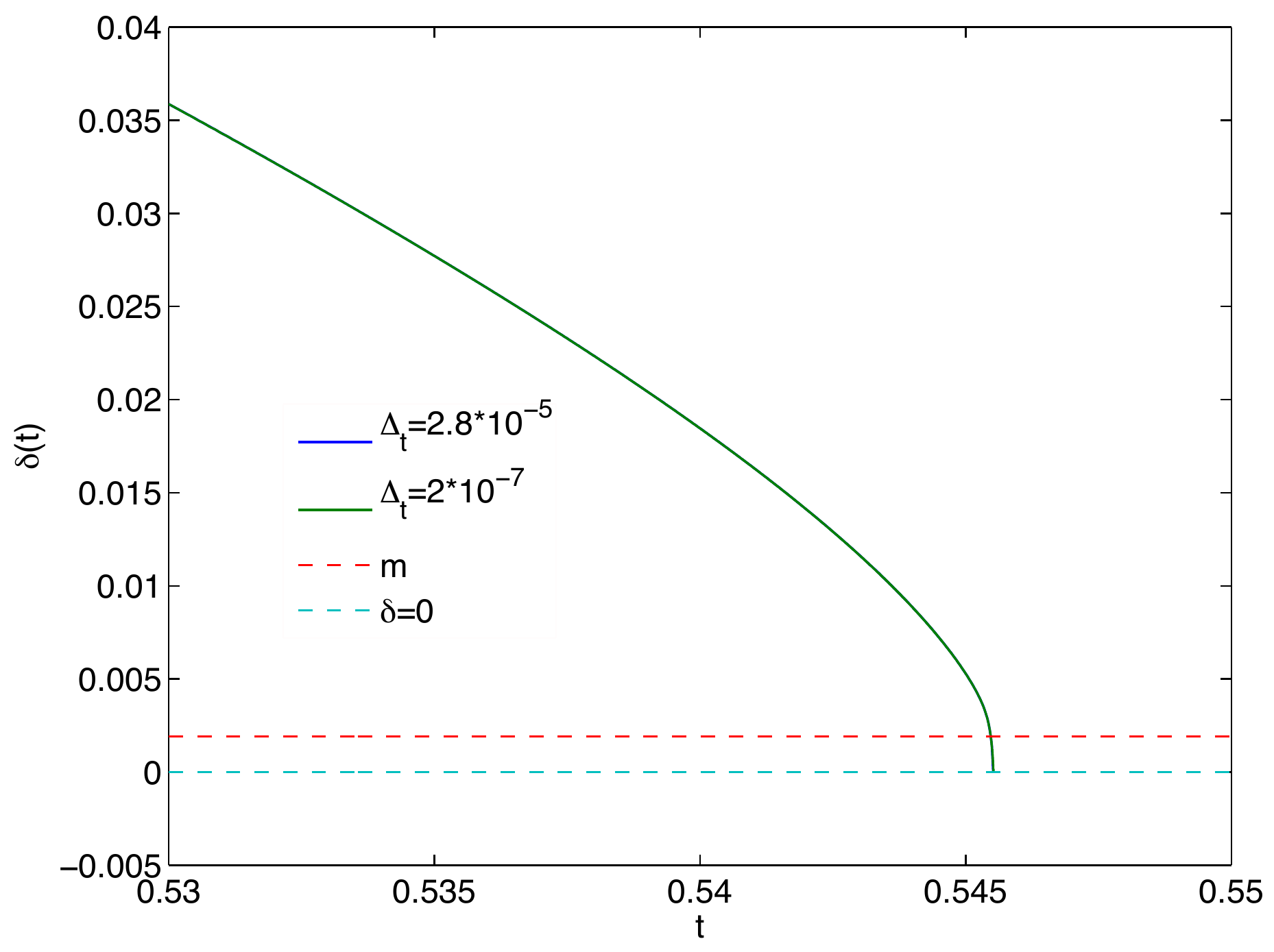} 
\includegraphics[width=0.32\textwidth]{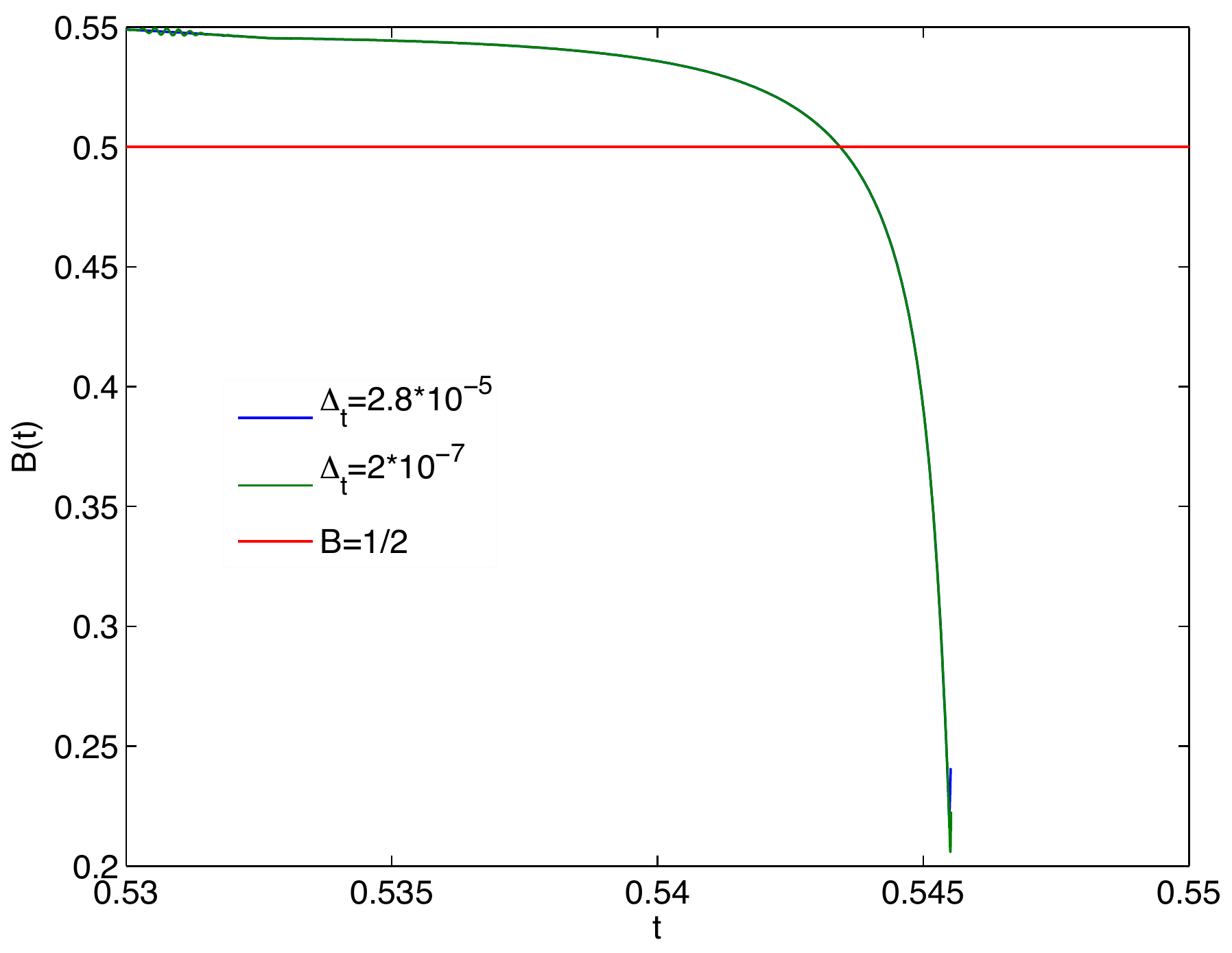} 
 \includegraphics[width=0.32\textwidth]{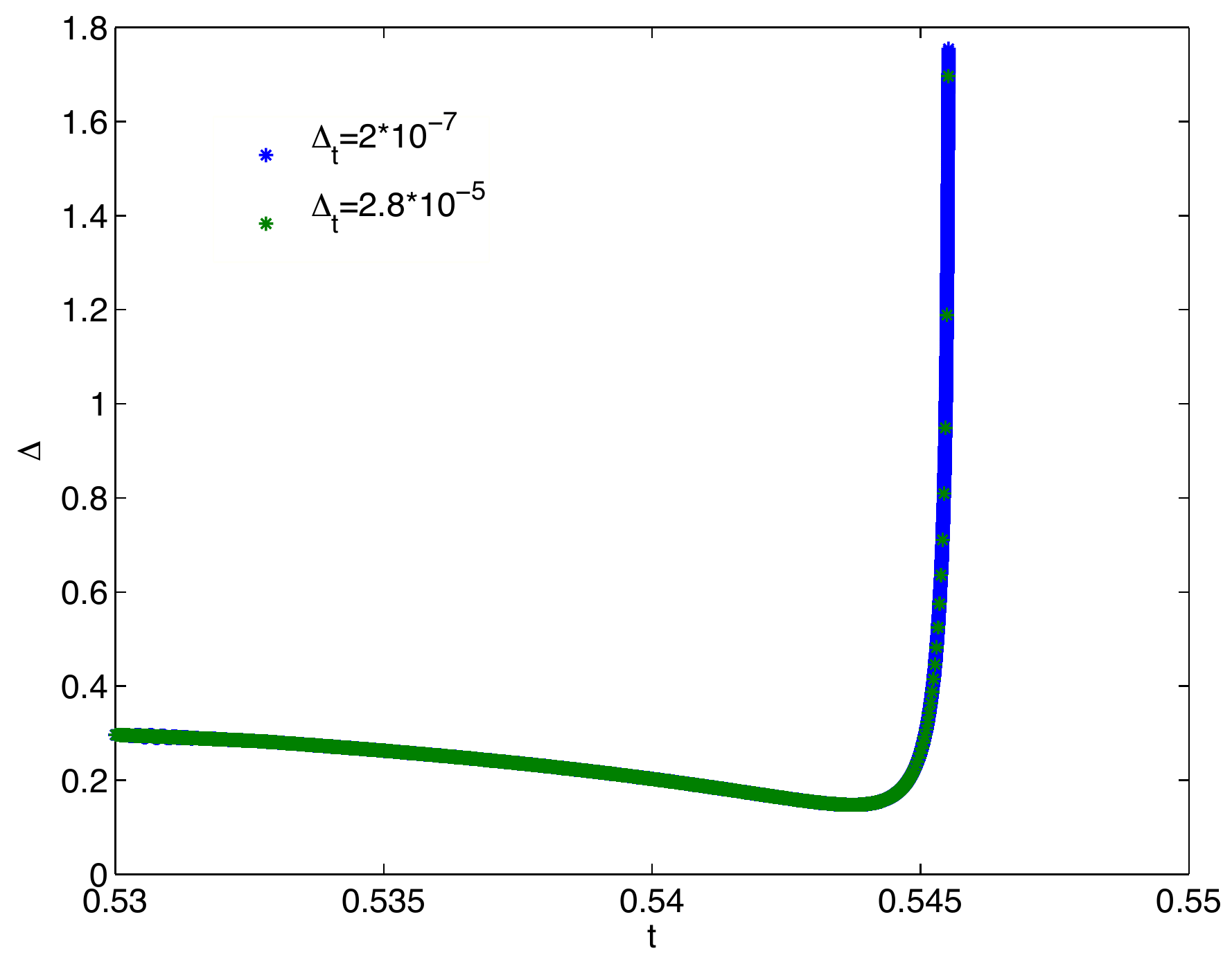} 
 \caption{Time dependence of the fitting parameters $\delta$ and $B$ 
(\ref{abd}) and of the fitting error $\Delta$ for $t$ close to $t^*$ for the numerical solution 
 to the focusing $1+1$-dimensional quintic NLS equation with $\epsilon=0.1$ 
 and initial data $\Psi_{0} (x) =\mbox{sech} \, x$. The fitting  is done for  $5<k<\max(k)/2$}
 \label{qsechmeshes}
\end{figure}

The situation is rather similar for the experiment in the 
semiclassical limit. We consider here initial data of the form 
$\Psi(x,0)= \mbox{sech} \, x$, $\epsilon=0.1$ and $2^{15}$ Fourier 
modes for 
$x \in [-10\pi, 10\pi]$. We find that the blow-up occurs at $t^* \sim 
0.5455$ as already observed in \cite{DGK13}. We show the time evolution of 
the fitting parameters in Fig. \ref{qsechmeshes} together with the time 
evolution of the fitting error $\Delta$, again for two different mesh sizes.
%, 
%mesh1: $\Delta_t=2.75*10^{-5}$ and mesh2: $\Delta_t=2*10^{-7}$.
%\begin{figure}[htb!]
%\centering
% \includegraphics[width=0.32\textwidth]{qNLS/del1qsech.pdf} 
% \includegraphics[width=0.32\textwidth]{qNLS/alp1qsech.pdf} 
%  \includegraphics[width=0.32\textwidth]{qNLS/errfitqsech.pdf} 
%\caption{Time dependence of the fitting parameters $\delta$ and $B$, 
%and of the fitting error $\Delta$ for the numerical solution to the 
%focusing $1+1$-dimensional quintic NLS equation with $\epsilon=0.1$ and initial data $u_{0} (x) =\mbox{sech} \, x$.
%The fitting  is done for  $5<k<\max(k)/2$. }
% \label{qsechfitpara}
%\end{figure}
The parameter $B$ (\ref{abd}) once more decreases rapidly close to the blow-up time, 
and the fitting error increases there. In this case, a 
mesh refinement close to $t^*$ does not have any influence.
%, see Fig. \ref{qsechmeshes}, 
%where we considered: mesh1: $\Delta_t=2.75*10^{-5}$ and mesh2: $\Delta_t=2*10^{-7}$. 

In both cases, we find, however, that the critical time can be 
roughly recovered from the fitting of the Fourier coefficients. The 
type of the singularity, here a blow-up, can also be identified
via the value of $B$ close to $0.5$ before the blow-up time. 
Typically, a fitting error smaller than $0.5$ can be achieved before 
the blow-up time, where the latter diverges. This  is a noticeable 
difference to  the case of a  singularity with finite $L_{\infty}$ 
norm (see section 3, and also \cite{dkpsulart}), where we did not 
observe such a behavior.

\subsection{Symmetric initial data for the DS II equation}

\begin{figure}[htb!]
\centering
 \includegraphics[width=0.49\textwidth]{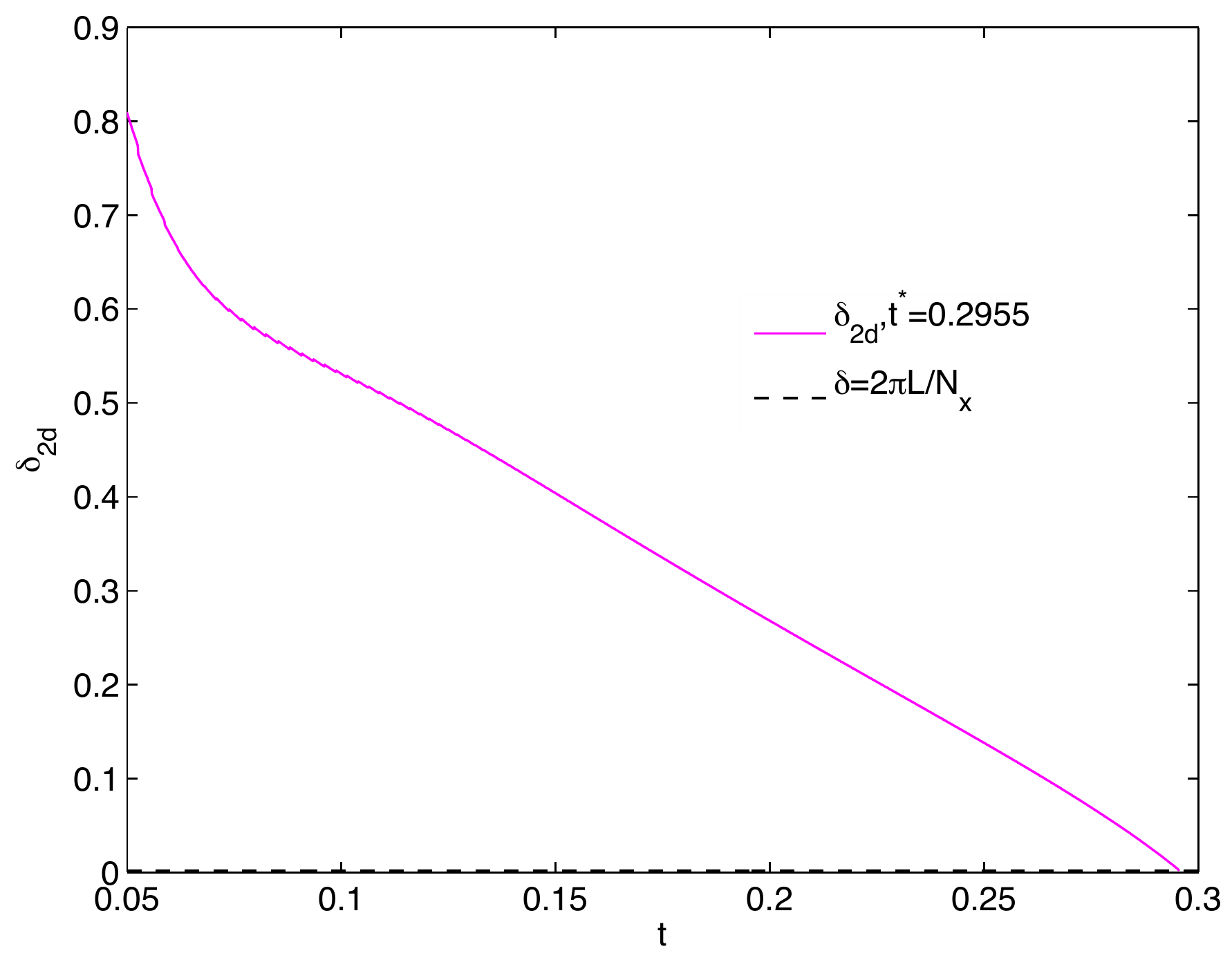} 
  \includegraphics[width=0.49\textwidth]{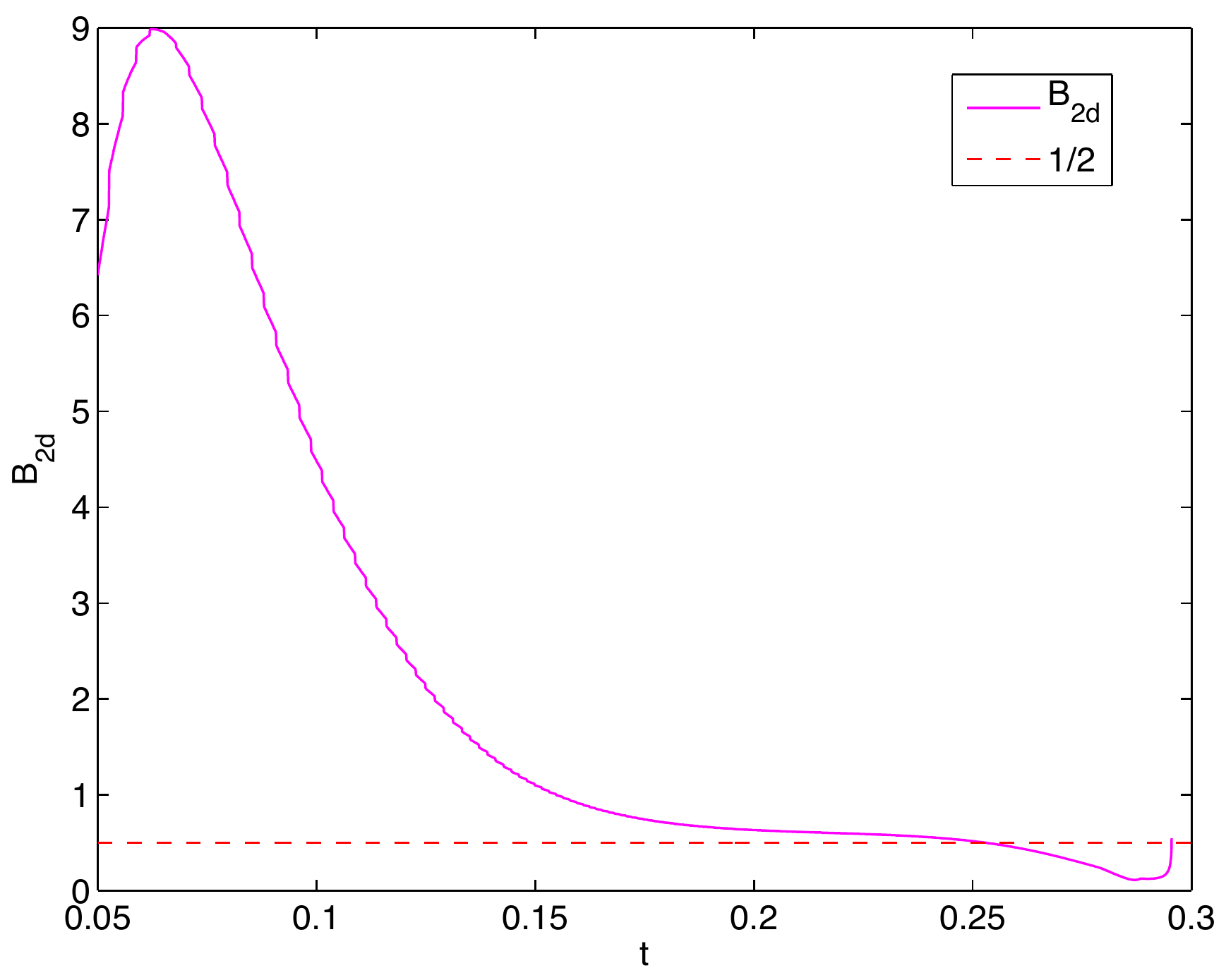} 
\caption{Time dependence of the fitting parameters $\delta_{2d}$ and $B_{2d}$ in (\ref{abd2}) for the solution of the focusing DS II equation with initial data (\ref{adinis}) and $\epsilon=0.1$.}
 \label{delfocblow}
\end{figure}

We now consider symmetric (with respect to an interchange of $x$ and 
$y$) initial data of the form 
\begin{equation}
\Psi_0(x,y) = e^{-R^{2}}, \,\,\mbox{with}\,\, R = \sqrt{ x^{2}+y^{2} },
\label{adinis}
\end{equation}
for the DS II equation, i.e., the second case studied in Sec. 3.2.2, and $\epsilon=0.1$.
As we will see, the behavior of the solutions for this initial condition is similar to the 
solutions of the $1+1$-dimensional quintic NLS equation in the
semiclassical limit above, i.e., a blow-up occurs. 
\\
The computation is carried out with $2^{14} \times 2^{14}$ Fourier 
modes for  $ x \times y \, \in \, [-5\pi,5\pi]\times[-5\pi,5\pi]$, and $\Delta_t=4.5*10^{-5}$.
By performing a two-dimensional fit of the Fourier coefficients (\ref{abd2}), we find 
that the solution develops a singularity at time $t^*=0.2955$, see 
Fig.~\ref{delfocblow}. The behavior of $B_{2d}$ is shown in the same 
figure.         

\begin{figure}[htb!]
\centering
 \includegraphics[width=0.49\textwidth]{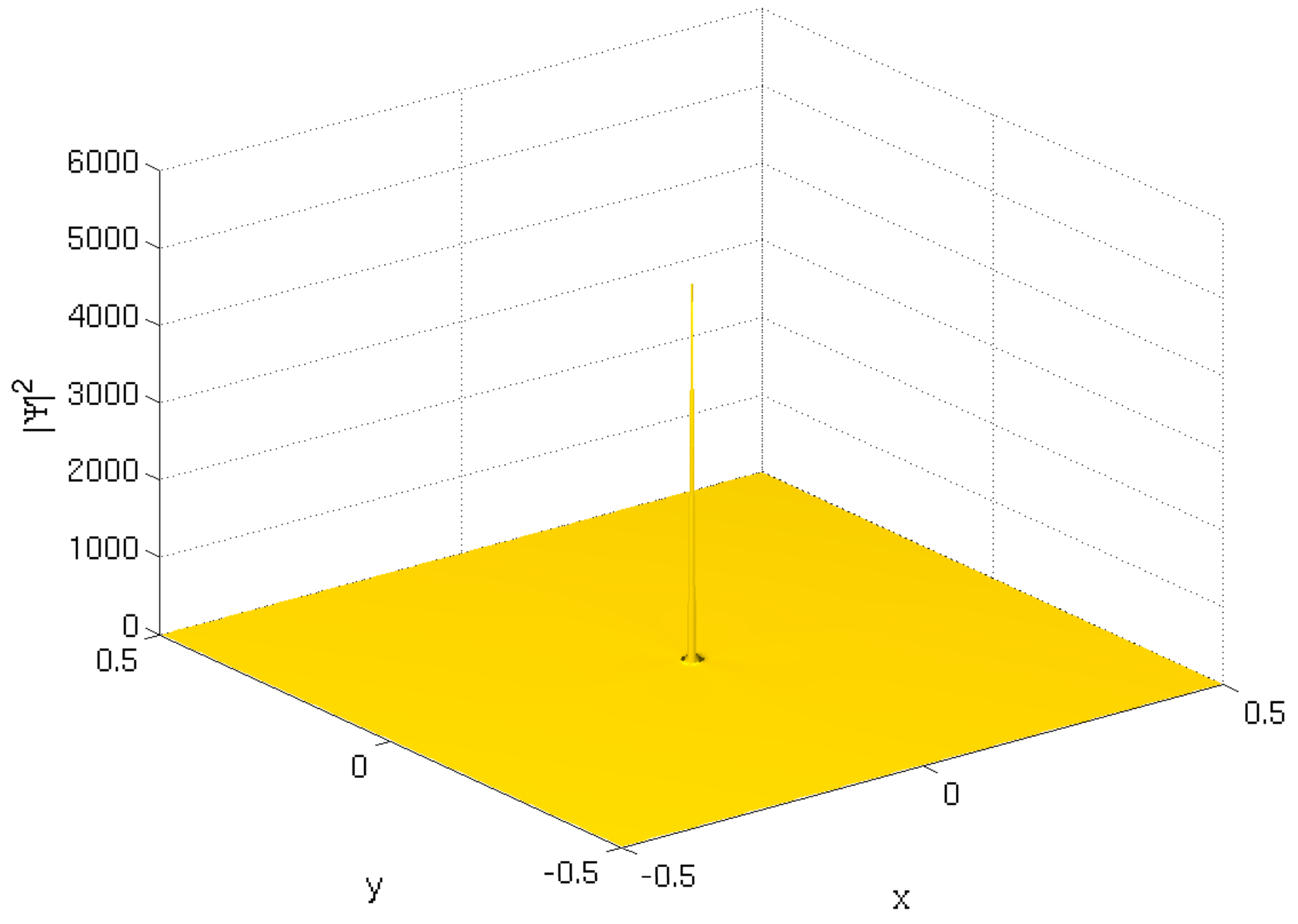} 
  \includegraphics[width=0.49\textwidth]{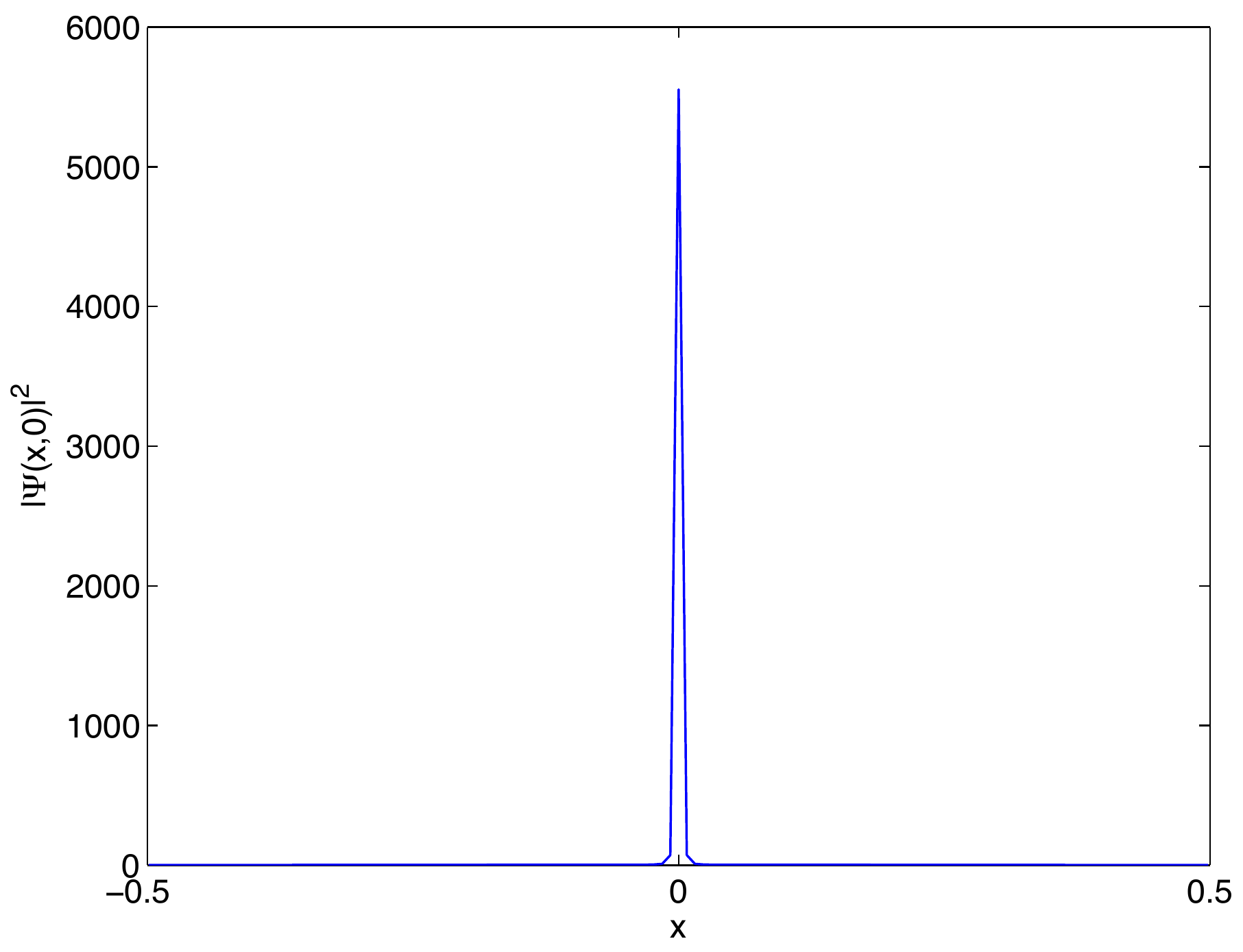} 
 \caption{Square of the absolute value of the solution of the focusing DS II equation with initial data (\ref{adinis}) and $\epsilon=0.1$ at $t=0.2955$.}
 \label{focdssdsyme01}
\end{figure}

\begin{figure}[htb!]
\centering
 \includegraphics[width=0.49\textwidth]{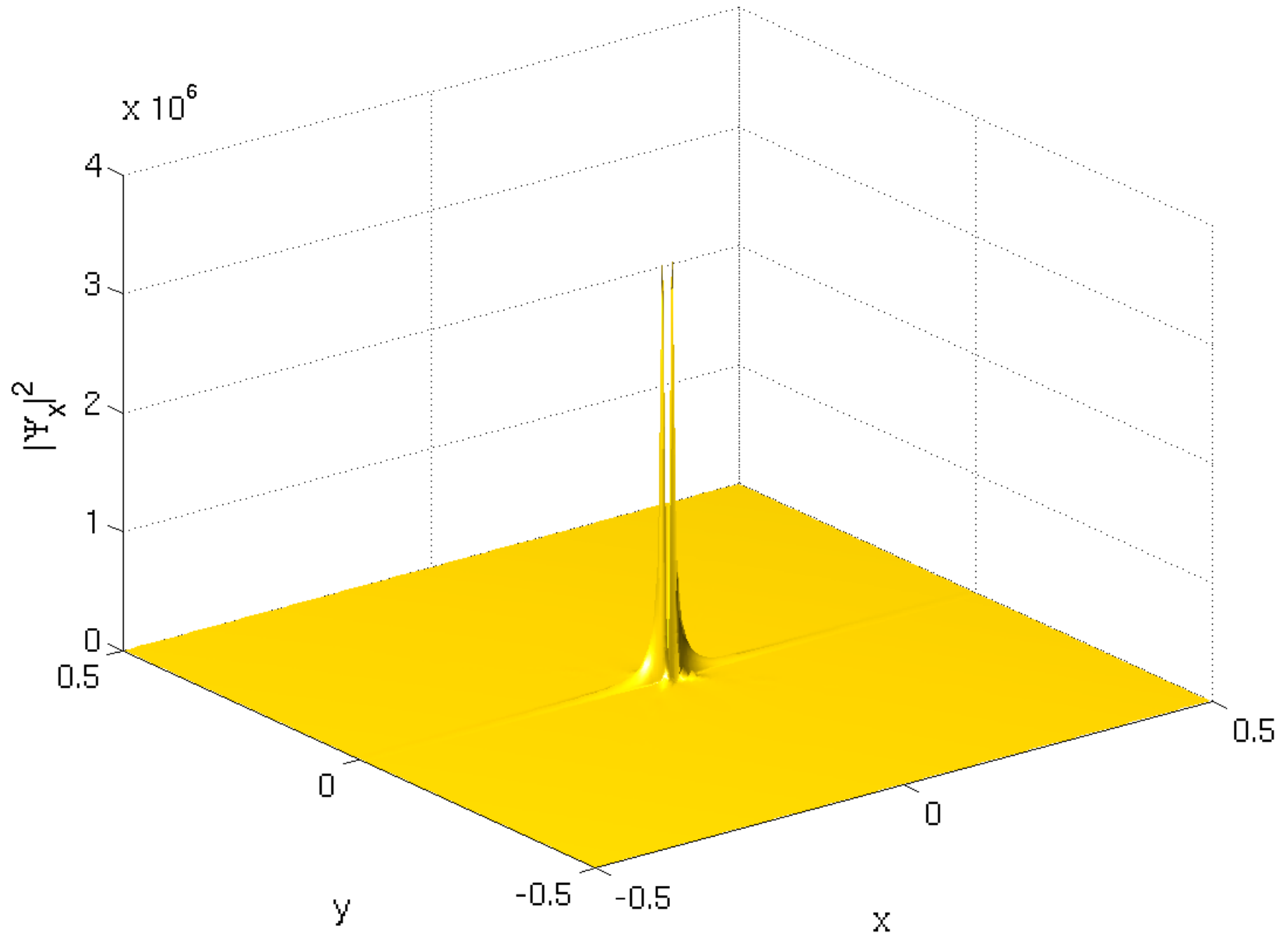} 
  \includegraphics[width=0.49\textwidth]{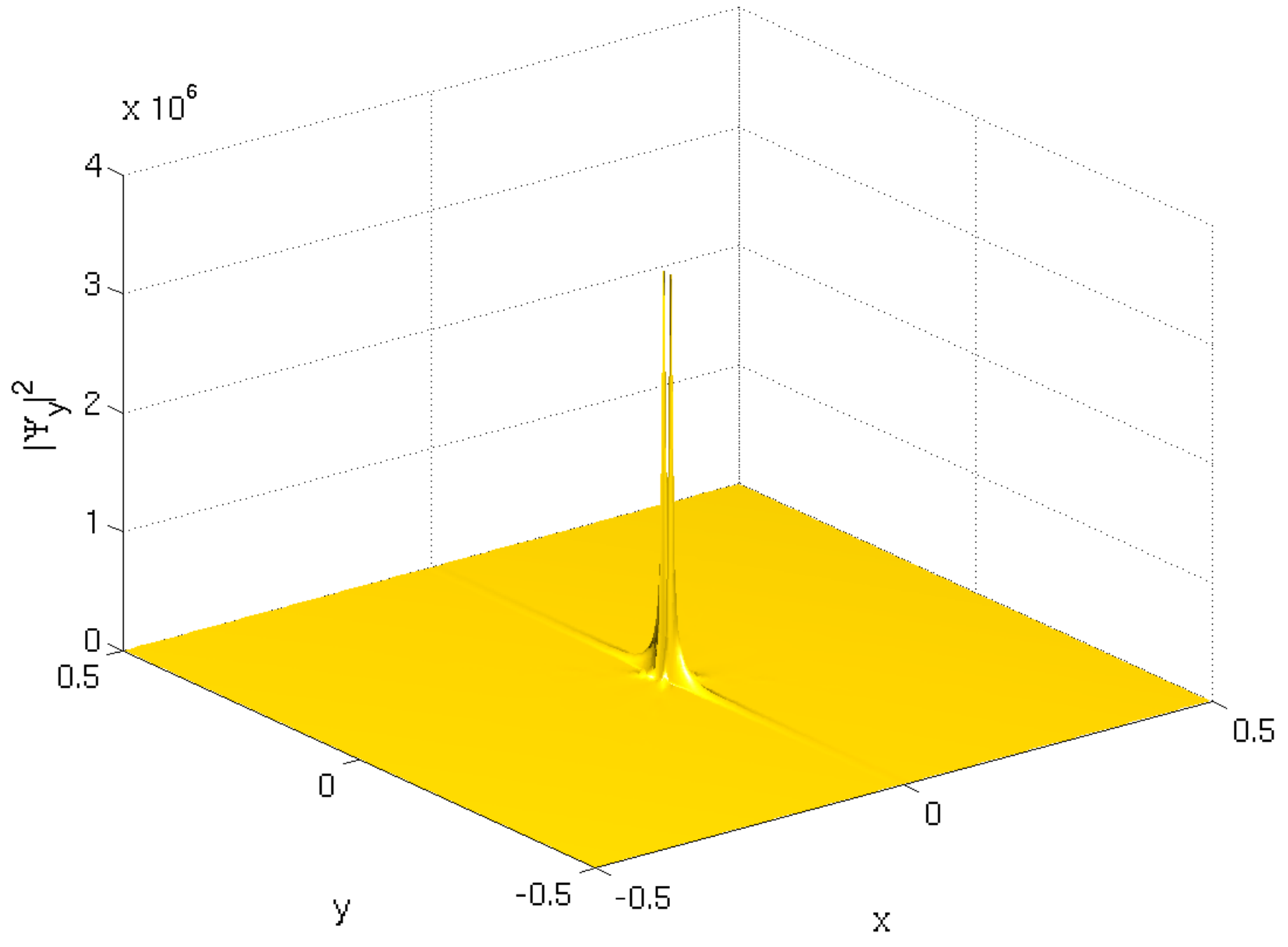} 
 \caption{Derivatives of the solution of the focusing DS II equation with initial data (\ref{adinis}) and $\epsilon=0.1$ at $t=0.2955$.}
 \label{focdssdsyme01d}
\end{figure}

The profile of  the solution at this 
time in  Fig. \ref{focdssdsyme01} also clearly indicates an 
$L_{\infty}$ blow-up, with $\|u\|_{\infty} \sim 65$, and 
this is also confirmed by the derivatives of $u$ in Fig.~\ref{focdssdsyme01d}, with 
$\|u_x\|_{\infty} \sim \|u_y\|_{\infty} \sim 120$. 

%To make this even more explicit, we show the $L_{\infty}$ norms of $\Psi$, $\Psi_x$ and $\Psi_y$ 
%in dependence of time in Fig.~\ref{ampls}.
%\begin{figure}[htb!]
%\centering
% \includegraphics[width=0.49\textwidth]{DSSD/ulinfbeta1e01.pdf} 
%  \includegraphics[width=0.49\textwidth]{DSSD/uxlinfbeta1e01.pdf} 
% \caption{Time evolution of $\|\Psi\|_{\infty}$ (left) and of 
% $\|\Psi_x\|_{\infty}$ (right) for the solution $\Psi$ of the focusing DS II equation with initial data (\ref{adinis}) and $\epsilon=0.1$. }
% \label{ampls}
%\end{figure}

\begin{figure}[htb!]
\centering
 \includegraphics[width=0.49\textwidth]{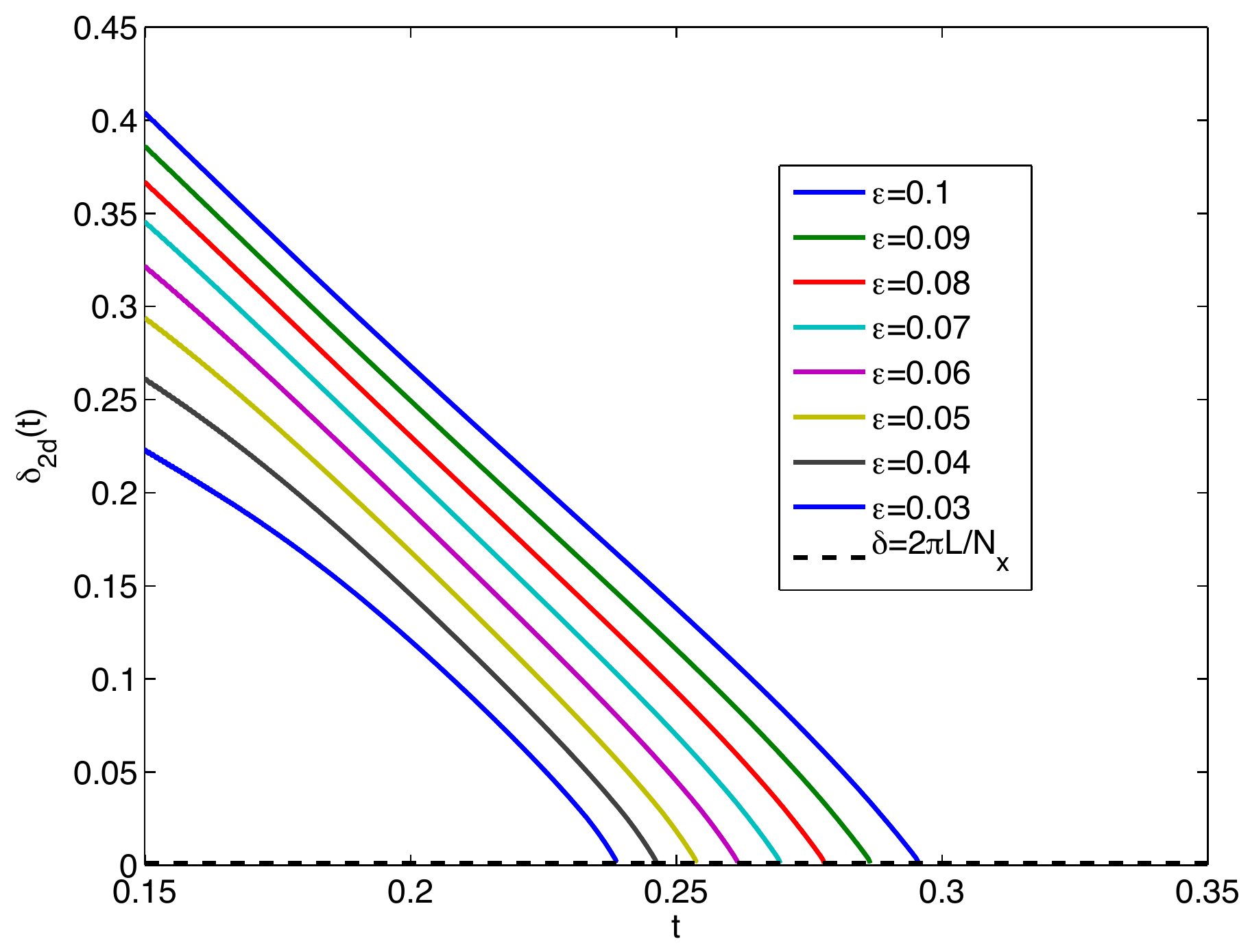} 
  \includegraphics[width=0.49\textwidth]{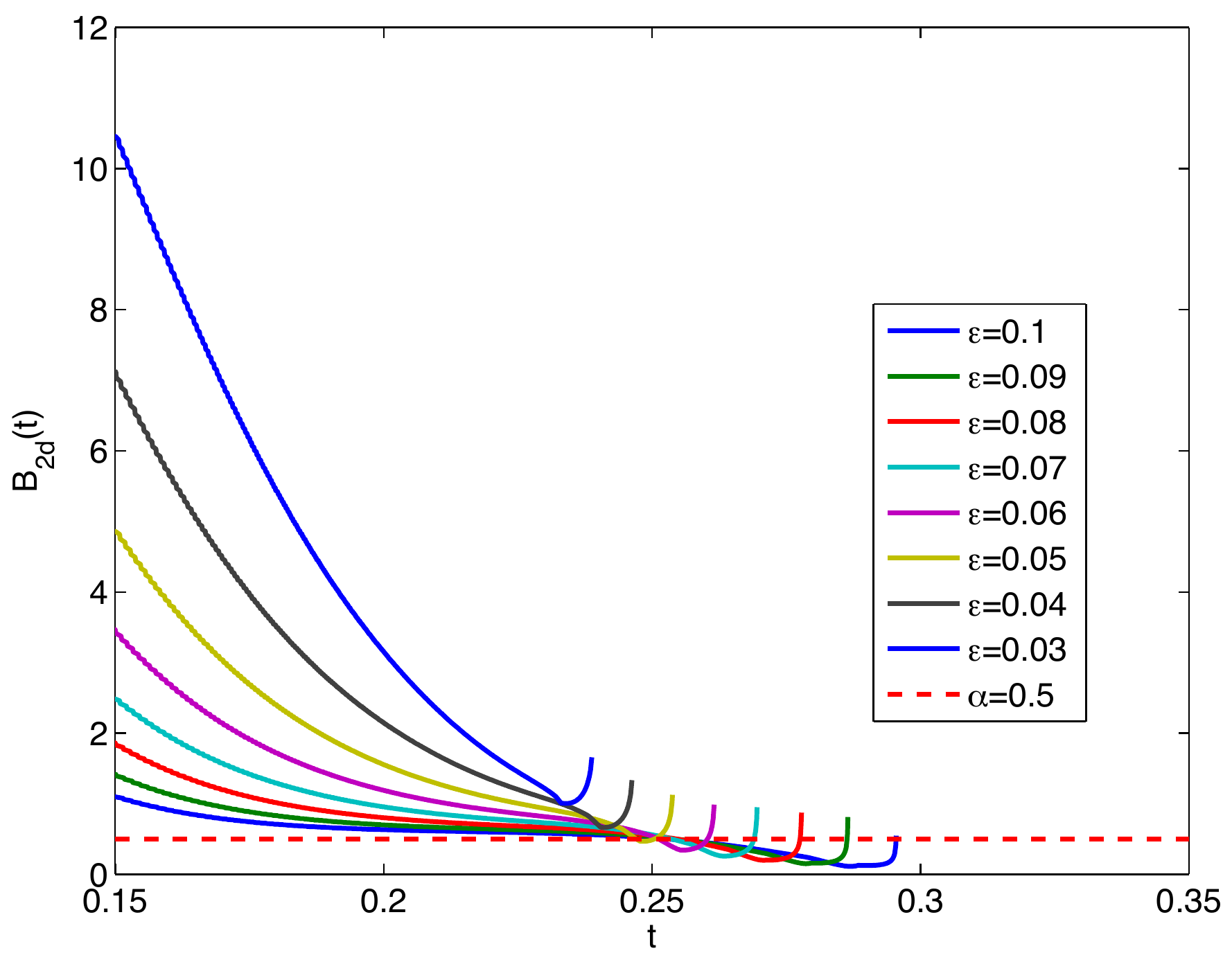} 
 \caption{Time dependence of the fitting parameters $\delta_{2d}$ and 
 $B_{2d}$ (\ref{abd2}) for the solution of the focusing DS II equation with 
 initial data (\ref{adinis}) for several values of $\epsilon$.}
 \label{deleps}
\end{figure}

As $\epsilon$ decreases, the blow-up time decreases as well, see Fig. 
\ref{deleps}, where we show the time dependence of the fitting parameters 
$\delta_{2d}$ and $B_{2d}$ (\ref{abd2}) for several values of $\epsilon \in [0.03, \, 0.1]$.

\begin{figure}[htb!]
\centering
 \includegraphics[width=0.4\textwidth]{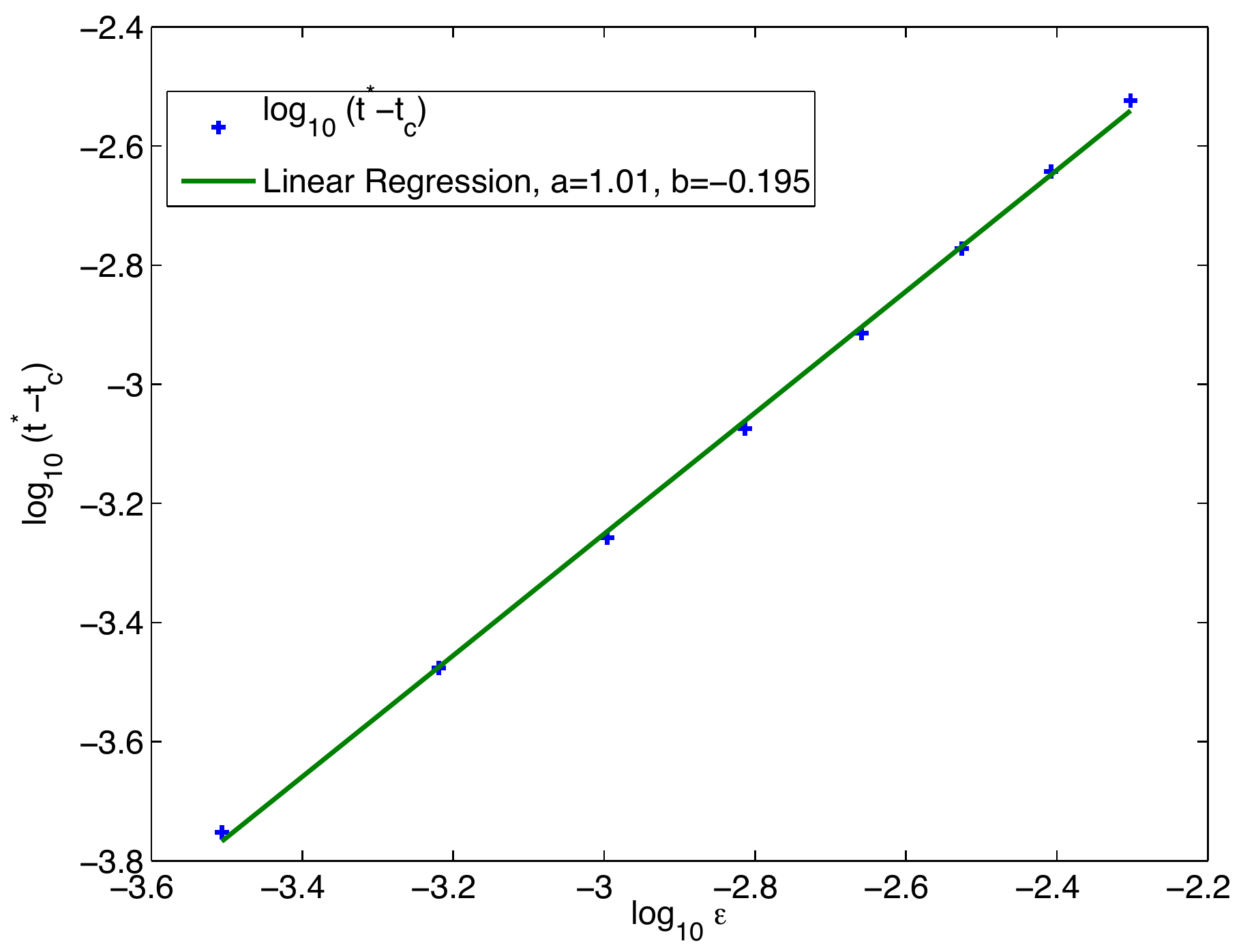} 
 \caption{Difference between the blow-up time $t^*$ and the 
 corresponding break-up time of the semicalssical DS II system $t_c=0.2153$ in dependence of $\epsilon$}
%  on the left, and
% $L_{\infty}$-norm of the difference between the focusing DS II and focusing sDS solution 
% at $t_c=0.2153$ in dependence of $\epsilon$ on the right.}
 \label{scalepsse1}
\end{figure}

The difference between the blow-up time $t^*$ and the corresponding 
break-up time of the dispersionless system $t_c=0.2153$ scales roughly  as $\epsilon$, see Fig. \ref{scalepsse1} (left). Indeed, a linear regression analysis, 
 ($\log_{10} (t^* -t_c) = a \log_{10} \epsilon + b$) gives 
\begin{equation*}
 a=1.0189,  \,\, b=  -0.1948 \,\,\mbox{and} \,\,  r= 0.999.
\end{equation*}
Since the determination of the blow-up time is numerically delicate, 
it is difficult to decide whether this scaling is really different 
from the $\epsilon^{4/5}$ scaling observed for the blow-up in 
$1+1$-dimensional NLS solutions. 

%\begin{figure}[htb!]
%\centering
% \includegraphics[width=0.4\textwidth]{DSSD/scalfoc1.pdf} 
% \caption{$L_{\infty}$-norm of the difference between the focusing DS II and focusing sDS solution 
% at $t_c=0.2153$ in dependence of $\epsilon$.}
% \label{scalfoce1}
%\end{figure}

\section{Conclusion}

In this paper we have shown that important information on the 
semiclassical limit of the DS II system can be obtained numerically. 
We considered localized initial data and used the asymptotic behavior 
of the Fourier coefficients to identify the points of gradient 
catastrophe in the semiclassical DS II system. This approach was 
shown to be very efficient for the defocusing semiclassical NLS in 
$1+1$ dimensions, and within a few percent accuracy for the focusing case.

In both the defocusing 
and focusing case we observe a \textit{hyperbolic blow-up}, i.e., a 
gradient catastrophe at points, where the solution stays finite. In 
the defocusing case, we find as for the defocusing semiclassical 
system of the NLS in $1+1$ dimensions a cubic behavior. This means 
that in our examples, there are four break-up points which due to 
the symmetry of the initial data were located on the coordinate 
axes. The break-up at each singular point is such that only one of 
the derivatives blows up, whereas the other stays finite (here for 
symmetry reasons, either the $x$- or the $y$-derivative). In the 
focusing semiclassical DS II system, the break-up is again similar to 
the $1+1$-dimensional focusing semiclassical NLS system. It appears 
to be  a square root type break-up. For generic initial data, 
only one of the derivatives (for symmetry reasons they coincide here 
again with the $x$ and $y$ derivatives) blows up, whereas the other 
stays finite. But for data with a symmetry with respect to an 
exchange of $x$ and $y$, these break-ups can happen at the same time 
and location. 

In a second step we solved the DS II equation for some finite small 
$\epsilon$ for the same initial data as before up to the previously 
identified critical time. We found that the difference between the 
semiclassical DS II solution $u_{SC}$, and the DS II system solution $u_{DS}=|\Psi|^2$ shows the same scaling in 
$\epsilon$ as the corresponding $1+1$-dimensional NLS solutions for 
which an asymptotic description was conjectured in \cite{DGK13}. This 
means we have in the defocusing case $$|u_{DS}-u_{SC}|_{|x-x_{c}|\ll 1}= 
\mathcal{O}(\epsilon^{2/7}),$$
and in the focusing case:
$$|u_{DS}-u_{SC}|_{|x-x_{c}|\ll 1}= 
\mathcal{O}(\epsilon^{2/5}).$$

Since in \cite{DGK13} an asymptotic description of NLS 
solutions in particular (the conjecture actually applies to a much larger 
class of equations) based on Painlev\'e transcendents was given, it 
is an interesting question whether the latter also play a role in 
this context. This will be the subject of further research. 

We also studied solutions to the DS II system for times much larger 
than the critical time $t_{c}$ of the corresponding semiclassical DS 
II 
system. It was found that generically dispersive shocks appear as in 
the case of the $1+1$-dimensional NLS equations which were documented 
in this paper for the first time. No asymptotic 
description of these shocks has been given so far, but we hope that 
our results stimulate analytical activies in this field. The 
numerical results clearly indicate that cusped zones appear which for 
small $\epsilon$ will delimit the oscillations. A first analytic 
progress in the asymptotic description of dispersive shocks in DS II 
solutions would be to determine the boundary of these zones. However such 
shocks were not observed for the focusing DS II system
for initial data with a symmetry with 
respect to the exchange of the spatial coordinates. In this case the 
break-up in the semiclassical DS II system happens in both coordinates at 
the same time and place. For small 
$\epsilon$, the corresponding DS II solution has a strong peak at the 
critical point $(x_{c},y_{c},t_{c})$ of the semiclassical DS II  system and  
continues to grow for $t>t_{c}$ up to a time $t^{*}$, where a blow-up 
is observed. We presented a careful study of this case also based on 
an asymptotic analysis of the Fourier coefficients. It indicates the 
same kind of blow-up known from the quintic NLS in $1+1$ dimensions 
which has the critical nonlinearity to allow blow-up for this dimension. 
Note that the type of  blow-up is different from 
the hyperbolic blow-up in the semiclassical DS II system. Here we clearly 
have an $L_{\infty}$ blow-up. 

As already mentioned, the reason for 
the blow-up appears to be the symmetry of the initial data with 
respect to the interchange of $x$ and $y$, a symmetry the equation 
(\ref{DSnonlocal}) visibly has as well if at the same time $\Psi$ is 
replaced by $\bar{\Psi}$. Note that due to the different dynamics 
between DS and NLS due to the operator $\mathcal{D}_{-}$ in 
(\ref{DSnonlocal}), the blow-up in DS systems is much less understood 
than in the latter. The only known criterion for DS is due to Sung 
\cite{Sun}, see Theorem~\ref{theosung}. Note that the Sung condition 
is not satisfied for any of the initial data for the focusing DS II 
we study, also for the cases, where we observe dispersive 
shocks and no blow-up. Thus the Sung criterion does not appear to be 
optimal, and an interesting question is what such criterion could be.

\section*{Acknowledgments}
We thank B.~Dubrovin, E.~Ferapontov and T.~Grava for helpful remarks and hints. This work has been supported by the project FroM-PDE funded by the European
Research Council through the Advanced Investigator Grant Scheme, 
the ANR via the program ANR-09-BLAN-0117-01, and the Austrian Science Foundation FWF, project SFB F41 (VICOM) and project I830-N13 (LODIQUAS). 
We are grateful for access to the HPC resources from GENCI-CINES/IDRIS (Grant 2013-106628) on which part of the computations in this paper has been done, the CRI (Centre de Ressources Informatiques) of the university of Bourgogne, and to the Vienna Scientific Cluster (VSC).

\bibliographystyle{plain}
\bibliography{bibliof}{}

\end{document}